\documentclass[11pt,a4paper,footinclude=true,oneside,headinclude=true]{report}
\usepackage{amsmath}
\usepackage{subcaption}
\usepackage{graphicx}
\usepackage{tabularx}
\usepackage{booktabs}
\usepackage{bold-extra}
\usepackage{xcolor}
\usepackage{float}
\usepackage{url}
\usepackage{titlesec}
\usepackage{hyperref}
\usepackage[a4paper,margin=2cm,headsep=24pt]{geometry}
\usepackage{fancyhdr}
\usepackage{afterpage}
\usepackage{listings}
\usepackage{longtable}
\usepackage{multirow}

\titlespacing{\section}{0pt}{7pt}{2pt}
\titlespacing{\subsection}{0pt}{5pt}{2pt}
\usepackage{tikz}
\usetikzlibrary{positioning}
\usetikzlibrary{arrows}
\usetikzlibrary{fmbepics}
\usetikzlibrary{automata}
\usetikzlibrary{fit}
\usetikzlibrary{decorations.pathreplacing}
\usetikzlibrary{backgrounds}
\usetikzlibrary{calc}
\usetikzlibrary{shapes}
\usetikzlibrary{shadows}

\begin{document}

\begin{titlepage}

    \centering
    \includegraphics[width=0.7\linewidth]{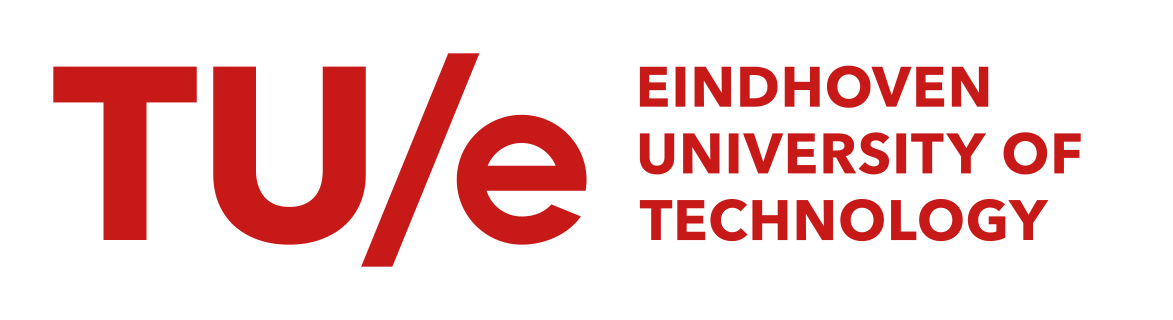}\par
    Department of Mechanical Engineering\\
    Control Systems Technology
    
    \vspace{1.5cm}
    {\LARGE\textbf{Development of a 3D Digital Twin of}}\\ \vspace{0.1cm}
    {\LARGE\textbf{the Swalmen Tunnel in the}}\\ \vspace{2mm} {\LARGE\textbf{Rijkswaterstaat Project}}\par\vspace{1cm}
    {\large\textit{Bachelor End Project Report}}\par\vspace{1.5cm}
    {\large{CST 2021.036}}\par\vspace{1cm}
    {\large Jeroen van Hegelsom}\par
    {\large 1370936}\par
      
    \vfill
    
    \emph{Supervisors}\par \vspace{3mm}
    dr. ir. J.M. van de Mortel-Fronczak \\ \vspace{1mm}
    ir. L. Moormann \\ \vspace{1mm}
    dr. ir. D.A. van Beek \\ \vspace{1mm}
    prof. dr. ir. J.E. Rooda \\ \vspace{1mm}
    \vspace{2cm}
    
    June 30, 2021

\end{titlepage}

\begin{abstract}
In an ongoing project, a cooperation between the TU/e and the Dutch Department of Waterways and Public Works (``Rijkswaterstaat" in Dutch, abbreviated to RWS) is established. The project focuses on investigating applicability of synthesis-based engineering in the design of supervisory controllers for bridges, waterways and tunnels. Supervisory controllers ensure correct cooperation between components in a system. The design process of these controllers partly relies on simulation with models of the plant (the physical system). A possible addition to this design process is digital twin technology. A digital twin is a virtual copy of a system that is generally much more realistic than the 2D simulation models that are currently used for supervisory controller validation.\\\\
In this report, the development of a digital twin of the Swalmen tunnel that is suitable for supervisory control validation is described. The Swalmen tunnel is a highway tunnel in Limburg, the Netherlands. This case study is relevant, because the Swalmen tunnel will be renovated in 2023 and 2028. These renovation projects include updating controlled subsystems in the tunnel, such as boom barriers and traffic lights, and updating the supervisory controller of the tunnel. The digital twin might be useful to aid the supervisory controller design process in these renovation projects.\\\\
The digital twin is developed in a 3D game engine, Unity, using the Prespective plugin for PLC signal communication. It includes 3D models, user-interaction and simulation of simple traffic streams and other test scenarios. Several methods are developed to simplify the modular setup of a digital twin for supervisory control validation. An example is automatic generation of logic signal objects based on variable names in the PLC code. In the test setup, the digital twin is controlled with a virtual PLC running in TwinCAT. This soft PLC works almost identically to the hardware PLC, but the cycle times are lower and the signals are fully digital. A simple operator interface is developed for testing.\\\\
This test setup is used to validate the digital twin of the Swalmen tunnel. Some further development of the digital twin is desired, but the first full-scale tests concerning supervisory controller validation are promising. This makes that digital twins are a useful addition in the supervisory controller design process. The main reason is that it offers more accessible and high-fidelity testing methods and scenarios, resulting in better supervisory controller validation earlier in the design process. With minor changes to the digital twin, other applications, such as operator training, are also possible. Furthermore, additional work going into development of other digital tunnel twins is relatively small, certainly with the modular setup that allows for evolvability and can be used to easily create digital twins for other tunnels. These reasons make the digital tunnel twin an attractive tool.
\end{abstract}

\tableofcontents{\protect\thispagestyle{empty}}
\newpage

\setcounter{page}{1}

\chapter{Introduction}\label{sec:introduction}

The Control Systems Technology (CST) group in the Mechanical Engineering department on the TU/e has been researching supervisory controller design. Supervisory controllers ensure cooperation between components in a system. Connected to this research, a collaboration is established with the Dutch Department of Waterways and Public Works ("\textit{Rijkswaterstaat}" in Dutch, abbreviated to RWS in this report). This collaboration project, referred to as the `RWS project' in this report, is focused on the implementation of synthesis-based controller design technology in the design of supervisory controllers for bridges, waterways and tunnels \cite{road_ahead_supervisory}.\\

The controller design process relies heavily on the use of simulation models of the plant (the physical system). A possible addition to the design process is digital twin technology. A digital twin is a virtual copy of a system that can be much more realistic than the 2D-simulation models that are currently used for controller validation in the controller design process.\\

\section{Case Study Details}

\begin{figure}[ht]
    \centering
    \includegraphics[width=.95\textwidth]{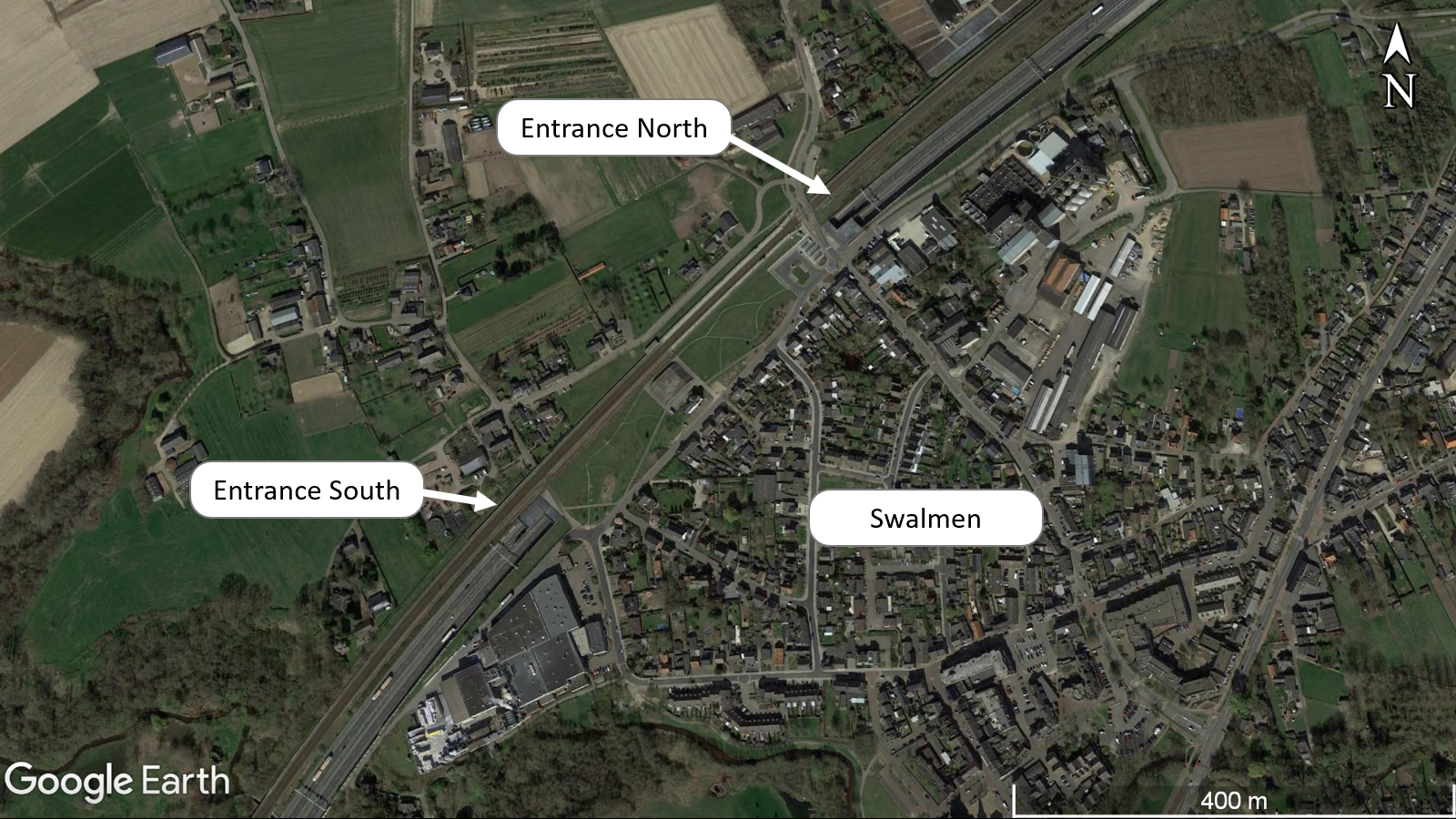}
    \caption{Overview of the Swalmen tunnel in its surroundings (Google Earth).}
    \label{fig:overview_tunnel}
\end{figure}

In this report, a case study of the Swalmen road tunnel ("\textit{Swalmentunnel}" in Dutch) is done with focus on the implementation of a digital twin into the control design process. The Swalmen tunnel, shown in Figure \ref{fig:overview_tunnel}, is situated in the village of Swalmen, which is near the city of Roermond in Limburg, the Netherlands. The tunnel tube is about 400 meters long and the whole tunnel road, including the entrance roads, measures approximately 1 kilometer. The highway A76 passes through this tunnel, which was opened in 2008 \cite{rws_overview_tunnels}. South of the Swalmen tunnel, the A73 passes through the Roer tunnel in Roermond. This tunnel has tight links to the Swalmen tunnel, since the two tunnels were built at the same time. The Swalmen tunnel has also been investigated in the RWS project at the TU/e. Figure \ref{fig:overview_tunnel_sections} below shows one of the entrances of the Swalmen tunnel.

\begin{figure}[ht]
    \centering
    \includegraphics[width=.6\textwidth]{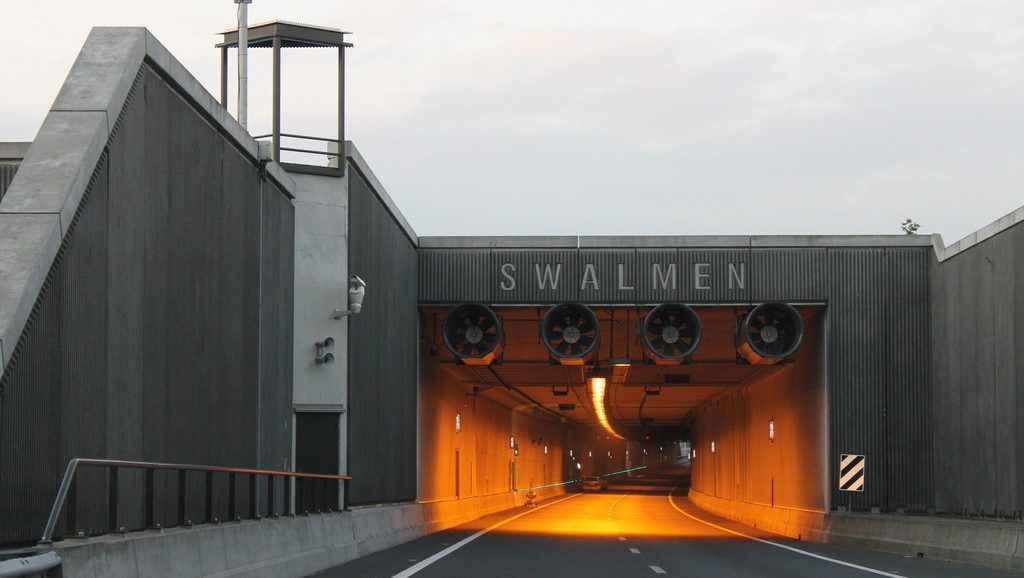}
    \caption{Entrance of the Swalmen Tunnel \cite{swalmentunnel_figure}.}
    \label{fig:swalmentunnel}
\end{figure}

Since the opening of the Swalmen road tunnel and the Roer tunnel, a lot of problems have occurred. Due to failures, the opening of the tunnels was delayed and when opened, the tunnels were closed frequently due to outages. Another problem was that some trucks that were too high were still entering the tunnels, which triggers the safety system that closes the tunnels. Because of this, the height detection system was altered and the ventilation units were raised to prevent collisions with high vehicles, leading to fewer occurrences of this problem \cite{tien_jaar_tunnels}. A large renovation project for both the Swalmen Tunnel and the Roertunnel is planned for 2023. In this renovation project, several technical installations will be replaced to meet the renewed tunnel standards and the operating system will be upgraded \cite{RWS_inkoopplanning_renovatie}. Furthermore, a complete overhaul of the whole tunnel is planned for 2028. The fact that these renovations are planned makes the Swalmen tunnel suitable for a case study on a digital tunnel twin for supervisory controller validation since the renovations include the development of a new supervisory controller. This means that a digital twin could potentially be implemented in the design process of this supervisory controller that is described in Chapter \ref{sec:controller_design}.\\

\section{Research Objectives}

The main goal of this project is to create a digital twin of the Swalmen tunnel that can be used in the controller design process of a supervisory controller. The focus is on validating the synthesized controller by using the digital twin. This immediately defines the first and main research objective:

\begin{itemize}
    \item[] \textit{1. Develop a digital twin of the Swalmen tunnel that can be used to test and validate a supervisory controller.}
\end{itemize}

This first objective includes evaluating the capability of the digital twin to effectively run the controller and assessing the quality of the digital twin. The quality assessment is based on the level of realism and implications of certain design choices in the development of the digital twin. The development process will largely be carried out using previous documented work of digital twins developed at the TU/e, as described in Section \ref{sec:prev_work} and with knowledge acquired from literature study on digital twins in general.\\

Since digital twins have a broad application field, it is interesting to determine what specific applications are useful for a digital twin of a tunnel system. Furthermore, it should be documented how these applications can be implemented. This raises the second research objective:

\begin{itemize}
    \item[] \textit{2. Investigate applications of digital tunnel twins.}
\end{itemize}

This second objective is approached by first conducting literature research in order to find interesting applications and to determine what applications are desirable for the digital tunnel twin. The implementation process of these desirable applications is then assessed based on insights gained when building the digital twin. Conclusions are then drawn on the feasibility of these desirable applications.\\ 

As will be discussed in this report, former development of digital twins on the TU/e was limited to systems at a much smaller scale than the tunnel system investigated in this project. Therefore, it is interesting to find out whether the same development methods can be used for the digital tunnel twin. This comparison also includes defining what modeling techniques are suitable for a system with the same characteristics as the tunnel. The third research objective thus is:

\begin{itemize}
    \item[] \textit{3. Describe the method of developing digital twins of systems similar to the Swalmen tunnel.}
\end{itemize}

\section{Previous Work}\label{sec:prev_work}

The most relevant previous work comes from two projects form the TU/e. The first is the RWS project that is mentioned at the start of this introductory chapter. For the case study of the Swalmen tunnel a supervisory controller is already synthesized in this project as part of an internship at RWS \cite{control_model_swalmentunnel}. The models created in \cite{control_model_swalmentunnel} are the basis of the design of the digital twin of the Swalmen tunnel and the digital twin is used to validate the controller synthesized in \cite{control_model_swalmentunnel}. In the other relevant project, digital twins have been developed for controller validation of workstations of a Festo production line \cite{DTs_4TC00}. Documentation of the design process of these digital twins is used as reference material for developing the digital twin of the Swalmen tunnel.

\section{Digital Twin Technology}\label{sec:DT_technology}

Because the research objectives are focused around the development process and applications of digital twins in tunnel systems, a background on the essence of digital twins is needed. A digital twin is most generally described as a digital copy of a physical system where the virtual system usually collects data from the real system. This data is simulated and analyzed with the goal of improving the physical system \cite{Pires2019}. Digital twins are seen as an important part of the Industry 4.0 initiative, which represents an industrial shift towards digitalization. Digital twins are expected to play a major role in this revolution by enabling performance boosts with the use of high-end simulations in development and maintenance of systems.\\

This chapter starts with a description of some relevant examples of digital twins, highlighting the different applications and ways in which the digital twins are used to improve systems. Afterwards, possible applications of digital twins for tunnels in particular are discussed. Finally, the tools that are used to develop the digital twin of the Swalmen tunnel are described.

\subsection{Applications of Digital Twins}

As is evident from the definition given in the introduction of this chapter, digital twins are very versatile. Therefore, the application field of digital twins is very broad. Some relevant examples are discussed here to illustrate different types of digital twins and to specify the type of digital twin suitable for validating the supervisory controller of the Swalmen tunnel.\\

The first type of digital twin is data-driven and usually runs in real-time with the real system. An example is the use of digital twins in information sharing technologies for connected and automated mobility \cite{Schranz2020}. Here, digital twins are used to model and monitor the lifecycle phases of smart vehicles. Data available in the cloud, such as information of car users and connected devices, is used as input for the digital twins. The digital twin then generates data that is used to provide stakeholders insights in security and safety in the lifecycle of a smart vehicle. Another traffic-related example is described in \cite{Kumar2018}. Digital twins that are called `virtual vehicles' are used to improve traffic flows by enabling communication between these vehicles based on all kinds of traffic-related data. \\

The second type of digital twins is based more on modeling 3D-simulations with visualizations, which can be used together with the physical system and even before the physical system is realized. These digital twins are not focused on real-time data gathering. An example is a simulation for self-driving cars, which can for instance be used for development of algorithms for autonomous driving \cite{DT_automonous_cars_atorf}. Another example is a digital twin for train operation and control \cite{Meng2020}. This digital twin is particularly interesting in this project, because it focuses on control and operation of a system. The paper describes the method of simulating physical entities that are important in the complex process of train operation. The goal is to improve the system by simulating important subsystems, such as train dynamics, weather influence and landscape interaction. Different goals require different subsystems to be modeled. In order to investigate energy saving for instance, all energy related components are modeled accurately, while interaction between passengers and the seats of the train is not important. The relevance of the visual geometric model is also evident here, as it is an important link in the interaction between the system and the users.\\

Of these two types of digital twins, the last one most closely resembles the goals of the digital twin for the Swalmen tunnel. The main reason for this is that control validation for large systems is ideally done before the real system is built, so the focus of the digital twin should not be on real-time cooperation with the physical system. Furthermore, the visualization and interaction components are vital for intuitive controller validation.

\subsection{Digital Tunnel Twins}\label{sec:digital_tunnel_twins_COB}

The \textit{`Centrum Ondergronds Bouwen'} (COB), Center for Underground Construction in English, is a network organization that gathers, creates and spreads knowledge about usage of underground spaces such as tunnels. This organization has written a document in which several types of digital aids in the field of underground construction are discussed with the focus on digital twin systems of tunnels \cite{COB_digitaal_aantonen}. Here, the digital tunnel twin is seen as the set of all digital tools used in all phases of the life cycle of the tunnel, which ideally are all connected and form one large simulation model. The digital twin technology in this document is based on 3D-BIM (Building Information Modeling) models, which are CAD-drawings of the tunnel structure with integrated information about other important aspects ranging from technical installation details to maintenance.\\

The report of COB also describes several advantages, disadvantages and applications of digital tunnel twins in a general sense. An important application of the digital tunnel twin lies in the design process. Small sections in the design can be verified and validated separately and early in the development process. After expanding and connecting these sections, additional verification tests can then be carried out to be sure that the whole system meets the requirements at every stage of the design process. This is especially important when multiple subsystems are integrated in the main system. This leads to a reduction in development time and costs, since verification happens earlier in the project, meaning that problems and integration problems are found sooner and the process can be adjusted at an early stage. Different implementation and design strategies can also be tested beforehand, leading to a more optimized process.\\

Due to the fact that testing the systems on all levels of detail for all desired scenarios can be included in the digital twin, on-site activities in construction can be limited to only installing and basic tests, since the system validation has already been carried out on the digital twin. This is particularly beneficial in the case of tunnel renovation, because closure of a major traffic route can have a large impact. With digital twins that are detailed extensively, an additional advantage is an improvement in communication and interaction between stakeholders. The digital twin model then functions as one place in which all information that is normally written down in many separate documents can be found in one place. This creates an overview and also makes the functionality of the tunnel in the digital twin unambiguous, since all details and functions can be included in the model.\\

Because of the overview created in the digital twin, the different aspects that are relevant in the design and renovation process of tunnels are accessible to relevant end-users such as tunnel managers, road traffic controllers, tunnel operators and emergency services. This means that they can more easily be involved in the tunnel development process and help in uncovering shortcomings in specifications and usability at an early stage.\\

Another application concerning end-users lies in training personnel, which can already start before the tunnel is constructed. Using digital twins for training purposes is often referred to as \textit{serious gaming}. In the case of tunnels, the digital twin can be used to train road traffic controllers and tunnel operators in different scenarios, such as emergencies and traffic jams. \\

\section{Report Structure}

The report starts with three background chapters after which the development process of the digital twin is described. Finally, the digital twin is tested and its applicability is assessed and its place in the controller design process is determined. \\

The first background chapter (Chapter \ref{sec:DT_technology}) describes digital twin technologies and their applications with some relevant examples. Additionally, the potential value of possible applications of the digital tunnel twin developed in this project is discussed. The tools used in the development process of the digital twin are also introduced here. The second background chapter, Chapter \ref{sec:controller_design}, describes the supervisory controller design process for systems in general with an emphasis on synthesis-based controller design and its application in tunnels. Chapter \ref{sec:overview_swalmen_tunnel}, the final background chapter, gives relevant information about the Swalmen Tunnel and the previous work concerning the synthesis of the supervisory controller for this tunnel. The models that are used to do this are discussed as well. Furthermore, a detailed description of all relevant components of the tunnel is given.\\

After describing and investigating all relevant background topics, the development process and setup of the digital twin of the Swalmen tunnel is described. Chapter \ref{sec:set_up_digital_twin} explains the methods used to set up the digital twin and serves as reference material for the development of digital twins of systems similar to tunnels that are targeted at controller validation. This includes a detailed description of the development of modular components and how these can be used to build a digital twin. The methods used for setting up the interface between the digital twin and the supervisory controller that runs on a PLC are described in Chapter \ref{sec:PLC_control}. This is done by first describing the generation process of the PLC code from the supervisor that is already synthesized. Also, the general setup of actuators and sensors is described and the method of connecting the digital twin and the PLC is explained.\\

Chapter \ref{sec:digital_twin_swalmentunnel} describes the application of the methods from the two previous chapters to the case study of the digital twin of the Swalmen tunnel. Here, all relevant components in the digital tunnel twin and their modeled behavior are discussed in detail. Chapter \ref{sec:validation} deals with the validation process of the digital twin and its results. The validation is done with the already synthesized supervisor. Based on the results, conclusions are drawn on the research questions posed in the previous section. These conclusions are discussed in Chapter \ref{sec:evaluation_conclusion}, where additional recommendations are given on the process of developing a digital twin and important choices that were made in this project. Recommendations on areas of the digital twin that are suitable for further development are treated as well. Finally, the applicability and place of the digital twin in the controller design process are assessed.

\chapter{Preliminaries}
Since the first research objective is to develop a digital twin of the Swalmen tunnel, this chapter starts with a description of tools that can be used in the development process. The other part of the first research objective concerns supervisory control, which is explained in the second part of this chapter. Here, the implementation of synthesis based engineering for supervisory controller development is also described. This development process is important, since the positioning of the digital twin in this process should be determined.

\section{Digital Twin Development Tools}\label{sec:tools_DT}

\begin{figure}[ht]
    \centering
    \includegraphics[width=.85\textwidth]{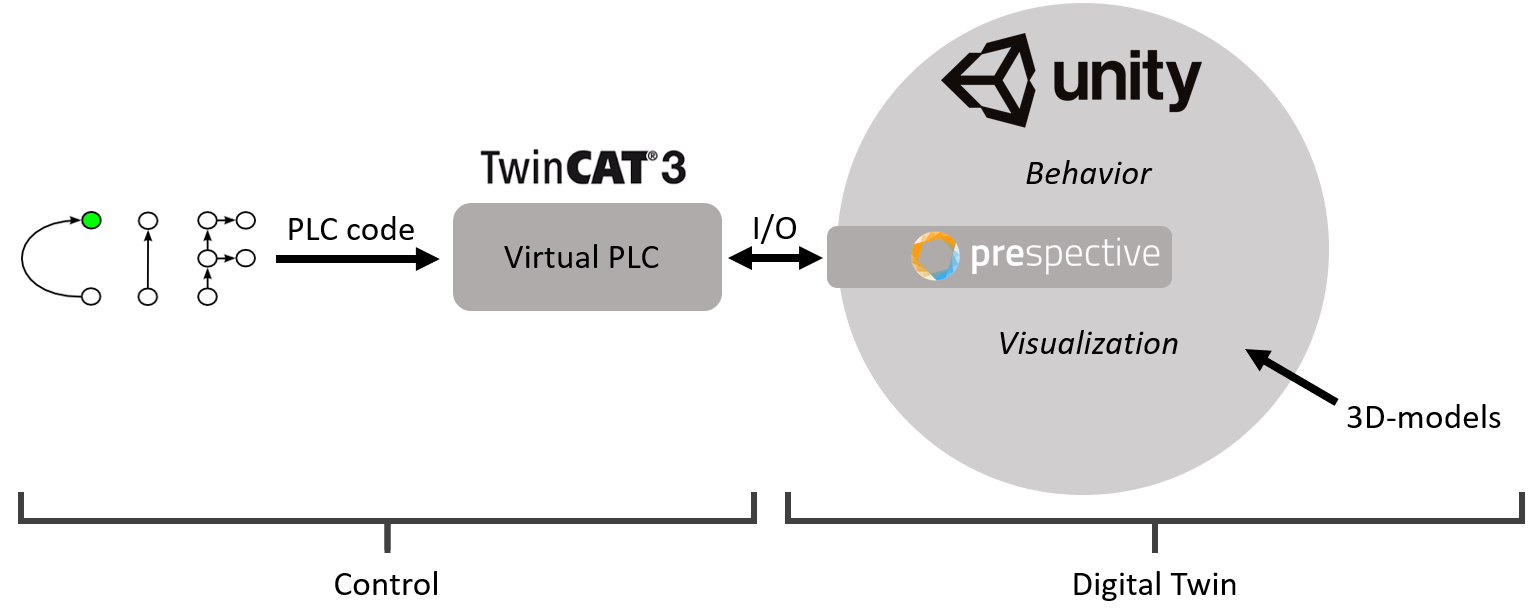}
    \caption{Overview of the tools used in the digital twin design process and the connection between these tools.}
    \label{fig:digital_twin_tools}
\end{figure}

For the development process of a digital twin, a multitude of programs can be used. Ideally, the same tool set as described in the previous projects at the TU/e in which digital twins have been developed is used in this project, since these tools are documented well \cite{digital_twin_festo_vissers} \cite{digital_twin_festo_verstegen} \cite{digital_twin_festo_ginneken} \cite{digital_twin_festo_vergeldt}. The most important programs and their function in the digital twin process are described below and an overview is given in Figure \ref{fig:digital_twin_tools}.\\

\vspace{-5mm}

\vspace{-5mm}
\subsection*{Unity}
In the previous digital twin creation process, the game engine Unity 3D \cite{Unity} was used to develop the digital twin. 3D-models are loaded into this game engine and put into the Unity environment. The behavior of different components and interaction between them is modeled using several tools included in the engine. Unity is directly responsible for the visualization aspect of the digital twin.\\

\vspace{-5mm}
\subsection*{Prespective}
Because Unity itself is a game engine, a plugin for Unity called Prespective \cite{Prespective} is used to implement essential tools for developing the digital twin. Prespective adds realistic standard physical components like motors and indicator lights. The most useful addition is that Prespective provides an interface that can be used to link Unity to a (virtual) PLC.\\

\vspace{-5mm}
\subsection*{TwinCAT}
In order to control the digital twin that is running in Unity with the Prespective plugin, the PLC code runs on a virtual PLC, for which TwinCAT \cite{Beckhoff_TwinCAT} is used. It is essential to note that the virtual PLC works exactly like a `real' hardware PLC because it uses the same signals and cycle times. The only difference is that it does not transfer the signals through cables. TwinCAT usually runs on a separate CPU-core on the same PC as the digital twin.\\

\vspace{-5mm}
\subsection*{CIF 3 Toolbox}
An additional relevant tool in this project is the CIF 3 Toolbox. This toolbox is based around CIF, which is an automaton-based modeling language developed at the Eindhoven University of Technology (TU/e) \cite{CIF}. This toolbox is used to design the controller and to generate the PLC code that is ran in TwinCAT. In this project in particular the controller code is already supplied as described in Section \ref{sec:prev_work}, so it only needs to be prepared and exported to PLC code and can then be executed in TwinCAT.\\

\section{Controller Design}\label{sec:controller_design}

Tunnels are systems that generally consist of a large number of subsystems, of which most are relevant for guaranteeing safety of road users. A supervisory controller is responsible for ensuring correct cooperation of all entities in the tunnel. The control setup for a tunnel system is discussed in this chapter. Furthermore, the design process of this controller as used in the RWS-project is introduced.\\

\subsection{Supervisory Control}\label{sec:supervisory_control}
\begin{figure}[ht]
    \centering
    \begin{tikzpicture}[node distance=1cm, auto, scale=.95, transform shape]  
    
    	\tikzset{
        mynode/.style={rectangle,rounded corners,draw=black, top color=white, bottom color=white!50,very thick, inner sep=0.75em, minimum size=2.5em, minimum width=16em, text 		centered},
        mynode2/.style={rectangle,rounded corners,draw=black, top color=white, bottom color=white!50,very thick, inner sep=0.5em, minimum size=1em, minimum width=6em, text centered},
        mynode3/.style={rectangle,rounded corners,draw=none, top color=white, bottom color=white!50,very thick, inner sep=0.5em, minimum size=1em, minimum width=1em, text centered},
        myarrow/.style={->, >=latex', thick},
        mylabel/.style={text width=7em, text centered}
    	}  
    	
    	\definecolor{airforceblue}{rgb}{0.36, 0.54, 0.66}
    	\definecolor{navyblue}{rgb}{0.0, 0.0, 0.5}
    	\definecolor{royalblue}{rgb}{0.25, 0.41, 0.88}
    
      	\node[mynode,scale=0.7] (q1) {Operator Interface}; 
    	\node[mynode3, left=0.25cm  of q1, scale=1]   (n5) {5};  
    	\node[mynode3, right=0.25cm of q1, scale=1]   (n5) {};  
      	\node[mynode,  below=0.75cm of q1, scale=0.7] (q2) {Supervisory Controller};   
    	\node[mynode3, left=0.25cm  of q2, scale=1]   (n4) {4};
      	\node[mynode, below=0.75cm of q2, scale=0.7] (q3) {Resource Controller(s)};
      	\node[mynode, below=0.75cm of q3, scale=0.7] (q6) {};
      	\node[mynode2, below=0.85cm of q3, xshift=-1.11cm, scale=0.7] (q4) {Actuators};  
    	\node[mynode2, below=0.85cm of q3, xshift= 1.11cm, scale=0.7] (q5) {Sensors};  
      	\node[mynode, below=0.75cm of q6, scale=0.7] (q7) {Mechanical Components};
      	\node[mynode3, left=0.25cm  of q3, scale=1]   (n3) {3}; 
      	\node[mynode3, left=0.25cm  of q6, scale=1]   (n2) {2};  
    	\node[mynode3, left=0.25cm  of q7, scale=1]   (n1) {1};  
    
    	\draw[myarrow] (q4.north |- q1.south) -| (q4.north |- q2.north);
    	\draw[myarrow] (q4.north |- q2.south) -| (q4.north |- q3.north);
    	\draw[myarrow] (q4.north |- q3.south) -| (q4.north);
    	\draw[myarrow] (q4.south) -| (q4.south |- q7.north);
    	
    	\draw[myarrow] (q5.north |- q2.north) -| (q5.north |- q1.south);
    	\draw[myarrow] (q5.north |- q3.north) -| (q5.north |- q2.south);
    	\draw[myarrow] (q5.north) -| (q5.north |- q3.south);
    	\draw[myarrow] (q5.south |- q7.north) -| (q5.south);	
    	
    	\node [fit=(q3) (q7)] (fit) {}; 
    	\node [mynode3, right=0.05cm  of fit, scale=.7]   (plant) {Plant};  
    	\draw [decorate, line width=1pt] (fit.south east) -- (fit.north east);
      	
      	\node [fit=(q2)] (fit2) {};
      	\node [mynode3, right=0.05cm  of fit2, scale=.7]   (controller) {Controller};       
      	\draw [decorate, line width=1pt] (fit2.south east) -- (fit2.north east);
    
        \node [fit=(q1)] (fit3) {};
  		\node [mynode3, right=0.05cm  of fit3, scale=.7]   (GUI) {GUI};    
  		\draw [decorate, line width=1pt] (fit3.south east) -- (fit3.north east);
    
    \end{tikzpicture}
    \caption{Supervisory control structure in a tunnel system (adapted from \cite{supervisory_controller_synthesis_tunnel}).}
    \label{fig:controller_stack}
\end{figure}
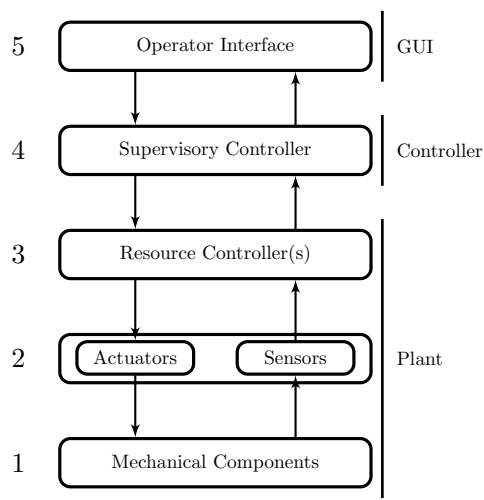

In the introduction, the concept of supervisory control was already introduced briefly. A supervisory controller controls all components in a system by deciding which actuators should be turned on based on information provided by sensors and manual input from tunnel operators.
This control structure is schematically shown in Figure \ref{fig:controller_stack}. Here, the difference between the controller and the plant in the view of supervisory control is depicted. The plant consists of all mechanical components (1) in the tunnel that are controlled through actuators (2) by, or provide data through sensors (2) to their respective resource controller (3). This can for instance be a controller that controls the angle of a boom barrier. These resource controllers are also part of the plant and are controlled on a higher level by the supervisory controller (4). The supervisory controller can then for example give a signal that a barrier should open or close. The final part of the control structure is the operator interface (5) that enables the operator of the system to send input signals to the controller by pressing buttons. The operator interface also gets feedback from the controller that provides information about the state of the system.

\vspace{2mm}

\subsection{Synthesis-Based Controller Design}\label{sec:synthesis_based_controller_design}

A novel way of controller design is called synthesis-based controller design, which has also been applied to the design of a tunnel supervisory controller and plays a large role in the RWS-project as explained in \cite{supervisory_controller_synthesis_tunnel}. Here, it is stated that traditionally the control strategy of a system is designed manually by engineers based on the requirements and is then directly tested on the real system. The downside of this way of working is that the quality of this method of designing supervisory controllers declines when more actuators and sensors are present in the system. This is due to the greater complexity and limitations to testing. These testing limitations can be partially overcome by using model-based controller design, because this allows the controller to be tested during the design process by means of simulation. This enhances the controller design, because errors can more easily be found early in the design process. A further improvement is the application of synthesis-based controller design. This means that the supervisory controller is created by an algorithm based on given requirements imposed on the uncontrolled behavior and models of the system behavior. The large benefit of this is that the controller is guaranteed to be correct when the requirements and model are correct. This further improves the controller quality \cite{Reijnen2020}. It is also mentioned here that synthesis-based supervisory controller design has been applied to several systems in the RWS project at the TU/e, which apart from tunnels \cite{supervisory_controller_synthesis_tunnel} \cite{moormann_tunnel_kwa} also includes bridges \cite{Reijnen2020} and water locks \cite{reijnen_waterway_synthesis_control} in The Netherlands.\\

Figure \ref{fig:controller_development_process} below shows a schematic overview of the model-based controller design process including controller synthesis. Starting on the left, first the high-level requirements of the system are defined. from which the requirements of the controller and the plant are extracted. From these plant requirements, a plant is designed and modeled. The controller requirements are formalized and from these formalized requirements and the modeled plant, a controller is synthesized. This controller is then usually tested first by \textit{model simulation}, where the controller directly controls an enriched plant model, called the hybrid model. The `enriching' step usually includes the addition of timers, a visualization and additional plant behavior. The next step is to take the hardware interface, which could be a PLC, into account by running the controller on a hardware controller and connecting this to the hybrid model (\textit{hardware-in-the-loop testing}). The final testing phase is carried out with a controller running on hardware which is controlling the real plant (\textit{system testing)}.

\begin{figure}[ht]
	\centering
	\resizebox{.85\textwidth}{!}{
	\begin{tikzpicture}[fmbe,scale=1,transform shape, node distance=3cm ]
		\node[document]                                    	(D)       	{$H_\mathrm{R}$};
		\node[document,above right=1.25cm and 2cm of D]		(R1)      	{$C_\mathrm{R}$};
		\node[model,right=2.9cm of R1]                     	(D1)      	{$C_\mathrm{R}$};
		\node[right of=D1,circle,draw,inner sep=2pt] 		(t)       	{};
		\node[model,right of=t]                           	(M1)     	{$C$};
		\node[realization,right of=M1]                 		(Z1)      	{$C$};
		\node[document,below right=1.25cm and 2cm of D]     (R2)      	{$P_\mathrm{R}$};
		\node[document,right=2.9cm of R2]                  	(D2)      	{$P_\mathrm{D}$};
		\node[model,right of=D2]                       		(M3)      	{$P$};
		\node[model,right of=M3]                 			(M4)      	{$P_\mathrm{H}$};
		\node[realization,right of=M4]                 		(Z2)      	{$P$};
				
		\definecolor{evalcolor}{named}{blue} \path (M4) -- node[eval,minimum width=2.85em,minimum height=10.25em,sloped,rotate=-90] 			(MZ1) {} (M1);
		\definecolor{evalcolor}{named}{teal} \path (M4) -- node[eval,minimum width=3.5em,minimum height=14.7em, xshift=.3em, sloped,rotate=-90] (MZ2) {} (Z1);
		\definecolor{evalcolor}{named}{red}  \path (Z1) -- node[eval,minimum width=2.85em,minimum height=10.25em,sloped,rotate=-90] 			(MZ3) {} (Z2);
		
		\draw[act] (D.20)  -- +(0:1em) |- node[actlb',pos=.25] {extract} (R1);
		\draw[act] (D.-20) -- +(0:1em) |- node[actlb',pos=.25] {extract} (R2);
		\path[act]
		([xshift=-4em]D.center) 
		      edge node[pos=.45]       {define}                   	(D)
		(R1)  edge node                {formalize}                	(D1)
		(R2)  edge node[below]         {design}            			(D2)
		(D1)  edge                                                	(t)
		(M3)  edge                                                	(t)
		(t)   edge                                                	(M1)
		(D1)  --   node                {synthesize}               	(M1)
		(M1)  edge node[align=center]  {generate and\\ implement} 	(Z1)
		(D2)  edge node[below]         {model}             			(M3)
		(M3)  edge node[below]         {enrich}            			(M4)
		(M4)  edge node[below]         {realize}           			(Z2)
		;
		
		\pgfusepath{use as bounding box}
			
		\end{tikzpicture}
		}
	\small{
		\begin{center}
			\tikz[fmbe,baseline=(x.base)]{\node[document] (x) {$\phantom{X}$};} = documents,
			\tikz[fmbe,baseline=(x.base)]{\node[model](x){$\phantom{Y}$};} = models,
			\tikz[fmbe,baseline=(x.base)]{\node[realization](x){$\phantom{Z}$};} = realizations.
			
			\vspace{1em}
			$H$ = high-level, $P$ = plant, $C$ = controller, $\mathrm{R}$ = requirement, $\mathrm{D}$ = design, $\mathrm{H}$ = hybrid.		
			\vspace{1em}
			
			\resizebox{.75\textwidth}{!}{
			\tikz[fmbe,baseline=(x.base)]{\node[realization, blue, rounded corners, thick] (x) {$\phantom{X}$};} = model simulation,
			\tikz[fmbe,baseline=(x.base)]{\node[realization, teal, rounded corners, thick] (x) {$\phantom{X}$};} = hardware-in-the-loop testing,
			\tikz[fmbe,baseline=(x.base)]{\node[realization, red, rounded corners, thick] (x) {$\phantom{X}$};} = system testing.
			}
		\end{center}
	}
    \caption{Supervisory controller development (adapted from \cite{supervisory_controller_synthesis_tunnel}).}
    \label{fig:controller_development_process}
\end{figure}
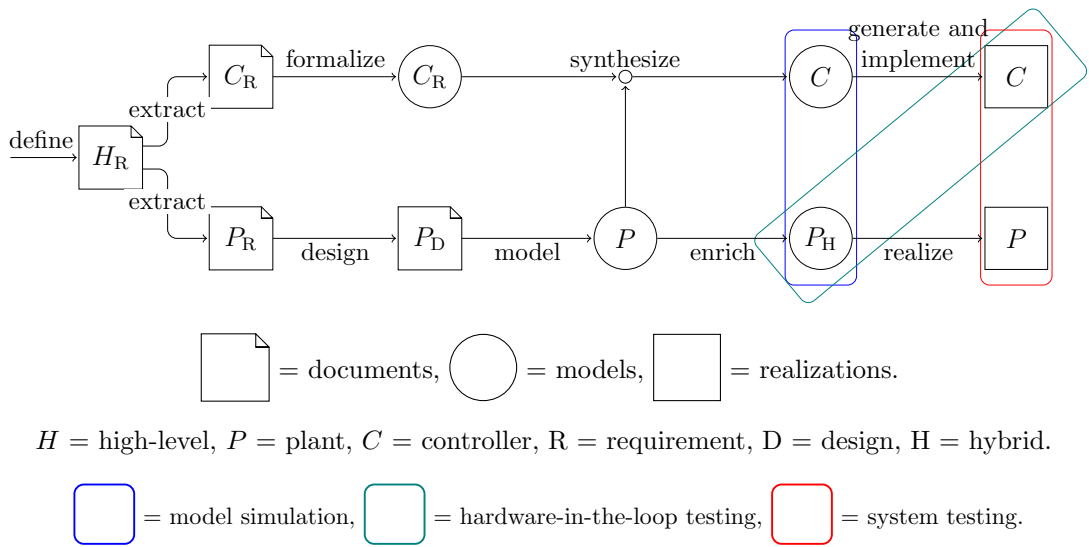

\subsection{Tunnel Control}
As mentioned in Section \ref{sec:synthesis_based_controller_design}, synthesis-based supervisory controller design has already been applied to tunnel systems. This method of designing controllers is suitable for tunnel systems maintained by RWS because of the existence of the National Tunnel Standard, \textit{'Landelijke Tunnelstandaard'} in Dutch, of RWS \cite{landelijke_tunnelstandaard}. This document includes the standard method of building and operating tunnels in The Netherlands and also includes all requirements and ways of validating the tunnel systems. This means that the high-level requirements can be defined relatively easily from the National Tunnel Standard documentation files. Additionally, it is the case that all tunnels maintained by RWS are constructed or will be renovated to meet the National Tunnel Standard in the future. This implies that the models and methods used for synthesizing the supervisory controller can easily be reused for all road tunnels maintained by RWS in the Netherlands.

\chapter{System Overview of the Swalmen Road Tunnel}\label{sec:overview_swalmen_tunnel}

As discussed, all components that have been modeled to create the existing supervisor should be implemented in the digital twin. This chapter gives a schematic overview of the different areas in the tunnel and a brief description of all controlled components in each area. As is also the case in the existing CIF model, some components of the tunnel are not taken into account. Reasons for this are that these systems have no direct effect on control of the tunnel itself. Omitted subsystems include electrical systems, manual communication systems and systems in the service buildings.

\section{Tunnel Sections}
\begin{figure}[H]
    \centering
    \includegraphics[width=.85\textwidth]{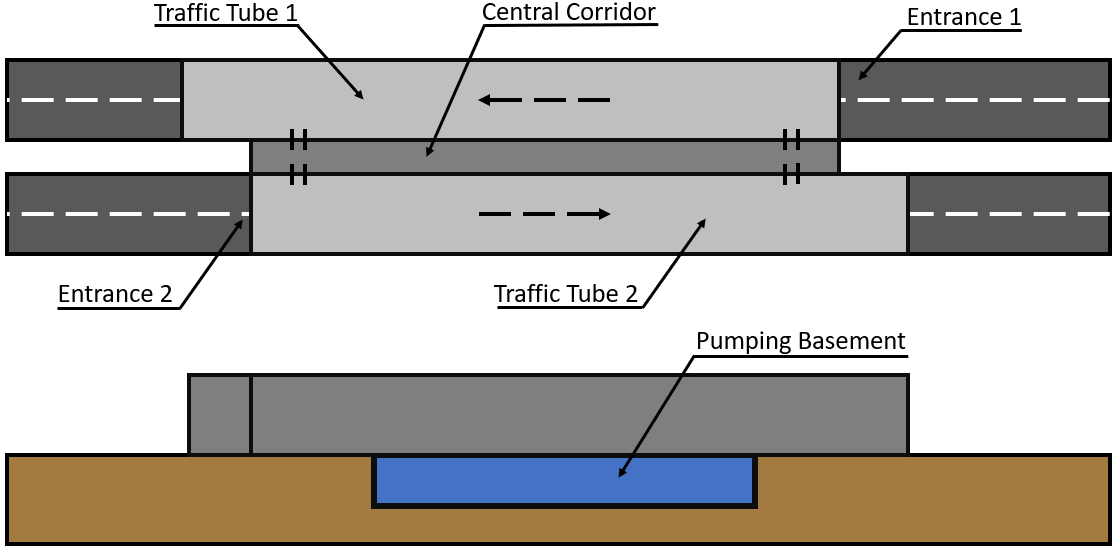}
    \caption{Main sections of the Swalmen road tunnel.}
    \label{fig:overview_tunnel_sections}
\end{figure}

Because the tunnel contains many subsystems, the tunnel controller is organized in different sections. At the highest level, the whole system can be divided into three different main sections. These are traffic tube 1 (North-West), traffic tube 2 (South-East) and the central corridor that lies between the traffic tubes. The entrance and exit roads belong to the corresponding traffic tubes. Apart from the subsystems in these main sections, some separate subsystems such as the pumping basement are also present. A schematic overview of all entities in the tunnel system that are controlled by the supervisor is given in Figure \ref{fig:overview_tunnel_sections}. 

\begin{figure}[ht]
    \centering
    \includegraphics[width=\textwidth]{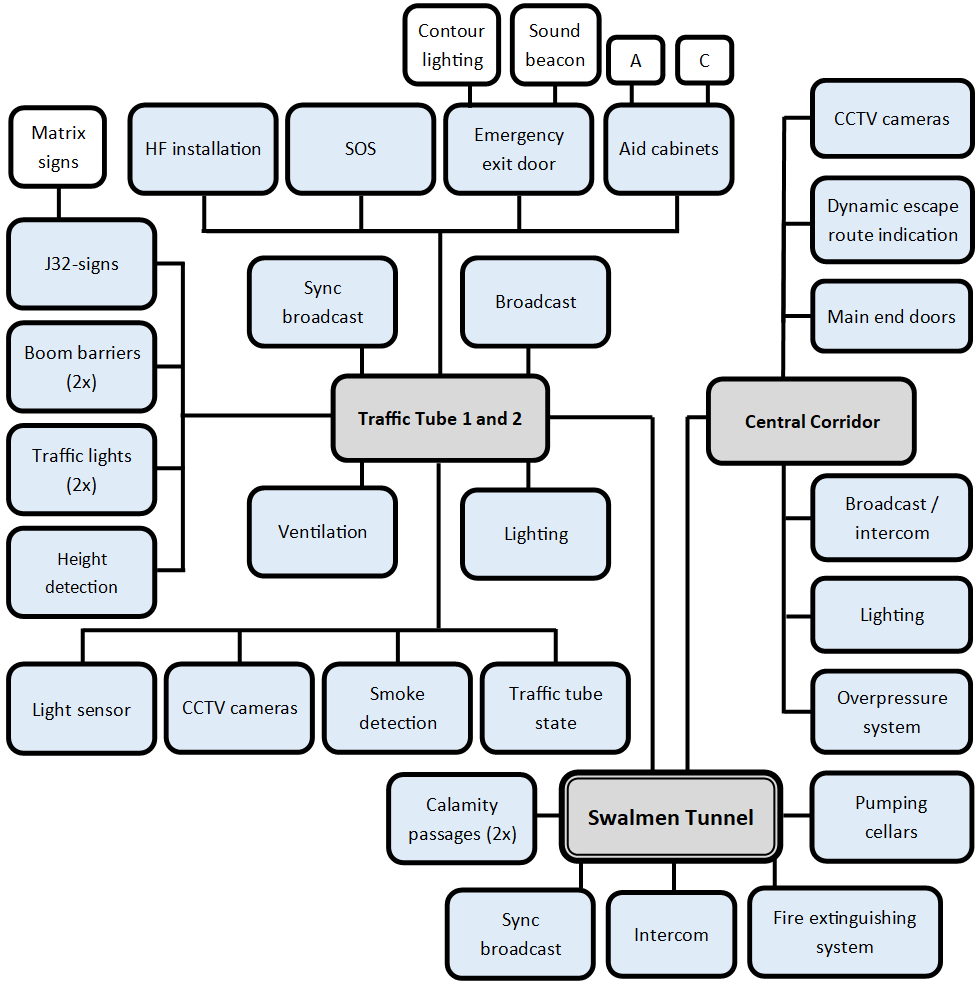}
    \caption{Main components of the Swalmen road tunnel.}
    \label{fig:overview_tunnel_components}
\end{figure}

The two traffic tubes of the tunnel are very similar in terms of components. The only difference is that traffic tube 1 has an elaborated height detection system with more height detection units. The reason for this is that the largest portion of the problems with high trucks entering the tunnel occurred in this traffic tube. An overview of the main sections in the tunnel with their important components is given in Figure \ref{fig:overview_tunnel_components}.

\renewcommand{\labelitemi}{-}
\section{Traffic Tubes 1 and 2}

The traffic tubes mainly contain subsystems that interact with the road users and the tunnel environment. The most important systems that are present in each traffic tube are listed here with their main function:

\begin{itemize}

    \item \textbf{Aid cabinets:} cabinets that house components that can be accessed by road users or emergency services in an emergency situation. There are two variants: \textit{A} and \textit{C}. Aid cabinet \textit{A} houses an emergency telephone, a fire extinguisher and a fire hose. Because aid cabinet \textit{C} is located at the right side of the road and the fire extinguisher pump is located below the central corridor on the left of the road, cabinet \textit{C} only houses an emergency telephone and a hand extinguisher.
    
    \item \textbf{Boom barriers:} road barriers that can be opened or closed to physically open or block a main road, a traffic lane, an emergency lane, a tunnel tube or an evacuation road for traffic.
    
    \item \textbf{CCTV (Closed Circuit TeleVision):} a camera system that is used to observe the traffic situation in the closed tunnel section, the area in which traffic measures are enforced and specific locations around the tunnel. In specific situations, for example when an intercom installation is active, the relevant camera view is automatically displayed to the operator in the traffic control station.

    \item \textbf{Emergency exits:} emergency doors in the left tunnel tube wall. These exits enable road users to access a safe evacuation route through the central corridor in case of an emergency. The emergency exits are equipped with contour lighting that make the doors easily visible and a sound beacon that helps people with poor visibility to find the door.
    
    \item \textbf{Height detection:} a system that can identify vehicles that are too high in order to prevent damage to the tunnel ceiling and systems mounted to the tunnel ceiling.
    
    \item \textbf{HF (High Frequency) installation:} a communication system that can be used to address road users in the closed tunnel section by means of FM radio transmitters.
    
    \item \textbf{Intercom:} a communication system that enables direct connection between persons in or outside the tunnel with a road traffic controller in case of calamities (independently of the public telephone network).
    
    \item \textbf{J32 signs:} electronically controlled traffic signs that can be turned on or off. J32 signs warn users of an active traffic light ahead. The J32 indication in Figure \ref{fig:overview_tunnel_components} also includes the matrix signs.
    
    \item \textbf{Light sensors:} sensors that keep track of the outside light sensitivity. The output of this sensor is used to determine the light level in the traffic tubes.
    
    \item \textbf{Matrix signs signaling:} matrix signs are positioned above the roads and give signals to road lanes separately. Possible signals include: speed limits, closed roadways and indications to switch roadways.
    
    \item \textbf{Public lighting:} street lights that illuminate the main entrance roads and specific locations outside the covered part of the tunnel at night, such that a safe situation is guaranteed for users of the tunnel and emergency services under all circumstances. This lighting system is coupled with the lighting system in the traffic tubes.
    
    \item \textbf{Smoke detection and visibility measurement installation:} together with a temperature sensing system, the smoke detection system is used to detect a fire. This happens by means of special cameras and visibility measurement systems. Additionally, the tunnel is equipped with a gas detection installation that provides data to the fire department concerning dangerous concentrations of explosive gasses. This system is directly linked to the ventilation system.
    
    \item \textbf{SOS system:} a traffic observation system that provides the tunnel operator with information on possible danger caused by wrong-way drivers, stationary vehicles or speeding vehicles.
    
    \item \textbf{Traffic lights:} traffic lights ahead of the tunnel entrance are used to visually close or open a tunnel tube.
    
    \item \textbf{Traffic tube state:} a virtual system that indicates the state of the traffic tube to the operator. Possible states are: operational, standby, emergency, evacuation, recovery, supportive and maintenance.     
    
    \item \textbf{Tunnel lighting:} a light system inside the tunnel tube that ensures a clear situation in the tunnel at all times. Additionally, the system provides a sufficient light level for fleeing road users and emergency services during an emergency and during maintenance work. The entrance and exit area of the traffic tube is lighted such that road users can adapt to the change in illumination.     

    \item \textbf{Ventilation systems:} groups of four ventilators are situated at the entrance and groups of two ventilators are situated at the halfway point of the tunnel. These ventilators keep the tunnels free from excess polluted air caused by exhaust gasses and keep the air flowing in case of emergencies to prevent a local gas cloud from forming. When an emergency occurs in one of the traffic tubes, the system should also prevent dangerous gasses from entering the adjacent tunnel.
    
\end{itemize}

\section{Central Corridor}

Between the two traffic tubes, the central corridor functions as a gateway mainly intended for handling evacuation situations. The most important systems that are present in the central corridor are listed here with their main function:

\begin{itemize}

    \item \textbf{CCTV (Closed Circuit TeleVision):} like the traffic tubes, the central corridor is equipped with a camera system that serves the same purpose as in the other tunnel sections.
    
    \item \textbf{Central corridor end doors}: at the entrance and exit areas of the tunnel, two main doors give access to the central corridor. These are also part of the evacuation routes.
   
    \item \textbf{Central corridor intercom:} a communication system that enables direct communication between persons in the central corridor and a road traffic controller.

    \item \textbf{Dynamic escape route indication:} a system of signs that indicate the best escape route to road users in an evacuation situation. This route is based on what areas in and around the tunnel are dangerous and safe.
    
    \item \textbf{Lighting:} lights illuminate the central corridor in normal situations, in maintenance situations and in case of emergencies such that fleeing persons can safely leave the tunnel.
    
    \item \textbf{Overpressure system:} a system that maintains an overpressure in order to keep the escape corridor free of smoke and other dangerous gases from the tunnel tube during an emergency.
    
\end{itemize}

\section{Other Systems}

Some systems do not belong to either the traffic tubes or the central corridor, but support the tunnel in its entirety. These are described here:

\begin{itemize}
    \item \textbf{Broadcast synchronization system:} in some emergency cases, all broadcasting systems in the different section in the tunnel should give the same message at the same time. The broadcast synchronization system controls whether these broadcasting systems are synchronized.
    \item \textbf{Emergency passages}: there are two barriers in the central reservation protection of the entrance roads, one at the Northern entrance and one at the Southern entrance. These barriers can be opened in the event of calamities to enable emergency services and normal road traffic to reach the other highway lane.
    \item \textbf{Fire extinguishing system:} the fire extinguishing system regulates the flow of water to the fire hoses in the aid closets and to the fire taps for emergency services. This is done using two pumps.
    \item \textbf{Pumping cellar:} below the tunnel, the pumping cellar with the main pumping installation is located. The purpose of this system is to store and pump away excess water coming from the tunnel construction. There are two separate basins for clean and polluted water.
\end{itemize}

\chapter{Digital Twin Setup}\label{sec:set_up_digital_twin}

    In this chapter the way of modeling and setting up the main components of a digital tunnel twin is described. This is done by addressing methods that are used to make 3D models for components in the digital twin and to set up components that can be controlled with PLC signals. This chapter does not elaborate on the details of the communication between the digital twin and the PLC, but the inputs and outputs are used in the behavior of the PLC controlled components. A detailed description of the PLC communication is given in Chapter \ref{sec:PLC_control}. Furthermore, the chapter focuses around the modular setup of components in the digital twin, which is vital for its evolvability. Throughout this chapter, a boom barrier component is taken as an example, since this component is present in the Swalmen tunnel, its function is generally known, and because it contains multiple actuators and sensors.

\section{3D Modeling and Importing Models to Unity}\label{sec:modeling_sketchup}

    The 3D models used in the previously created digital twins for the TU/e course 4TC00 were designed in Siemens NX \cite{Siemens_NX}. Therefore, these models are CAD-type files, which cannot be imported into Unity directly. In order to use these models in 
    Unity, a third party program called PiXYZ \cite{PiXYZ} was used. The previous digital twins were made for machine set-ups at millimeter to centimeter scale. This is very different from the scale of the tunnel and its components, since these are generally much larger. Furthermore, the dimensions of the tunnel are approximate, so this does not require a level of precision in the geometry that lies in the range of millimeters. For this reason, an alternative 3D modeling environment is used, which is SketchUp.

\subsection{SketchUp}

    SketchUp \cite{sketchup_homepage} is a 3D modeling software that is targeted to be \textit{"The most intuitive way to design, document and communicate your ideas in 3D."} \cite{sketchup_homepage}. This program is frequently used for prototyping in the field of architecture and the gaming industry and Unity can directly import .skp-files (SketchUp model files).\\
    
    The main difference between CAD and SketchUp is in the way of modeling. In SketchUp, a flat outline of a shape can easily be drawn in a few seconds, after which 3D shapes can just be pulled out by dragging them with an extrusion tool. This means that it is possible to easily make rough 3D models with photo references for instance. It is also possible to insert desired lengths and distances in [mm], but it is much harder to work on details and small curved surfaces in SketchUp. CAD programs such as Siemens NX allow for precise parametric control over dimensions and the application of several features which are linked to each other, which is more suitable for small and detailed models of technical components.\\
    
    As explained, it is not necessary to develop models at a high level of detail for a digital tunnel twin. This is due to the large scale of the system and the fact that the goal of the digital twin in this project is supervisory controller validation. This, together with the user-friendliness and the direct compatibility with Unity makes SketchUp a useful tool for drawing 3D models to be used in a digital twin of a tunnel system that is targeted at supervisory controller validation. In this project, SketchUp 2017 is used, which is the latest non-pro standalone version.\\

\subsection{General 3D Modeling Workflow}
    
\begin{figure}[H]
        \centering
        \includegraphics[width=\textwidth]{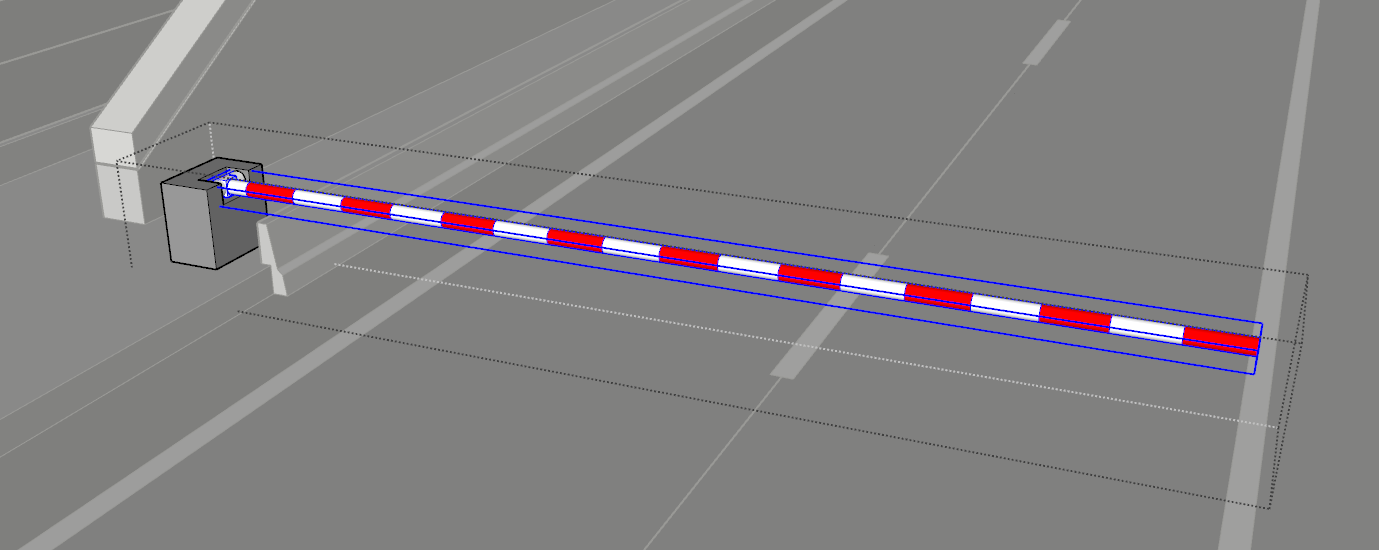}
        \caption{The boom barrier entity in the SketchUp model. The whole group is represented with the dotted box and the barrier itself is its own group, shown by the blue box.}
        \label{fig:boom_barrier_groups}
    \end{figure}

    For large systems, it can be useful to draw all components in one large SketchUp file in order to get an idea of the relative proportions of all components. In the process, the model parts are grouped using the group function in a strategic manner. This means that all separate entities have their own group and that all parts with a specific role in the digital twin should be in a separate group. For instance, a traffic light should have its own group, since this is a separate entity. The parts that represent the lights in the traffic light should then each be in their own group inside the traffic light group, because each light has its own function, which is lighting up. Another example is the boom barrier. The whole group of this entity consists of the base and the barrier itself. The barrier should then be in a separate group within the boom barrier group, as shown in Figure \ref{fig:boom_barrier_groups}, because this part has its own function, which is rotating relative to the base.
    
    Having modeled all parts of the model, the separate entities, such as the main boom barrier group, which can still have subgroups, are then saved in separate SketchUp files. The advantage of saving all components separately is that the models can be loaded into Unity separately, allowing for a modular way of setting up the tunnel. This is helpful, since a small part of the system can be drawn only once and then be used repeatedly to build up the whole tunnel inside Unity. The general workflow following the guidelines described above is:
    
    \begin{enumerate}
        \item Model the whole system with the relevant parts of the environment in SketchUp. Use the grouping functionality in SketchUp to group the geometry of each different part or section of the model together. It is important to group all parts with a separate function in the digital twin. Modeling the whole system in one SketchUp file is not a necessity, but this gives a better sense of scale and whether all models fit in the system.
        \item For each entity in the SketchUp file, make a new separate Sketchup file and copy the base group of the entity to be exported to this empty file. Make sure to position the part near the origin to make it easier to find the model in Unity. (This step and the next step are meant to make the models easily accessible in Unity. It is also possible to import the SketchUp file of the whole system and then extract the groups out of the large model.)
        \item Save the file containing the separate entity as a .skp file.
        \item This SketchUp file can then be imported into Unity by putting it in a folder in the assets directory belonging to the Unity Project and can then be used.
        \item Use all separate entity models to build the digital twin environment.
    \end{enumerate}
    
    An additional advantage of using SketchUp models instead of CAD models is the ease of changing models after importing them into Unity. CAD models need to be reloaded into Unity via PiXYZ when a change to a model is desired. This then means that behavior linked to the objects in unity that use meshes that is imported via PiXYZ needs to be transferred to the new model, which can be a lot of work. Using SketchUp models gives more flexibility to this process, since the files in the Unity project folder are \textit{.skp}-files, which can be opened in SketchUp and can easily be altered. The changes to the model then automatically update in Unity when the altered model in SketchUp is saved. One important factor to note is that this only works when the whole model (with its subgroups) is grouped in a main group in SketchUp and when the file that is being altered is inside the Unity project folder.
    
\section{Modular Component Structure}\label{sec:mod_comp_str}
    
    \begin{figure}[ht]
        \centering
        \includegraphics[width=.25\textwidth]{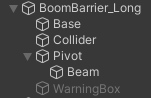}
        \caption{Prefab setup for the boom barrier.}
        \label{fig:boom_barrier_prefab}
    \end{figure}
    
    As described in the previous section, modularity of the digital twin is taken into account when creating the 3D models. This way of modeling can be exploited in Unity by combining the modeled meshes with the desired behavior. This combination of the mesh and behavior, which is implemented in the form of scripts, Prespective components and Unity components, can be saved as a prefab \textit{GameObject}. This is a reusable component in the digital twin.\\
    
    As an example, the structure of the long boom barrier tunnel component is shown in Figure \ref{fig:boom_barrier_prefab}. As shown in Figure \ref{fig:boom_barrier_groups}, the 3D model of this component consists of several groups. In the prefab, these groups are still present, but several GameObjects are added in-between with some components. GameObjects are objects in Unity with a position and rotation that can have several \textit{components}. Components can be scripts, meshes or colliders for example. In the prefab, the upper parent GameObject is \textit{BoomBarrier\_Long}, which is used as a `group' that determines the position and orientation of the whole barrier. The way this works is that all child GameObjects are oriented relative to their parent GameObject. \textit{BoomBarrier\_Long} also contains all scripts and components that enable control and determine the behavior of GameObjects in the barrier. \\ 
    
    The first child GameObject, \textit{Base}, is the mesh of the base of the boom barrier, which does not have any components except for the mesh, since it is just there for the visualization. This GameObject is taken directly from one of the model groups shown in Figure \ref{fig:boom_barrier_groups}. The second GameObject in the barrier is \textit{Collider}. This GameObject contains one collider component that is essentially an invisible mesh that can detect collisions with other colliders. This GameObject is used to stop vehicles when the barrier is closed. The next GameObject, \textit{Pivot}, is the parent object of the barrier beam. The pivot GameObject makes sure that the child object \textit{Beam} rotates around the right point in space. This again works because the child object rotates relative to the parent and because GameObjects can easily be rotated around their center. Therefore, the rotation point of the barrier beam can be changed by altering the location of \textit{Pivot} inside \textit{BoomBarrier\_Long} and the beam can be rotated by rotating \textit{Pivot}. \textit{Beam} contains one component, which is the mesh of the beam. The last Child component of \textit{BoomBarrier\_Long} is \textit{WarningBox}. This is a red transparent box around the barrier component that is used to indicate whether something is wrong with the control signals in the boom barrier. This GameObject is displayed in gray text in Figure \ref{fig:boom_barrier_prefab}, which indicates that the GameObject is disabled and therefore it is not shown inside Unity and its components can not do anything. GameObjects can be disabled and enabled from scripts, which is for instance useful in hiding geometry in the scene.\\
    
    This example shows that a system component in Unity can contain several GameObjects that all have different functions ranging from visualization, movement and interaction with other entities. While the structure of such a system component can be more compact, the usage of multiple GameObjects, with some even being empty, makes the setup of the entity much more clear and structured. Another reason is that for instance the meshes are in separate GameObjects, because these can then more easily be replaced by other meshes if desired.


\section{Controlled Entities}\label{sec:controlled_entities_setup}

    The main difference between non-controlled and controlled entities in the digital twin is that a controlled entity acts based on Boolean actuator values or set Boolean sensor values based on the state of entity in the digital twin. Boolean variables can have one of two values, either \textit{true} or \textit{false}. In general also non-Boolean values such as integers can be communicated to and from the PLC, but the supervisory controller of the Swalmen tunnel only uses Boolean values, so other PLC variable types are not discussed in this report. Sensor Booleans are called input variables, since they are input of the PLC, and actuator Booleans are called output variables, since these are output from the PLC. For each entity, for instance the boom barrier, an IO-script is written that handles the input and output variables for this entity.
    
    \subsection{General Setup of an IO-Script}\label{sec:IO_script_explanation}
    
    In Listing \ref{listing:controlled_components_structure} below the general way of working with Boolean signals in IO-scripts in Unity is shown. The script below shows the way to assign Boolean values to sensor variables (line 2) and the way to act on the state of Boolean actuator variable values (lines 5-7) in IO-scripts. Here, a Boolean value is referred to by \textit{IO\_dict[``s\_sensorName''].Boolean} for instance. This is a reference to an input variable with the name \textit{``s\_sensorName''} in the IO-script. The exact details of the communication between the PLC and the digital twin and how the logic values are handled in Unity are described in Chapter \ref{sec:PLC_control}.\\

    \begin{lstlisting}[caption={General way of defining the value of input variables and acting on actuator variables for controlled components in an IO-script in Unity.}, label={listing:controlled_components_structure},basicstyle=\scriptsize,frame=single]
// Set a value of a sensor boolean
IO_dict["s_sensorName"].Boolean = // boolean expression;
        
// Act based on an actuator boolean
if (IO_dict["a_actuatorName"].Boolean){
   // Do something;
}
    \end{lstlisting}

    \subsection{IO-Script Example: Boom Barrier}
    As an example of an IO-script, a part of the IO-script of a boom barrier, \textit{IO\_Barrier.cs}, is shown in Listing \ref{listing:IO_barrier_short}. Here, the actuator signals change the rotation direction of the motor that rotates the \textit{Pivot} GameObject as explained in Section \ref{sec:mod_comp_str}. For instance when the signal \textit{a\_close} becomes \textit{true}, the rotational direction of the motor is set to -1, meaning that the barrier closes. \\
    
    From line 28, sensor Boolean values are set. These are determined based on the position of the barrier beam for the states \textit{opened} and \textit{closed} and on the rotational direction of the motor for the other states.


    \begin{lstlisting}[caption={\textit{IO\_Barrier.cs} (incomplete and simplified).}, label={listing:IO_barrier_short},basicstyle=\scriptsize,frame=single]
public class IO_Barrier : MonoBehaviour
{
private void Update()
    {
        // Check whether multiple actuators are on and activate a red box signal
        if (Static_Functions.CountTrueActuators(IO_dict) > 1)
        {
            WarningBox.SetActive(true);
        }

        // --- Enable events based on the actuator states ---
        if (IO_dict["a_noChoice"].Boolean)
        {
            // Do nothing
        }
        else if (IO_dict["a_open"].Boolean)
        {
            motor.RotationDirection = 1;
        }
        else if (IO_dict["a_close"].Boolean)
        {
            motor.RotationDirection = -1;
        }
        else if (IO_dict["a_stop"].Boolean)
        {
            motor.RotationDirection = 0;
        }
        else
        {
            // Default state
            motor.RotationDirection = 0;
        }

        // --- Set the sensor values based on states of the components ---
        IO_dict["s_opened"].Boolean = 
            Mathf.Abs(motor.OpenRotation - motor.currentRotation) < SensorOffsetPos;
        IO_dict["s_opening"].Boolean =
            Mathf.Abs(motor.ClosedRotation - motor.currentRotation) < SensorOffsetPos;
        IO_dict["s_stopped"].Boolean =
            motor.RotationDirection == 0;
        IO_dict["s_closing"].Boolean =
            motor.RotationDirection == 1;
        IO_dict["s_closed"].Boolean =
            motor.RotationDirection == -1;

        // Move the collider out of the way when the barrier is opened
        if (IO_dict["s_opened"].Boolean)
        {
            BarrierCollider.transform.position = 
            	Vector3.Lerp(transform.position, colliderStartPosition + 
            	Vector3.up * 20, Time.deltaTime * 100);
        }
        else
        {
            BarrierCollider.transform.position = 
            	Vector3.Lerp(transform.position, colliderStartPosition, Time.deltaTime * 100);
        }
    }
}
    \end{lstlisting}

    In Unity, when a GameObject is selected, all components on the GameObject are shown in the \textit{inspector}. Here, the settings of a component can be changed. Figure \ref{fig:IO_barrier_inspector} below shows the inspector view of the script \textit{IO\_Barrier.cs}, that is put on an instance of a boom barrier. The settings that are displayed here are defined in the script, but this part was left out in the listing for readability. The complete script is shown in Listing \ref{listing:IO_Barrier} in Appendix \ref{sec:appendix_scripts}.

    \begin{figure}[H]
        \centering
        \includegraphics[width=.75\textwidth]{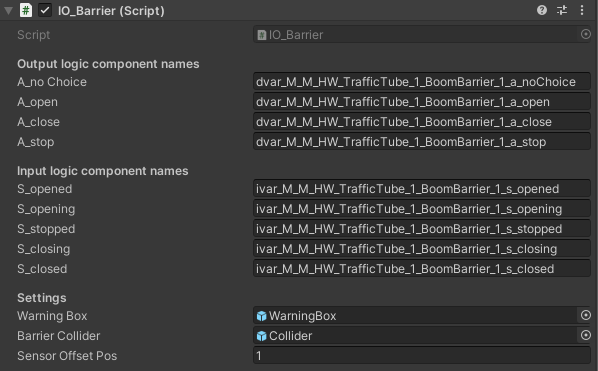}
        \caption{Inspector view of the IO-script of a barrier.}
        \label{fig:IO_barrier_inspector}
    \end{figure}
    
    The inspector view first shows the names of all actuator and sensor signal names that are filled in here. How these are then exactly linked to the input and output signals of the PLC is described in Chapter \ref{sec:PLC_control}. It is however important to note that the names can be filled in here, since this is essential for the modular nature of the instances in the digital twin. For instance, when two barrier-type instances are put in the digital twin, these usually use different signals for their actuators and sensors. These can then be set up separately for each component, such that the same IO-script can be used for both barrier instances.\\
    
    The final section in the inspector view of \textit{IO\_Barrier} shows the component settings. Here, initial conditions, variable values and references to other components and GameObjects are set. In this case, a reference to the warning box child GameObject in the Barrier GameObject structure is made as discussed in Section \ref{sec:mod_comp_str} is referenced. This means that the script can be used to enable or disable the warning box. The vehicle stopping collider is also referenced in order to regulate when vehicles are stopped. The next setting is a \textit{float} (real number) variable that determines the sensor offset that is allowed in measuring whether the barrier is opened or closed.

    \subsection{Control Validation Tools}

    Because the target of the digital twin made in this project is to validate the supervisory controller, it is useful to implement some features to aid the user in validating the controller. One such feature is already touched upon in Section \ref{sec:mod_comp_str}, where the \textit{WarningBox} GameObject was described briefly. This GameObject can be used to signal the user that something is wrong with the control signals in a component for instance. An implementation of a situation in which the warning box should be activated is given in Listing \ref{listing:warning_box_activation} below. This code snippet is placed in the IO-script of an entity in the digital twin. A function called \textit{Static\_Functions.CountTrueActuators()} counts how many output variables for the component are \textit{true}. When this value is larger than 1, the warning box is activated. This is useful, since a resource controller should not get conflicting signals. For example, a barrier should not get the command to open and close simultaneously.

    \begin{lstlisting}[caption={Code snippet that activates the warning box when multiple output variables in the controllable component are true.}, label={listing:warning_box_activation},basicstyle=\scriptsize,frame=single]
// Check whether multiple actuators are on and activate a red box signal
if (Static_Functions.CountTrueActuators(IO_dict) > 1)
{
   WarningBox.SetActive(true);
}
    \end{lstlisting}

\section{Non-Controlled Entities}

    Other entities in the digital twin are not controlled by the supervisory controller. These entities are environmental objects and objects that interact with controlled entities. They are modeled using the same scripts and components as the controlled entities, but do not use IO-scripts with PLC signals. The environment is important to get an idea of the layout of the plant and the position of controlled entities within the system, while other entities, such as vehicles or smoke, are used to create test scenarios for validation of the supervisory controller.

\section{User Interaction}

    For most entities in the digital twin that are used in the testing scenarios, it is important that interaction with the user of the digital twin is possible. Because the digital twin in this project is created in a game engine, it is fairly easy to add interactive components. Some examples of interactive components that are useful in a digital twin are discussed in this section.
    
    \subsection{GUI}\label{sec:GUI_button_example}
    The most straightforward way of interacting with the digital twin program is through buttons on a GUI (Graphical User Interface) on the screen. These buttons can for instance be used to navigate to different locations in the scene or to enable certain events. The example of setting the camera position and orientation via a button is explained here.\\
    
    In Unity, a \textit{canvas} first needs to be made, which is a GameObject in which a GUI can be made. As a child object of this canvas, a \textit{button} GameObject can be added directly, which can be placed on the desired place on the screen. A function in a script can be coupled to this button, such that this function is executed when the button is pressed. In this case, a script with the function of moving the camera is put on a separate GameObject. Multiple location hopping buttons can reference to this function. Listing \ref{listing:camera_moving_button_function} below shows this function, \textit{Button\_MoveFreeCamTo()}. It is important that this function is \textit{public} in order for it to be accessed by the buttons. The function takes in a string that should contain six real numbers that are separated by spaces. The first three determine the new $x-$, $y-$ and $z-$position of the camera in space and the other three determine the Euler angle rotation around the $x-$ $y-$ and $z-$axes, which sets the camera orientation.
    
    \begin{lstlisting}[caption={Script with a public function that moves the camera to a specified position and orientation.}, label={listing:camera_moving_button_function},basicstyle=\scriptsize,frame=single]
using UnityEngine;

public class GUI_CameraManager : MonoBehaviour
{
   [Header("Settings")]
        
   [Tooltip("Camera GameObject")]
   public GameObject FreeCamGO;
        
   public void Button_MoveFreeCamTo(string NewPose)
   {
      string[] parametersTextArray = NewPose.Split(" "[0]);
      float[] parameters = new float[parametersTextArray.Length];
        
      for (int i = 0; i < parametersTextArray.Length - 1; i++) {
        parameters[i] = float.Parse(parametersTextArray[i]);
      }
        
      FreeCamGO.transform.position =
        new Vector3(parameters[0], parameters[1], parameters[2]);
      FreeCamGO.transform.rotation =
        Quaternion.Euler(new Vector3(parameters[3], parameters[4], parameters[5]));
   }
}
    \end{lstlisting}
    
    The left side of Figure \ref{fig:Button_component} below shows the inspector view of a button component that uses the function from the script above. This component and the target graphic, which is the graphic that can be clicked, are created automatically when a \textit{button GameObject} is created. The script component containing the desired function is dragged to the box on the bottom left in the \textit{On Click ()} area. The function \textit{Button\_MoveFreeCamTo()} is then assigned on the right of \textit{Runtime Only}. On the bottom right, the six numbers that are input to the camera positioning function are given. These numbers can be extracted from a camera position that is desired by dragging the camera around and writing down the numbers that belong to this position and orientation. The right side of the figure shows the visual appearance of three buttons that use the method described in this section to move the camera to a desired position in the digital twin.
    
    \begin{figure}[H]
        \centering
        \includegraphics[width=.8\textwidth]{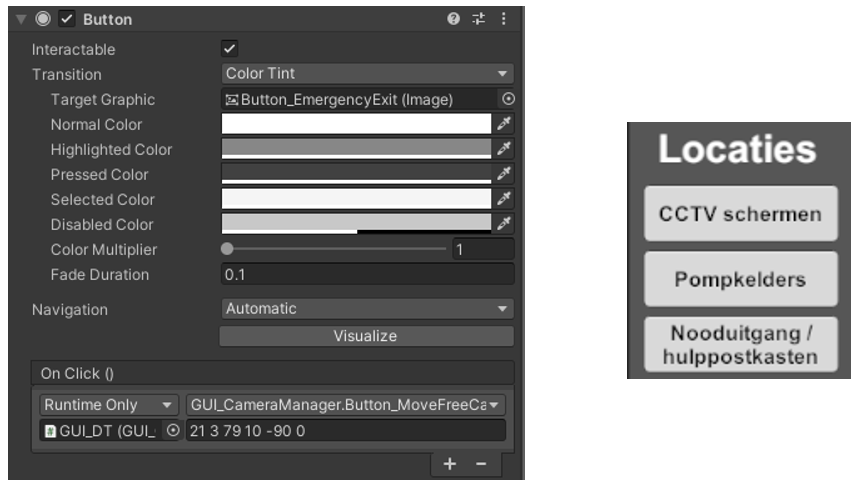}
        \caption{Inspector view of a button component that uses the function from Listing \ref{listing:camera_moving_button_function} (left) and the visual appearance of buttons that use this function on the GUI (right).}
        \label{fig:Button_component}
    \end{figure}
    
    \subsection{Mouse and Keyboard}\label{sec:interaction_mouse_keyboard}
    
    Apart from interaction via the GUI, the mouse and keyboard can also be used. One example is moving the camera around freely in the digital twin just like in a first-person video game. The mouse can then be used to look around. This is particularly useful for inspection of all areas in the digital twin, because it is not feasible to make buttons to go to each desired position in the system.\\
    
    The mouse can also be used to click on meshes of entities in the scene directly. In order for this to work correctly, colliders should be assigned to the GameObjects that are clickable. The technique that is then used is \textit{raycasting}, which means that a ray is cast from the camera viewpoint in the direction of the cursor when the left mouse button is clicked. A function then returns the first collider that is hit by this ray. Information about the object that is hit can then be used to start a certain event. This technique ensures that an object can only be clicked directly and not through other colliders. Listing \ref{listing:raycasting} below shows a piece of code that sends out a ray and can be used to carry out an action based on the name of the GameObject that is hit by the ray. The condition \textit{if (!(EventSystem.current.IsPointerOverGameObject()))} ensures that a ray is not cast when the GUI is clicked. This is to prevent an object in the scene from being clicked when the user intends to click a button on the GUI. This clicking method can for instance be used to open a door when the door mesh is clicked.
  
\begin{lstlisting}[caption={Sample of C\#-code that uses raycasting for interaction.}, label={listing:raycasting},basicstyle=\scriptsize,frame=single]
if (Input.GetMouseButtonDown(0))
{
	if (!(EventSystem.current.IsPointerOverGameObject()))
    {
    	if (Physics.Raycast(ray, out RaycastHit hit))
        {
        	if (hit.transform.name == "ObjectName")
            {
            	// Do something
            }
        }
	}
}
\end{lstlisting}
    
\section{Conclusion}

Using the methods described in this chapter, a digital twin can be set up in a modular way. 3D models can be constructed in SketchUp and then imported into Unity. Here, behavior scripts or other components are added to make controlled and uncontrolled entities that can be used in a modular fashion inside the digital twin. The controlled entities require IO-scripts that are also described in detail and the exact PLC signal details are described in the next chapter (Chapter \ref{sec:PLC_control}). To make the digital twin visually attractive, realistic material textures can be added to the 3D models inside Unity. Another visual improvement is to use post-processing techniques for a better graphical representation of the scene.

\chapter{PLC Control}\label{sec:PLC_control}

In this chapter, the method of setting up a PLC connection between the controller made with the CIF 3 Toolbox and the digital twin is explained. First, the PLC code generation process of the supervisory controller is described, after which the implementation process in TwinCAT is treated. Next, the PLC logic preparation in the digital twin in Unity with Prespective logic components is explained and finally the connection between the digital twin and the controller PLC running in TwinCAT is made. To prevent confusion, input and output signals are defined from the viewpoint of the PLC. Output signals go from the PLC to the digital twin and activate actuators, and input signals go from the digital twin to the PLC and represent button and sensor signals. All PLC signals used in the communication with the digital twin are Boolean, so they are either \textit{true} or \textit{false}.\\

Throughout this chapter, one of the boom barriers, more specifically boom barrier 1 in traffic tube 1, is taken as an example, because the function of this component is simple and the component has multiple inputs and outputs. This boom barrier is also present in the digital twin of the Swalmen tunnel. An emphasis is put on the automatic generation of logic components and how these logic components are referenced by IO-scripts in the digital twin. These methods have been developed to enhance the modular nature of the setup of the digital twin.

\section{CIF PLC Code Generation}\label{sec:PLC_code_generation}

As explained in Section \ref{sec:prev_work}, the supervisory controller that is created from the CIF3 model of the Swalmen Tunnel is used for controlling the digital twin and serves as reference material for the digital twin. In order to generate PLC code of a controller created with CIF3, a hardware mapping CIF specification is needed that defines the input and output signals of the supervisor. The combination of the supervisory controller and this hardware mapping undergoes some transformations and is finally converted to PLC code.

\subsection{Hardware Mapping}\label{sec:hardware_mapping}
The term hardware mapping comes from the fact that this CIF specification defines which virtual signals are mapped to the signals that usually exist in the PLC. The `signals' in CIF specifications are modeled with edges and locations in the controller. In CIF, an \textit{automaton}, which can represent a boom barrier for example, contains several locations. Examples of locations of a boom barrier model are \textit{opening} and \textit{closing}, that can represent whether the barrier is opening or closing. \textit{Edges} represent transitions between these locations, so an example of an edge is \textit{open} which can be the transition from the \textit{closing} to the \textit{opening} location.\\

Listing \ref{listing:hw_cif_barrier} below shows the part of the hardware mapping script for the supervisory controller of the Swalmen Tunnel that maps the inputs and outputs for the boom barrier to input and output signals. This boom barrier hardware mapping is an automaton that contains only one location. All edges can potentially be taken when they are not blocked by their \textit{guard}. The output signals are \textit{discrete bools} and are defined on line 4. The value of the inputs is based on the location of the boom barrier (``afsluitboom'' in Dutch) automaton in the supervisor. The value of each output variable is updated in the first four edges of the location (lines 11 to 18). The input signals are defined as \textit{input bools} on line 7. These \textit{input bools} block certain edges from being taken in lines 20 to 26 when the input Booleans are \textit{false}. This means that the values of the input variables determine which uncontrollable edges can be taken in the supervisor. The Dutch terms in this script are present due to references to the existing CIF model that is written in Dutch.

\begin{lstlisting}[caption={Boom barrier 1 section of traffic tube 1 in the hardware mapping script for the supervisory controller.}, label={listing:hw_cif_barrier},basicstyle=\scriptsize,frame=single]
automaton BoomBarrier_1:

// --- OUTPUTS ---
disc bool   a_noChoice, a_open, a_stop, a_close;

// --- INPUTS ---
input bool  s_opened, s_opening, s_stopped, s_closing, s_closed, s_obst_on, s_obst_off;

location: initial;
// --- UPDATING OUTPUTS ---
edge when a_noChoice  != Tunnel.Verkeersbuis1.Afsluitboom1.AansturingBedrijftoestand.geenKeuze
	do a_noChoice  := Tunnel.Verkeersbuis1.Afsluitboom1.AansturingBedrijftoestand.geenKeuze;
edge when a_open      != Tunnel.Verkeersbuis1.Afsluitboom1.AansturingBedrijftoestand.op
	do a_open      := Tunnel.Verkeersbuis1.Afsluitboom1.AansturingBedrijftoestand.op;
edge when a_stop      != Tunnel.Verkeersbuis1.Afsluitboom1.AansturingBedrijftoestand.stop
	do a_stop      := Tunnel.Verkeersbuis1.Afsluitboom1.AansturingBedrijftoestand.stop;
edge when a_close     != Tunnel.Verkeersbuis1.Afsluitboom1.AansturingBedrijftoestand.neer
	do a_close     := Tunnel.Verkeersbuis1.Afsluitboom1.AansturingBedrijftoestand.neer;
// --- ENABLING INPUT EVENTS ---
edge Tunnel.Verkeersbuis1.Afsluitboom1.TerugmeldingBedrijftoestand.u_op         when s_opened;
edge Tunnel.Verkeersbuis1.Afsluitboom1.TerugmeldingBedrijftoestand.u_opgaand    when s_opening;
edge Tunnel.Verkeersbuis1.Afsluitboom1.TerugmeldingBedrijftoestand.u_stop       when s_stopped;
edge Tunnel.Verkeersbuis1.Afsluitboom1.TerugmeldingBedrijftoestand.u_neergaand  when s_closing;
edge Tunnel.Verkeersbuis1.Afsluitboom1.TerugmeldingBedrijftoestand.u_neer       when s_closed;
edge Tunnel.Verkeersbuis1.Afsluitboom1.ObstakelDetectie.u_aan                   when s_obst_on;
edge Tunnel.Verkeersbuis1.Afsluitboom1.ObstakelDetectie.u_uit                   when s_obst_off;

end
\end{lstlisting}

In general, it is preferred to use definitions in CIF specifications to reduce the length of the specification and to improve readability. Definitions can be seen as 'blueprints' for automatons that are used multiple times without repeating the code for the whole automaton. The inputs for an instance of a definition can be names of locations and edges. While it is the case in the model of the Swalmen Tunnel that some components are used multiple times, the choice is made to write the hardware mapping without definitions, as most of them would only be used two times. Furthermore, looking at the hardware mapping of the boom barrier in Listing \ref{listing:hw_cif_barrier}, it is clear that there are a lot of references to the supervisory controller that contain many characters. In a definition, these long references to edges and locations would have to be inserted as variables, which would make the benefit of these definitions small since they are not used frequently. In other CIF models, such as the one used in \cite{reijnen_waterway_synthesis_control}, the model is set up such that these references are automatically made in a structured way, but applying this to the CIF model of the Swalmen Tunnel that has been made before, would mean that all CIF specifications in the model need to be rewritten. This is not done due to time limitations and because this is not the goal of this project.

\subsection{Transformations and PLC Code Generation}
Having defined all input and output variables in the hardware mapping, the PLC code can be generated. Before generating PLC code, first some CIF transformations are applied to the combination of the supervisor and hardware mapping files. These transformations are carried out using the \textit{cif3cif()}-function from the CIF3 toolbox.\\

First, the event exclusion invariants are eliminated and secondly the algebraic variables are removed. These two transformation are necessary because the PLC generation function in the CIF toolbox does not support these attributes. Thirdly, the model is linearized, which eliminates the use of locations and expressions and simplifies the model before generating PLC code. These transformations together transform the CIF specification of the controller with its hardware mapping from a specification with many different locations and exclusion invariants into a shorter but hardly readable specification where all edges are linked to the states of all components. An example for the edge $x$ that may or may not be taken based on the state of all Booleans $a,~b,~c,~ d,~...$ is:

\begin{lstlisting}[]
    edge x when a AND b OR c AND NOT d OR ...
\end{lstlisting}

Finally, PLC code is generated from the transformed CIF specification to an \textit{xml}-file using the \textit{cif3plc()}-function from the CIF Toolbox. A \textit{tooldef2}-script (a script that can carry out functions from the CIF 3 Toolbox) that contains the CIF transformations and PLC generation code is given in Listing \ref{listing:ST_DT_PLC.cif} in Appendix \ref{sec:appendix_scripts}.

\section{Virtual PLC in TwinCAT}\label{sec:plc_cif_twincat}

The \textit{xml}-file that is generated from the supervisory controller is loaded into TwinCAT in order to be able to run and read the generated PLC code. A CIF folder inside a TwinCAT PLC project is shown in Figure \ref{fig:CIF_in_TwinCAT} below. The generated \textit{xml}-file is imported into this folder. Two of the imported files are important for the inputs and outputs. The first one is the GVL-file (\textit{Global Variable List}) \textit{INPUTS} that contains all input variable names of which a small portion is shown in Listing \ref{listing:input_var_names_boombarrier}.

\begin{figure}[H]
    \centering
    \includegraphics[width=.275\textwidth]{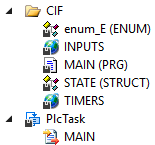}
    \caption{Result of loading in the generated \textit{xml}-file from the CIF specifications in TwinCAT.}
    \label{fig:CIF_in_TwinCAT}
\end{figure}

\begin{lstlisting}[caption={Part of the INPUTS GVL that contains the boom barrier input variable names.}, label={listing:input_var_names_boombarrier},basicstyle=\scriptsize,frame=single]
...
ivar_M_M_HW_TrafficTube_1_BoomBarrier_1_s_opened AT %I*: BOOL;
ivar_M_M_HW_TrafficTube_1_BoomBarrier_1_s_opening AT %I*: BOOL;
ivar_M_M_HW_TrafficTube_1_BoomBarrier_1_s_stopped AT %I*: BOOL;
ivar_M_M_HW_TrafficTube_1_BoomBarrier_1_s_closing AT %I*: BOOL;
ivar_M_M_HW_TrafficTube_1_BoomBarrier_1_s_closed AT %I*: BOOL;
ivar_M_M_HW_TrafficTube_1_BoomBarrier_1_s_obst_on AT %I*: BOOL;
ivar_M_M_HW_TrafficTube_1_BoomBarrier_1_s_obst_off AT %I*: BOOL;
...
\end{lstlisting}

Comparing this with the hardware mapping file section belonging to this component (Listing \ref{listing:hw_cif_barrier}), it is observed that these are indeed the input variables defined in the hardware mapping. The prefix \textit{"ivar"} indicates that the variable is an input, the part \textit{"M\_M\_"} is a byproduct of the CIF transformations and \textit{"HW\_TrafficTube\_1\_BoomBarrier\_1"} is a reference to the component in the CIF controller hardware mapping structure.\\

The other important file created in TwinCAT is \textit{STATE (STRUCT)}, the file that contains the output variables as partially shown in Listing \ref{listing:dvar_names_boombarrier} below.

\begin{lstlisting}[caption={Part of STATE (STRUCT) with the boom barrier output variable names.}, label={listing:dvar_names_boombarrier},basicstyle=\scriptsize,frame=single]
...
dvar_M_M_HW_TrafficTube_1_BoomBarrier_1_a_noChoice AT %Q*: BOOL;
dvar_M_M_HW_TrafficTube_1_BoomBarrier_1_a_open AT %Q*: BOOL;
dvar_M_M_HW_TrafficTube_1_BoomBarrier_1_a_stop AT %Q*: BOOL;
dvar_M_M_HW_TrafficTube_1_BoomBarrier_1_a_close AT %Q*: BOOL;
dvar_M_M_HW_TrafficTube_1_BoomBarrier_1 AT %Q*: enum_E;
...
dvar_M_M_Tunnel_Verkeersbuis1_Afsluitboom1_AansturingBedrijftoestand AT %Q*: enum_E;
dvar_M_M_Tunnel_Verkeersbuis1_Afsluitboom1_ObstakelDetectie AT %Q*: enum_E;
dvar_M_M_Tunnel_Verkeersbuis1_Afsluitboom1_TerugmeldingBedrijftoestand AT %Q*: enum_E;
...
\end{lstlisting}

These variables are not all of the same type, as some are Booleans and some are of the type \textit{enum\_E}. This is because this file contains all discrete variables, which are not all output variables from the hardware mapping. The \textit{enum\_E} type entries are the automatons or plants in the original CIF specification of the controller and the BOOL type entries that contain \textit{``HW''} in their name are the output variables that correspond to the discrete Booleans in Listing \ref{listing:hw_cif_barrier}. The most important file in Figure \ref{fig:CIF_in_TwinCAT} is \textit{MAIN (PRG)}. This file is the main PLC program that includes all logic.

\section{Digital Twin Logic Components}

\begin{figure}[H]
    \centering
    \includegraphics[width=.6\textwidth]{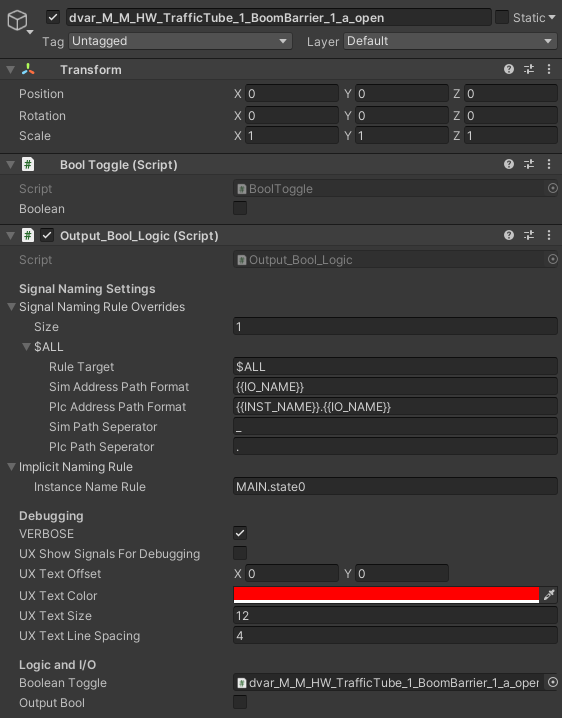}
    \caption{Example of a logic component GameObject.}
    \label{fig:logicComponent_output_a_open}
    \vspace{-3mm}
\end{figure}

In the digital twin in Unity, components should be able to react to output signals from, and to send input signals to the PLC running in TwinCAT. This is done using logic component GameObjects, of which the inspector view of an example is shown in Figure \ref{fig:logicComponent_output_a_open}. These GameObjects can be seen as Booleans that are used for communication with a PLC.\\ 

\subsection{Logic Component Prefabs}\label{sec:logic_component_prefabs}

A logic component GameObject contains two scripts, of which one is the BoolToggle script that is shown in Listing \ref{listing:boolToggleScript} below. This script contains a Boolean variable that is used for communication with component GameObjects such as the boom barrier in the digital twin as explained in Subsection \ref{sec:ref_logic_comps_IO}. The second script is either a logic component output script that changes the Boolean in the BoolToggle script based on a PLC signal, or a logic component input script that reads the BoolToggle Boolean value and sends it to the PLC. A detailed description of these scripts is given in Subsection \ref{sec:logic_component_generation}. A logic component GameObject thus communicates with a PLC and with other GameObject components in the digital twin.\\

In short, the BoolToggle script holds a Boolean value, and the logic component script communicates this Boolean value to the PLC in case of an input, or changes this Boolean value according to a PLC output signal. The logic components for an input and output can be saved as a \textit{prefab} in Unity. Prefabs can be used as templates for GameObjects.

\begin{figure}[H]
    \begin{minipage}{.45\textwidth}
\begin{lstlisting}[caption={BoolToggle script used for connecting Boolean in- and outputs to the components in the digital twin.}, label={listing:boolToggleScript},basicstyle=\scriptsize,frame=single]
public class BoolToggle : MonoBehaviour
{
//Boolean Value of the toggle
public bool Boolean = false ;
public void Toggle( bool oToggle )
{
Boolean = oToggle ;
}
}
\end{lstlisting}
    \end{minipage}
    \qquad
    \begin{minipage}{0.45\textwidth}
        \vspace{-3mm}
        \includegraphics[width=\textwidth]{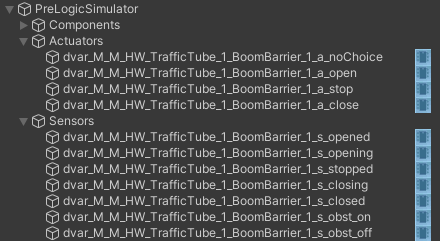}
        \vspace{-3.5mm}
        \caption{Logic component structure in Unity (only the logic components for the boom barrier are shown).}
        \label{fig:logic_components_boom_barrier}
        \vspace{-3mm}
    \end{minipage}
\end{figure}

The logic components for a boom barrier are shown in Figure \ref{fig:logic_components_boom_barrier}. These are sorted under \textit{empty GameObjects}. This means that these GameObjects do not contain any components, but they can have child objects. \textit{Actuators} contains all output variables and \textit{Sensors} and \textit{Buttons} contain the input variables. In the case of the single boom barrier, 4 actuators and 7 sensors are present. The logic component containers are child objects of a \textit{PreLogicSimulator} component that is included in the Prespective plugin in Unity. This component establishes the connection between the digital twin in Unity and the PLC in TwinCAT and is described in more detail later in Section \ref{sec:prelogicsimulator}. It is important that the logic components have the same names as the PLC variables in the generated PLC code that is described earlier in this chapter.

\subsection{Referencing Logic Components}\label{sec:ref_logic_comps_IO}

\begin{figure}[H]
    \centering
    \vspace{-4mm}
    \includegraphics[width=.45\textwidth]{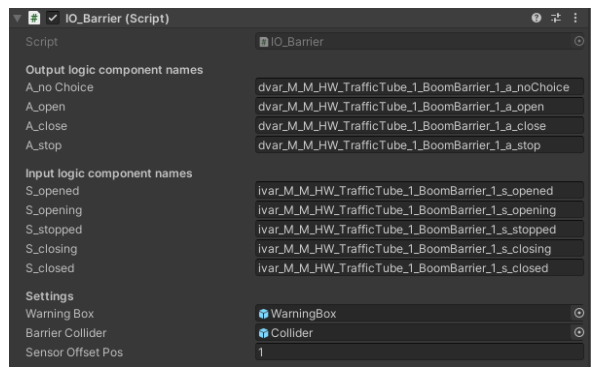}
    \vspace{-1mm}
    \caption{Inspector view of \textit{IO\_Barrier.cs}.}
    \label{fig:IO_Barrier_cs_inspector_2}
    \vspace{-1mm}
\end{figure}

As described, the BoolToggle Booleans in the Logic components are used to communicate with components in the digital twin. To do this, the IO-scripts introduced in Section \ref{sec:IO_script_explanation} reference the logic components. This is explained here in detail for the boom barrier example. The inspector view of \textit{IO\_Barrier.cs} is shown again in Figure \ref{fig:IO_Barrier_cs_inspector_2}. This component contains references to the logic component GameObjects from Figure \ref{fig:logic_components_boom_barrier} by their names.\\

A simplified version of the script \textit{IO\_Barrier.cs} is shown (that shows an other part than the version shown in Section \ref{sec:IO_script_explanation}) in Listing \ref{listing:IO_Barrier_referencing_logiccomponents} below.

\vspace{3mm}    
\begin{lstlisting}[caption={Incomplete version of \textit{IO\_Barrier.cs} that shows the way at which BoolToggle scripts of logic components are referenced.}, label={listing:IO_Barrier_referencing_logiccomponents},basicstyle=\scriptsize,frame=single]

public class IO_Barrier : MonoBehaviour
{
    [Header("Output logic component names")]
    
        [SerializeField, Tooltip("Open the barrier")]
        private string a_open;

    [Header("Input logic component names")]

        [SerializeField, Tooltip("The barrier is fully opened")]
        private string s_opened;
        
    public Dictionary<string, BoolToggle> IO_dict = new Dictionary<string, BoolToggle>();

    private void Start()
    {
        // Filling the I/O dictionary
            // Actuators
            IO_dict.Add("a_open", Static_Functions.FindBoolToggle(a_open));

            // Sensors
            IO_dict.Add("s_opened", Static_Functions.FindBoolToggle(s_opened));
    }

    private void Update()
    {
        // --- Enable events based on the actuator states ---
        if (IO_dict["a_open"].Boolean)
        {
            motor.RotationDirection = 1;
        }
        else
        {
            // Default state
            motor.RotationDirection = 0;
        }

        // --- Set the sensor values based on states of the components ---
        IO_dict["s_opened"].Boolean = 
            (Mathf.Abs(motor.OpenRotation - motor.currentRotation) < SensorOffsetPos);
    }
}
\end{lstlisting}

The script starts with the declaration of string variables for an actuator and a sensor (lines 4-12). The strings that are declared here can be filled in in the inspector view that was shown before. The strings that are entered in the inspector view of the script should be equal to the names of the logic components that need to be used in the controlled entity.\\

Then, in line 14, a dictionary is declared. A dictionary is a data container that is essentially a list where (usually) a string is assigned to each entry in the list. The string is then called the key of the entry (the value) in the list. The key can also be of another type, such as a float or int. The key is then used to refer to the value in the dictionary that belongs to it. It can be compared to a real-life dictionary, where a word (key) links to a certain meaning (value). In this case, the dictionary has keys of type string and values of type BoolToggle. Then, in the start function in lines 16-24, the dictionary is filled. Line 21 for example adds in item to the dictionary with key \textit{``a\_open''} and as value the BoolToggle that has the exact name of the string value assigned to the variable \textit{a\_open}. Finding this BoolToggle is done with the function \textit{FindBoolToggle()} that is located in a script that is called \textit{Static\_Functions.cs} and can be found in Appendix \ref{listing:Static_Functions}. This function loops through the actuator and sensor logic component containers (Figure \ref{fig:logic_components_boom_barrier}) in the scene and finds the logic component with the name given as the argument in the function. The reason that the function is in a separate script is that this enables reuse of the function such that it does not need to be copied to each IO-script. After the dictionary is filled, the BoolToggles can be used to change input Booleans or to act on output Booleans as explained in Section \ref{sec:IO_script_explanation}.\\

The main reason that the BoolToggles are stored in a dictionary is readability. Initially an array was used for storing the BoolToggle references, but using indices in the \textit{Update()}-loop can be confusing, since numbers do not say anything. By using the dictionary it is immediately clear what the triggered action of the \textit{"...a\_close"} BoolToggle is and how the sensor values are defined. This is especially helpful in \textit{IO} scripts with a large number of inputs and outputs.\\

\section{Linking the Digital Twin and TwinCAT} \label{sec:logicDT_PLC}

In the beginning of this section, the functionality of logic component GameObjects are explained and the BoolToggle script is described. Here, the Boolean Logic scripts for both the input and output are described in detail.

\subsection{Output Logic Components}\label{sec:output_logic_component_script}

Prespective enables communication with a PLC through generation of an \textit{xml}-file in Unity. This file contains all inputs and outputs in the digital twin. TwinCAT uses these signal definitions to communicate with the digital twin in Unity. In the first section of the output logic component script, the Boolean output signal definition is created. Its name is equal to the name of the GameObject. The signal is defined as:

\begin{lstlisting}[basicstyle=\scriptsize,frame=single]
new SignalDefinition(gameObject.name, PLCSignalDirection.OUTPUT, 
    	SupportedSignalType.BOOL, "", "Value", onSignalChanged, null, false)
\end{lstlisting}

The script then runs the \textit{onSignalChanged()} function from the next listing in each frame. The identifier string of each incoming signal \textit{<\_signal>} is compared to the name of the logic component GameObject. When these strings are the same, the Boolean variable \textit{OutputBool} changes its value to the value \textit{<\_newValue>} of the signal \textit{<\_signal>}. Then, the BoolToggle script on the logic component GameObject changes its value when this is necessary. In this way, the Boolean in BooleanToggle changes when the corresponding PLC variable value changes.

\begin{lstlisting}[basicstyle=\scriptsize,frame=single]
void onSignalChanged(SignalInstance _signal, object _newValue, DateTime _newValueReceived, 
		object _oldValue, DateTime _oldValueReceived)
{
	if (_signal.definition.defaultSignalName == gameObject.name)
    {
    	OutputBool = (bool)_newValue;
        if (BooleanToggle.Boolean != OutputBool)
        {
        	BooleanToggle.Boolean = OutputBool;
        }
    }
}
\end{lstlisting}

\subsection{Input Logic Components}

For the input logic components, a similar script is used. First, the signal is defined as a Boolean input value with the name of the logic component GameObject as its identifier.

\begin{lstlisting}[basicstyle=\scriptsize,frame=single]
new SignalDefinition(gameObject.name, PLCSignalDirection.INPUT, SupportedSignalType.BOOL, "",
	 gameObject.name, null, null, false)
\end{lstlisting}

The \textit{onSimulatorUpdated()} function is then called each frame. The word \textit{``simulator''} refers to the digital twin in Unity. This function calls \textit{readComponent()} that checks whether the BooleanToggle Boolean on the logic component GameObject has changed. When this is the case, its new Boolean value is written to the defined PLC signal and sent to the PLC. This means that when the Boolean in BooleanToggle changes in the digital twin, the corresponding PLC value will also change. The full scripts for the input and output logic components can be found in Appendix \ref{sec:appendix_scripts}.

\begin{lstlisting}[basicstyle=\scriptsize,frame=single]
protected override void onSimulatorUpdated(int _simFrame, float _deltaTime, 
	float _totalSimRunTime, DateTime _simStart)
{
	readComponent();
}
    
void readComponent()
{
	if (BooleanToggle.Boolean != InputBool)
    {
    	InputBool = BooleanToggle.Boolean;
        WriteValue(gameObject.name, InputBool);
	}
}
\end{lstlisting}

\subsection{Logic Component Naming Rules}

In the earlier digital twins (Section \ref{sec:prev_work}), an additional signal mapping file was used inside TwinCAT to link the variables from the digital twin to the PLC variables. To overcome this, the names of the variable names of the digital twin are already equal to the PLC variable names as described in earlier sections of this chapter. An additional step needed for this to work is that the logic components refer to the location of the input and output variables in the PLC. This is done with \textit{naming rules} as shown below in Figure \ref{fig:logiccomponents_namerulings}.\\

\begin{figure}[H]
    \centering
    \includegraphics[width=\textwidth]{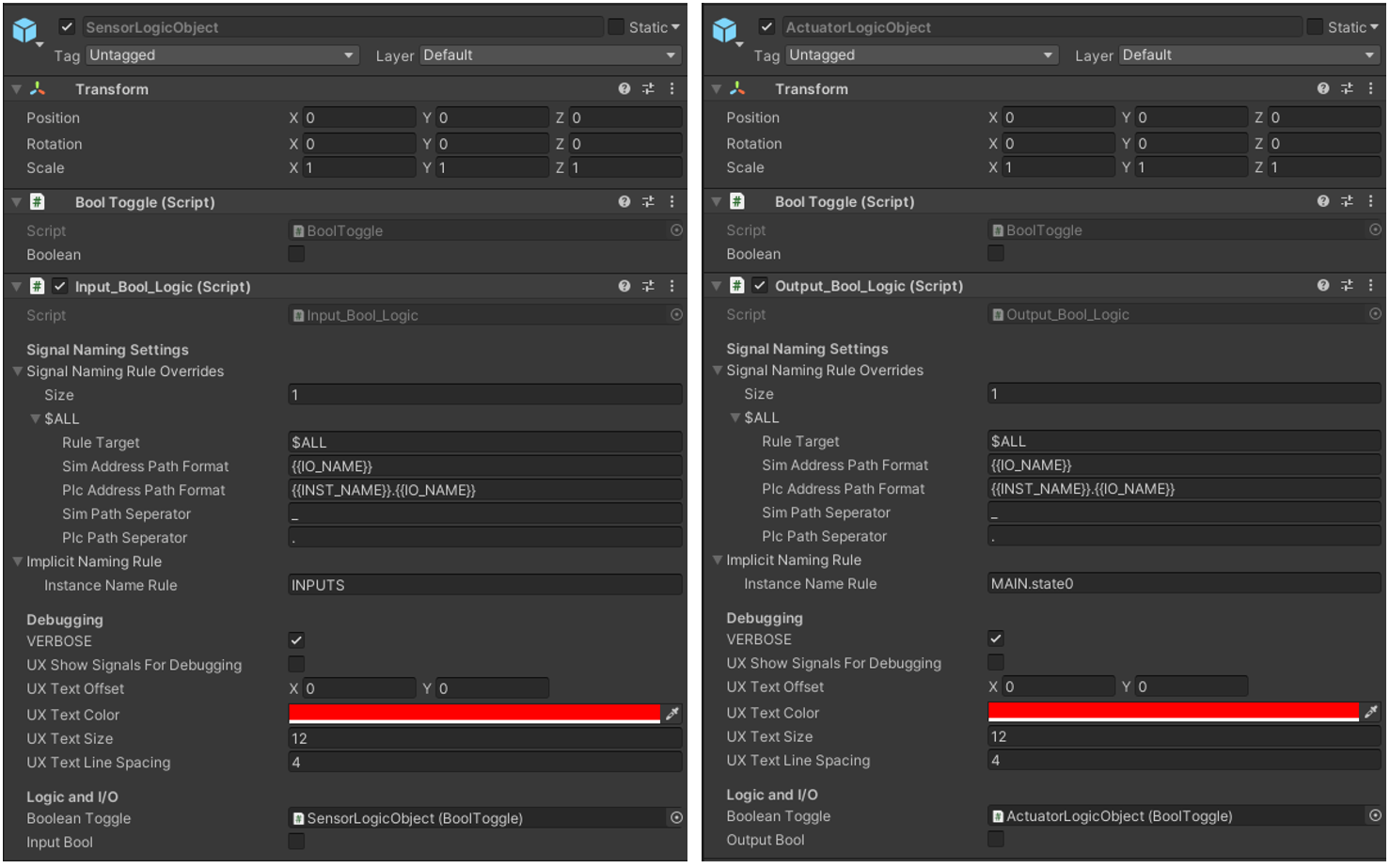}
    \caption{Settings of the logic components for the inputs (left) and outputs (right).}
    \label{fig:logiccomponents_namerulings}
\end{figure}

As shown in the figure above, the naming rule that should be set for the output logic components is \textit{"MAIN.state0.\{\{IO\_NAME\}\}"}, where \textit{"\{\{IO\_NAME\}\}"} refers to the input or output signal name, which is equal to the name of the logic component GameObject. These strings are filled separately in under the \textit{Signal Naming Settings}. The part "MAIN.state0" refers to the MAIN (PRG) file in TwinCAT, where the addresses for the output variables are defined and updated. The naming rule for the input logic components is \textit{"INPUTS.\{\{IO\_NAME\}\}"} that corresponds to the file name for the input variables in TwinCAT.

\subsection{Logic Component Generation} \label{sec:logic_component_generation}

The main differences with respect to the digital twins developed on the TU/e earlier (as discussed in Section \ref{sec:prev_work}) is that the Swalmen Tunnel PLC code contains many more Boolean PLC input and output signals. The highest quantity for the digital twins of the Festo systems is 73 Boolean signals and the supervisory controller of the Swalmen Tunnel communicates through more than 500 Booleans (including the operator interface buttons). With such a large number of logic components, it is a tedious task to create and name all logic component GameObjects for all input and output signals by hand. In order to overcome this, a method for the automatic generation of all logic components based on the input and output variable lists in TwinCAT is developed as shown in Listings \ref{listing:input_var_names_boombarrier} and \ref{listing:dvar_names_boombarrier}. This generates a logic component structure, such as in the example shown in Figure \ref{fig:logic_components_boom_barrier}.\\

It is also possible to work with one large logic component script that contains all input and output signal definitions, but this script would become very long. Also, when the controller is still in development, the number of inputs or outputs can change, meaning that these large scripts then need to be altered manually. This is not the case when the logic components are generated from the PLC input and output lists from TwinCAT, since these are directly generated from the PLC code belonging to the updated controller.\\

\begin{lstlisting}[caption={Simplified and incomplete actuator generation function.}, label={listing:actuator_generation},basicstyle=\tiny,frame=single]
private void Generate_Actuators()
{
    // Find the file path and read in the name list
    outputNames = System.IO.Path.Combine(Application.dataPath, "_outputNames.txt");
    outputList = System.IO.File.ReadAllLines(outputNames);

    foreach (string outputLine in outputList)
    {
        // varName is the first 'word' on the the line
        string varName = outputLine.Trim().Split((" ").ToCharArray()[0])[0];

        // Instantiate an actuator if the line contains a correct variable name of the correct type
        if (varName.StartsWith("dvar") && outputLine.Contains("_HW_"))
        {
            InstantiateLogicComponent(varName, actuatorPrefab, actuatorParent);
        }
    }
    
    void InstantiateLogicComponent(string ComponentName, GameObject Prefab, Transform ComponentParent)
    {
        // Instantiate the logic object, set the desired name and assign it to the desired parent
        GameObject newLogicComponent = Instantiate(Prefab, Vector3.zero, Quaternion.identity);
        newLogicComponent.name = ComponentName;
        newLogicComponent.transform.parent = ComponentParent;
    }
}
\end{lstlisting}

\begin{figure}[H]
    \centering
    \includegraphics[width=\textwidth]{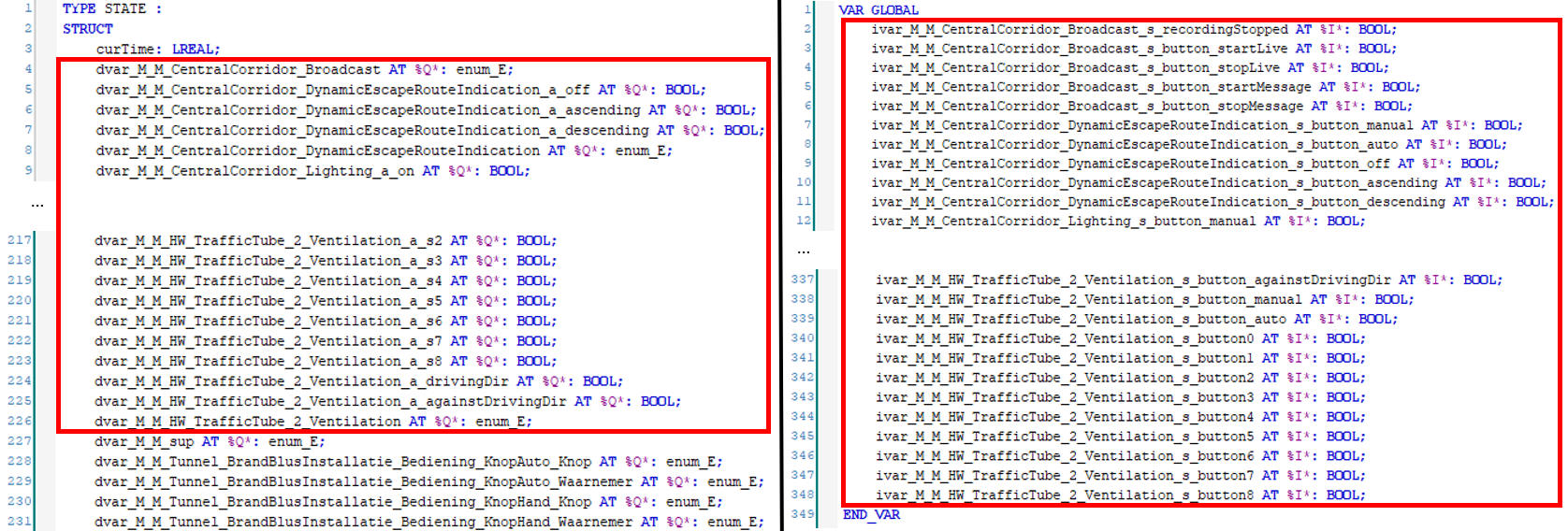}
    \caption{Relevant lines of the output variable list in STATE (STRUCT) (left) and INPUTS (right).}
    \label{fig:Relevant_parts_variable_lists_IO}
\end{figure}

In Listing \ref{listing:actuator_generation} above, a simplified part of the generation script is shown. Only the actuator generation part is shown here, as the sensor generation is carried out in the same manner. The actuator names are first read from a text file that is called \textit{"\_outputNames.txt"}. The full contents of the \textit{STATE (STRUCT)} file in TwinCAT (see Figure \ref{sec:plc_cif_twincat}) should be copied to \textit{"\_outputNames.txt"}. A small part of this file in TwinCAT is shown in Listing \ref{listing:dvar_names_boombarrier}. The relevant parts from TwinCAT for both the input and output variables are shown in Figure \ref{fig:Relevant_parts_variable_lists_IO}. The relevant part contains all Boolean type \textit{dvar} variables and contains all hardware mapped outputs that are indicated with \textit{HW} in the variable names. As can be seen, there are also variables of type \textit{enum\_E} in these lines that should not be selected. This is solved in the generation process as described below. The relevant lines in the INPUTS file are all lines containing code, except for the first ("\textit{VAR\_GLOBAL}") and last ("\textit{END\_VAR}") lines.\\

In line 5 of Listing \ref{listing:actuator_generation} above, all lines of the outputs text file are put in a list of type string. The next part of the function (line 7-17) loops through this list and takes the first `word' of each line. This `word' is the variable name when the line contains an output variable. The `word' is only used if it starts with \textit{``dvar''} and contains the string \textit{``BOOL''}, since only the Boolean variables are outputs. Additionally, \textit{``\_HW\_''} should be present, because all hardware mapping variables contain this string. This ensures that discrete Booleans used in the supervisor which are not output variables are not selected. This selection process ensures that only relevant variable names are used.

Then, in line 15 of Listing \ref{listing:actuator_generation}, an output logic component GameObject is generated from a logic component prefab for each actuator variable name. This GameObject is then put under the \textit{Actuators} GameObject. Because the logic component scripts take the name of the GameObject as their PLC variable name, the same variable names are used in communication between TwinCAT and the Digital Twin and in the PLC code of the supervisory controller. This eliminates the need of writing a mapping file in TwinCAT that maps the control variable names of the digital twin to the variable names in the PLC code of the supervisory controller.\\

An additional step that is necessary in order for this method to work is that the naming rules, as defined in the Bool Logic Component scripts, match the naming rules for the input and output variables from CIF. This is applied in the prefab logic component GameObjects that are copied and renamed each time when an output or input logic component GameObject is generated as described in Section \ref{fig:logiccomponents_namerulings}. As was shown in Figure \ref{fig:logic_components_boom_barrier}, there are two groups of logic components, being the actuators and sensors. Here, two different logic component prefabs are used. As described earlier, these logic components are generated in the same way as the example of the actuators in this section.


\section{Setting up the Prelogic Simulator}\label{sec:prelogicsimulator}

When all logic components are generated, all input and output signals are present inside Unity. In order to enable communication between the digital twin and the PLC, an \textit{xml} policy is needed that specifies all PLC variables present in the digital twin with their addresses. This is what the \textit{Prelogic Simulator} from the Prespective plugin does. Below in Figure \ref{fig:prelogicsimulator_inspector} the inspector view of the Prelogic Simulator component is shown.

\begin{figure}[H]
    \centering
    \includegraphics[width=.79\textwidth]{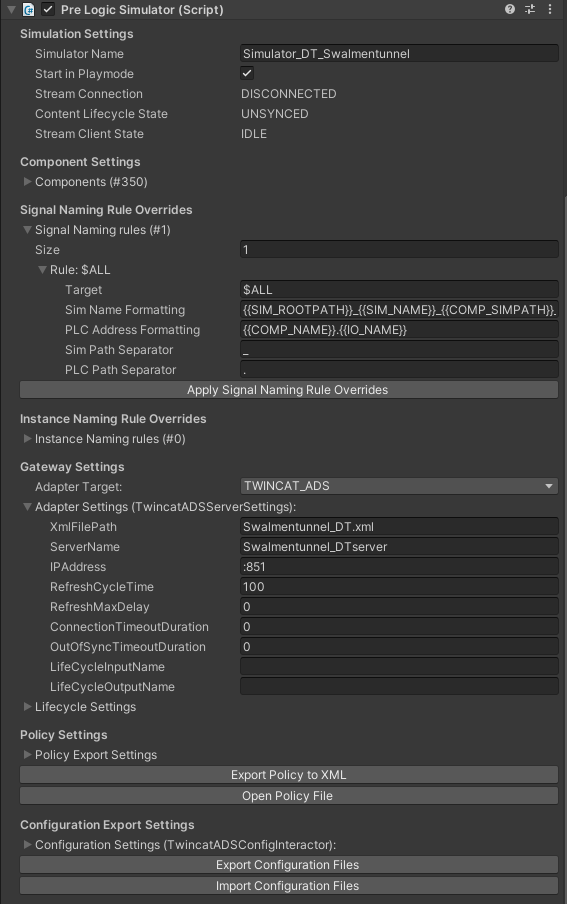}
    \caption{Inspector view of the prelogic simulator with the correct settings.}
    \label{fig:prelogicsimulator_inspector}
\end{figure}

The important settings here are in the Gateway Settings section. The remaining settings all use the default values, except for the Simulator Name that has been set to a recognizable name. In the Gateway Settings, the first important setting is the Adapter Target. This is set to \textit{``TWINCAT\_ADS''} and specifies the PLC type that is used to control the digital twin. Then, the \textit{XmlFilePath} contains the name of the \textit{xml} that will be exported and can be chosen freely. Then the \textit{ServerName} should be set to a recognizable name and the \textit{IP Address} should be set to `:851', which is the port used by TwinCAT. Only the port is specified here, and not the IP-address, since generally the digital twin runs on the same IP-address as the PLC.\\

When all settings are correct, the \textit{xml}-file for the policy should be generated each time when something changes in the logic component names in the digital twin, since this then changes the contents of the \textit{xml}-file and is necessary for all logic components to function correctly.

\section{Conclusion}

Using the methods described in this chapter, PLC code for a supervisory controller can be generated in the CIF3 toolbox where the inputs and outputs are defined by a hardware mapping file. A method for automatic generation of logic component GameObjects in Unity based on the inputs and outputs defined in the PLC code is developed. This reduces manual work in setting up the signals for the digital twin and reduces the chance of errors when naming the logic component GameObjects. A way of setting up naming rules for the logic component such that no additional mapping is needed in TwinCAT is described and the settings for the Prelogic Simulator are discussed.\\

In the development of a digital twin of a water lock for RWS \cite{jelmer_digital_twin_sluis}, which is done at the same time as the project described in this report, a different way of generating logic components is applied. With the other method, logic components are not generated from the input and output variable list from the PLC code, but from the instances of controlled objects in the digital twin itself. For instance, when a traffic light is put in the digital twin, the logic components for this traffic lights can be generated by pressing a button. Furthermore, the naming rules for the logic component GameObjects do not directly reference to the correct files in TwinCAT in \cite{jelmer_digital_twin_sluis}, but a mapping file is automatically generated from the logic component GameObjects present in the digital twin. This mapping file is then loaded into TwinCAT. In terms of reducing manual work, both methods of logic component generation do this equally well.\\

The method applied in this chapter has the advantage that all logic components from the hardware mapping are generated in the digital twin, from which the digital twin can be set up. This also ensures that no errors occur when the digital twin is connected with TwinCAT, because when logic components are present in the digital twin for which no signals are defined in the PLC, the connection fails. This means that the method presented in this chapter is more suitable when a digital twin is made for a supervisory controller that has already been fully designed. Furthermore, it is not necessary in the method from this chapter to generate the new logic components each time when a new object is added, since all logic components are generated at the same time. The method presented for the digital twin of the water lock however, can be used before the controller is designed fully and the system for which the digital twin is made is fully standardized and modular. An early version of the plant can then be designed in the digital twin, from which an overview of all inputs and outputs is generated.

\chapter{Digital Twin of the Swalmentunnel}\label{sec:digital_twin_swalmentunnel}
    
    Now that the methods for development of a digital twin have been described, the digital twin for testing the supervisory controller that is synthesized for the Swalmen Tunnel can be created. In this chapter, all components in the digital twin are described in detail with motivation of the most important modeling choices. A distinction is made between non-controllable components, which are not influenced or connected to the PLC, and controllable components, which communicate with the PLC.

\section{Simplifications and Assumptions}\label{sec:simplifications_assumptions}

    As is the case in every model, some simplifications of, and assumptions about the system are needed to design the digital twin. The first important fact is that the main goal of the digital twin is supervisory controller validation. This means that the systems that are not controlled by the supervisory controller, or that do not interact with controlled entities, are not essential in the model. Therefore, the service building belonging to the tunnel system and all systems that are omitted in the supervisory controller as described in the introduction of Chapter \ref{sec:overview_swalmen_tunnel} are not present in the digital twin. Another simplification is that only one instance of each controlled entity in a tunnel tube is needed. The aid cabinets for instance are positioned every few meters in the real tunnel, but in the digital twin, only one aid cabinet of each type (\textit{A} and \textit{C}) is present in each tunnel tube. This simplification is valid since all aid cabinets have the same sensors and actuators, so the argument of symmetry can be used to show that the simplification is sufficient for validation of the supervisory controller.\\
    
\begin{figure}[H]
        \centering
        \includegraphics[width=.6\textwidth]{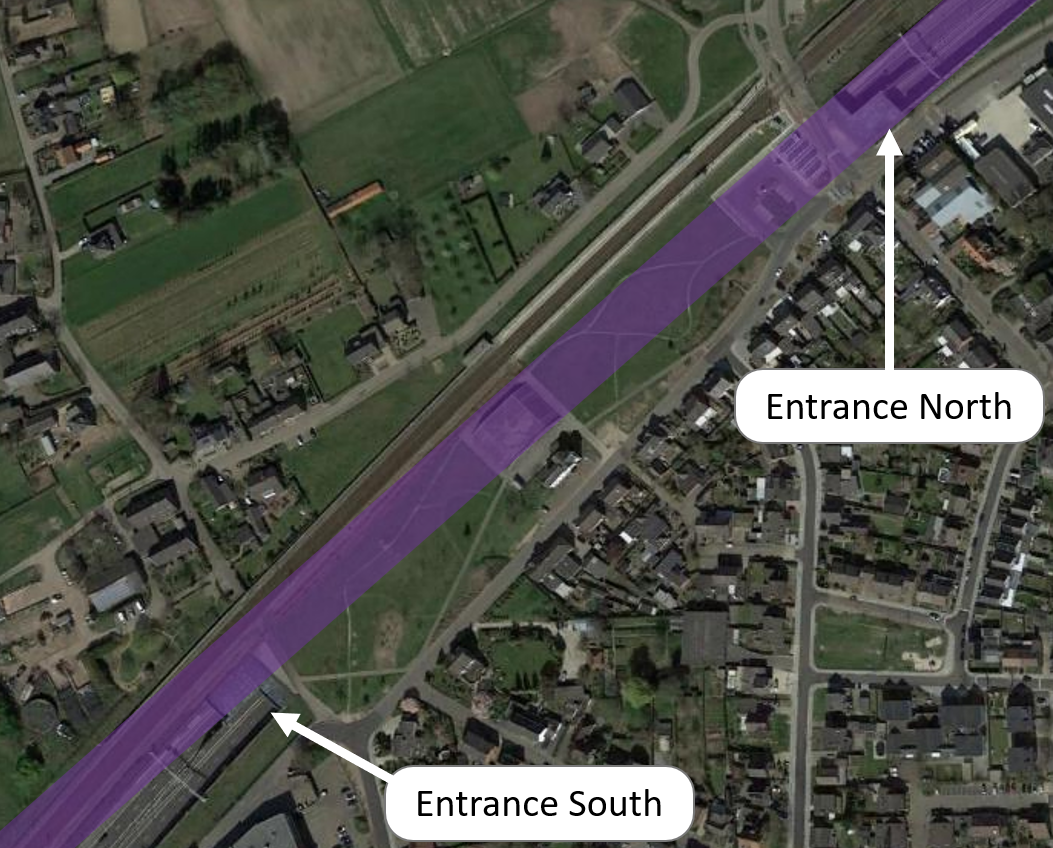}
        \caption{Aerial view of the Swalmen tunnel showing the difference between the tunnel modeled as a straight line (purple) and the real tunnel.}
        \label{fig:bend_tunnel}
    \end{figure}

    Other noticeable simplifications are done in the geometry of the tunnel. Firstly, the tunnel has a gentle bend, as shown in Figure \ref{fig:bend_tunnel}, and the road level goes down and up at the entrance and exit of the tunnel. These characteristics make it hard to model traffic driving through the tunnel, so the 3D model of the tunnel is totally straight and the road is at a constant height level. This enables the car models to just move straight forward through the tunnel. These simplifications do not interfere with the goal of supervisory controller validation, but for other applications the simplifications can be a problem. For instance, the camera view of the CCTV cameras is slightly different since cameras might be able to see through the whole tunnel in the digital twin, while this is not the case in reality because of the slight bend in the road trajectory and the height difference in the road surface. Furthermore, the dimensions of the tunnel geometry are not exact on the level of millimeters or centimeters. This can have the same implications for the digital twin as for the straight road simplifications, but therefore does not pose a problem for the main goal of the digital twin. Additionally, some entities, such as boom barriers and matrix signs, are not at the correct location, but closer to the entrance. This is to keep the digital twin scene compact and this is allowed since the supervisory controller validation is not hampered by this simplification.\\
    
    Finally, a clarification is given about the parts of the control stack modeled in the digital twin. As explained in Section \ref{sec:supervisory_control}, the mechanical components, their actuators and sensors, and the resource controllers are included in the plant from the viewpoint of supervisory control. This means that these entities are all modeled in the digital twin. The actuators and sensors as mentioned in the development of the digital twin in this chapter are signals that are used for communication between the resource controllers and the supervisory controller. For example, on the plant level, a boom barrier can have a motor that rotates the beam (actuator) and a decoder that measures the angle of the beam (sensor). These actuators and sensors are modeled in the digital twin and no logic components are present for these sensors and actuators, because they are internal. On the supervisory control level, a sensor signal tells the supervisory controller whether the barrier is opening for instance. This is determined by the boom barrier resource controller based on signals from a physical sensor, for instance a decoder. An actuator signal is then sent to the digital twin that tells what the boom barrier should do. These actuator and sensor signals are then implemented as logic components in the digital twin.
    
\section{Static Environment}\label{sec:static_environment}

\begin{figure}[H]
        \centering
        \includegraphics[width=\textwidth]{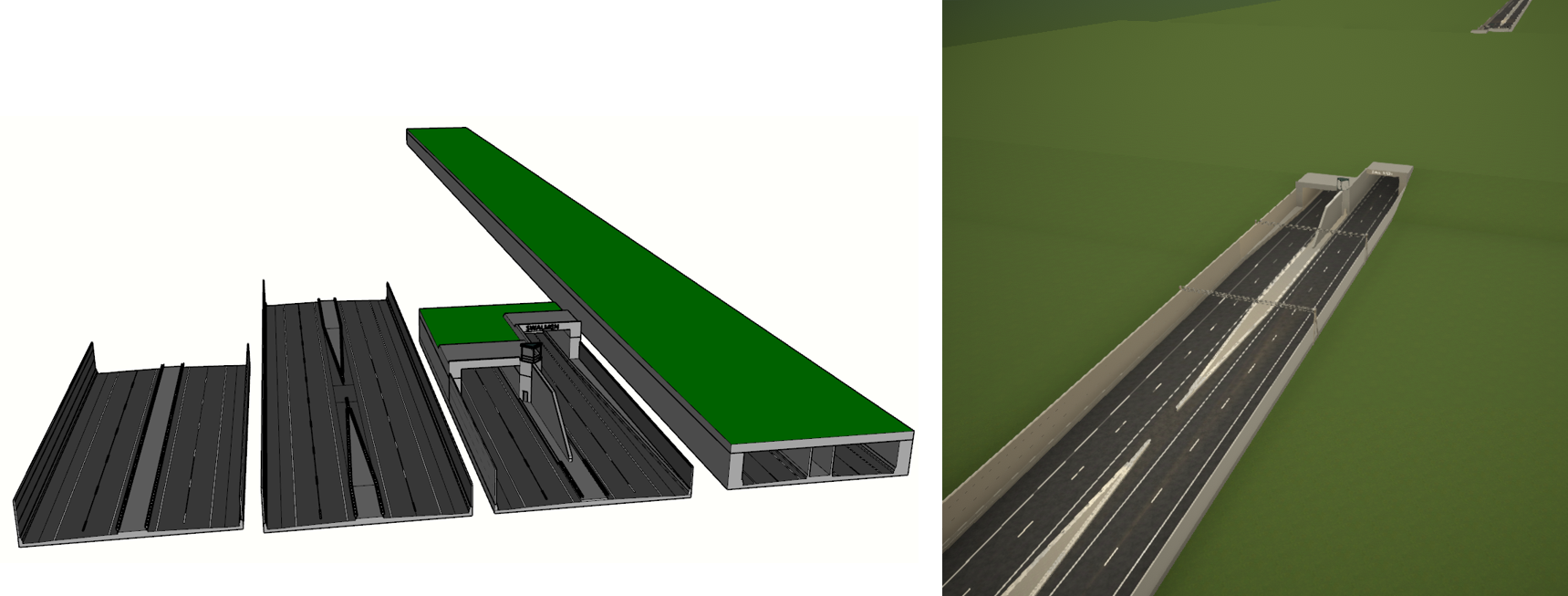}
        \caption{Modeled tunnel parts in SketchUp (left) and the tunnel environment built from these parts in Unity (right).}
        \label{fig:tunnel_models_static}
    \end{figure}

    The largest models are the geometry of the tunnel itself with the highway road and its environment. The model of the tunnel and the road is made in a modular way, as can be seen on the left in Figure \ref{fig:tunnel_models_static}, and the sections are saved separately as discussed in Section \ref{sec:modeling_sketchup}. The tunnel parts shown here are a generic road section, a road section with room for a emergency passage, a tunnel entrance section and a tunnel middle section. These parts are then used to build the environment in Unity. The main advantage of this way of assembling the tunnel is that the road before and after the tunnel and the tunnel itself can be made as long as desired, without having to change the models. Also, the number of models are smaller, since the entrance can be used twice for example. The main sources used for making the 3D models of the tunnel and its surroundings are as-built construction drawings provided by RWS and images from Google Maps.\\
    
    
\section{Non-Controllable Entities}

     The main non-controllable entities in the digital twin are traffic, environmental lighting and smoke. These components all have a predetermined behavior and can interact with controllable components in some cases.

\subsection{Vehicles}\label{sec:traffic}

    Since the Swalmen tunnel is a tunnel for road traffic, the traffic stream is the most important non-controllable entity that has a lot of interaction with the controlled components in the tunnel. The traffic stream is composed of a collection of cars and trucks shown in Figure \ref{fig:overview_vehicles} below. The vehicles are described from left to right in detail below the figure.

    \begin{figure}[H]
        \centering
        \includegraphics[width=.4\textwidth]{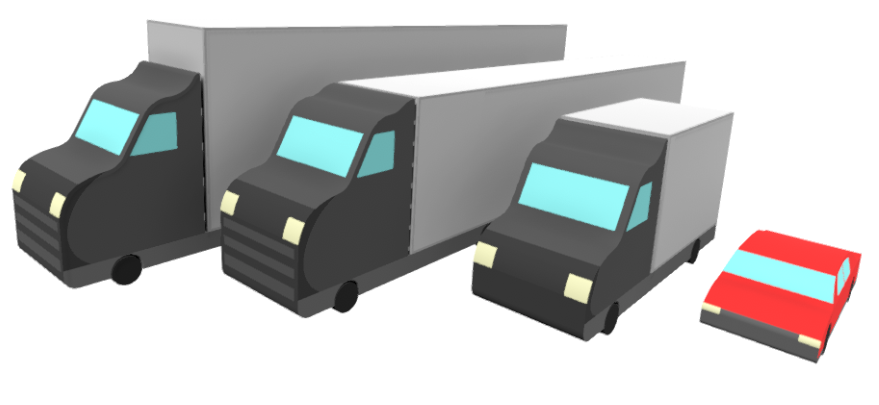}
        \caption{The four modeled vehicles.}
        \label{fig:overview_vehicles}
    \end{figure}
    
    \begin{itemize}
        \item \textbf{High large truck:} this vehicle should not enter the tunnel, as it is too high ($>4.1$ [m]) and can damage equipment mounted to the ceiling of the tunnel or the ceiling itself. The model is 16.6 [m] long and 4.65 [m] high. 
        \item \textbf{Low large truck:} this long vehicle can safely enter the tunnel. Its dimensions are 6.6 [m] long and 3.65 [m] high.
        \item \textbf{Small truck:} this vehicle is also allowed to enter the tunnel and is mainly there to get more variety in the traffic stream. The small truck is 6.6 [m] long and 3.25 [m] high.
        \item \textbf{Car:} most of the vehicles on the highway are standard cars. The cars in the digital twin can have different colors. They are 4.2 [m] long and 1.4 [m] high.
    \end{itemize}

\begin{figure}[H]
        \centering
        \includegraphics[width=.8\textwidth]{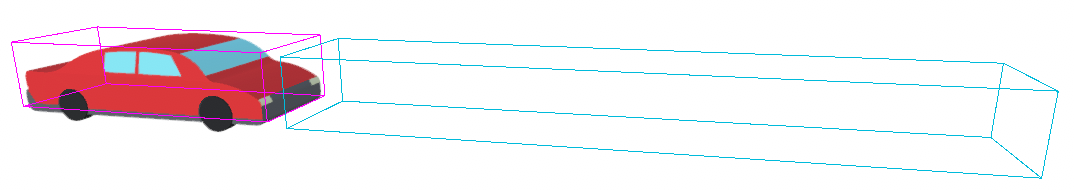}
        \caption{Colliders for the behavior of vehicles.}
        \label{fig:colliders_vehicle}
    \end{figure}

    In essence, there is no difference between the three vehicles that are not too high, but these are added to create a more realistic-looking traffic flow with more diversity. As mentioned, the tunnel is modeled as a straight tube, so the driving behavior of the vehicles becomes one-dimensional. This simplifies the driving behavior a lot. Three distinct variables are used to realize the driving behavior, which are the maximum speed in [km/h], the acceleration in [km/h/s] and the (braking) deceleration in [km/h/s]. The vehicles always accelerate to their maximum speed as long as there is no obstacle in their way. In order to sense the presence of an obstacle or other vehicles, each vehicle has two different colliders, that are shown as the two boxes in Figure \ref{fig:colliders_vehicle}.
    
    The pink box is the 'physical' collider of the vehicle that is put on a child object of the vehicle GameObject. This child object is tagged as \textit{"vehicle\_stop"}. A tag can be given to a GameObject in Unity to identify it. GameObjects can also be identified by their names, but by using a tag, several objects can be categorized without them having the same name. This pink collider enables vehicles to `see' each other. The blue collider in front of the vehicle is the sensing collider, which functions a trigger that mimics a proximity sensor. When a collider from a GameObject with the tag \textit{"vehicle\_stop"} enters this blue collider, the vehicle starts to decelerate and it starts accelerating again when the obstacle leaves the trigger. Because of the pink colliders, this causes vehicles to stop when a vehicle is stationary in front of it, meaning that a traffic jam will form behind a vehicle that has broken down for instance. The tag \textit{"vehicle\_stop"} is also used for the boom barrier collider in order to stop the vehicles from driving through the barriers. The red lights also have a collider that stops vehicles. The behavior script of the vehicles can be found in Appendix \ref{listing:vehicle_behavior}. All vehicles use the same behavior scripts, but the settings are different for the different vehicle types.\\
    
    \begin{figure}[ht]
        \centering
        \includegraphics[width=.42\textwidth]{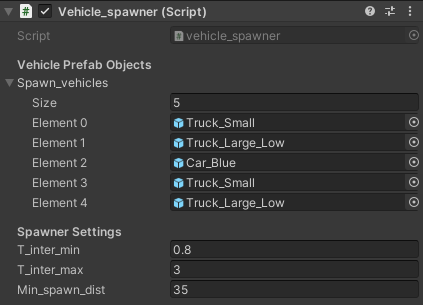}
        \caption{Inspector view of the vehicle spawner script.}
        \label{fig:vehicle_spawner_script_inspector}
    \end{figure}
    
    To create the traffic flows, vehicle spawners are created. These GameObjects only contain one script that spawns vehicles based on settings that are shown in Figure \ref{fig:vehicle_spawner_script_inspector}. The first setting is an array of vehicle prefabs that are spawned by the spawner. This is used to let the left lane spawner create more cars and the right lane spawner to create more trucks, which creates a realistic highway traffic flow. The frequency of certain vehicle types can also be manipulated by adding a certain prefab to the array multiple times, because each time a vehicle is spawned, a random index of this array is chosen. This is used to create more trucks in the right lane and more cars in the left lane for instance. To determine the spawn frequency of a spawner, a timer is set with a duration determined by a uniformly distributed random value ranging from \textit{T\_inter\_min} to \textit{T\_inter\_max}. The final setting is used to stop the spawner from instantiating new vehicles when a traffic jam is forming in front of the spawner. This is done by keeping track of the position of the last spawned vehicle and calculating the distance between the spawner and this vehicle. If this distance is smaller than \textit{Min\_spawn\_dist}, the spawner will not create a new vehicle. The spawning location of a new vehicle is determined by the position of the spawner, as shown in Figure \ref{fig:vehicle_spawners} below. The pink boxes are destroyed when the digital twin is started and are only meant for development purposes. The vehicle spawning script can be found in Appendix \ref{listing:vehicle_spawning_script}.\\
    
    \begin{figure}[H]
        \centering
        \includegraphics[width=.5\textwidth]{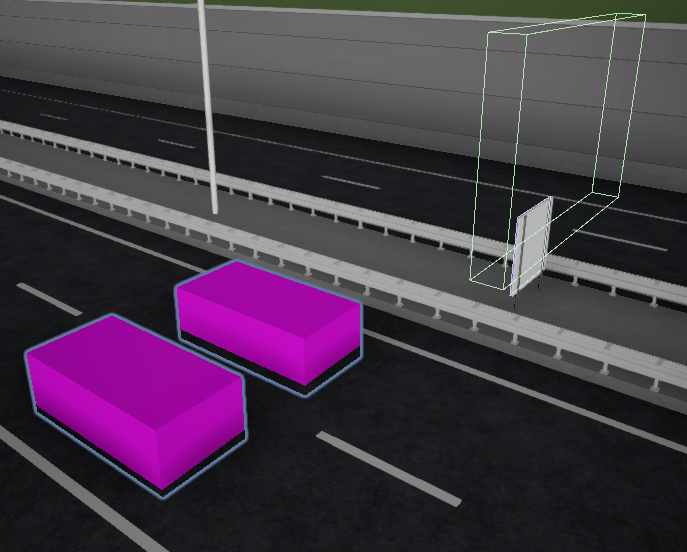}
        \caption{Two vehicle spawners (pink) and a vehicle destroyer (green box).}
        \label{fig:vehicle_spawners}
    \end{figure}
    
    Because many moving vehicles can cause the digital twin program to become computationally intensive, the vehicles are destroyed after having driven through the tunnel. For this purpose, a collider is put on the road shown in Figure \ref{fig:vehicle_spawners}. When a vehicle senses this collider, the vehicle is deleted from the scene. An important thing to note is that a vehicle stops when an obstacle enters the sensing collider and that it starts driving only when the obstacle fully physically exits the collider. This means that a vehicle being destroyed does not count as a vehicle exiting the sensing collider, which leads to vehicles that do not start driving again when they should. To solve this problem, vehicles first move forward with a very high speed (> 1000 [km/h]) when they are triggered by the vehicle destroyer and are deleted after a short time. This trick is also carried out when vehicles are destroyed in another manner.

\subsection{Lighting}

\begin{figure}[H]
        \centering
        \includegraphics[width=\textwidth]{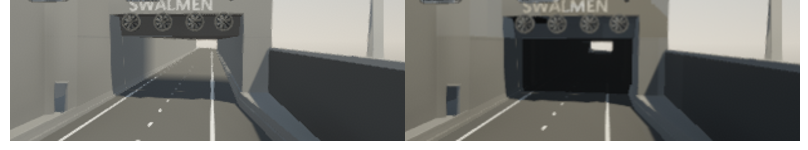}
        \caption{The difference between unbaked (left) and baked lighting (right).}
        \label{fig:normal_vs_static_baked_lights}
    \end{figure}

    Since some of the controlled entities in the tunnel are lights, lighting is an important component of the digital twin. As Unity is a game engine, it already includes many lighting options and techniques. Firstly, the sun is represented by a directional light, which is a light source for which all light rays travel in the same direction and come from a plane infinitely far from the objects in the scene. The lights in the tunnel are modeled with point lights, for which light rays come from one point and travel outward in a sphere. The road lights on the highway are modeled using spot light sources, for which the light rays travel from a point in a cone shape for which the angle can be set. For all light sources, the intensity and light color can be altered. An important technique for inside lighting in the tunnel is \textit{baked lighting}. This essentially means that lighting data, so shadows and light reflections, are ‘baked’ (painted) onto the texture of a model, such that the lighting calculations in the game itself are less intense, which improves the look of the game and its performance. This technique also allows for more realistic lighting in covered areas, such as in the tunnel tubes. Figure \ref{fig:normal_vs_static_baked_lights} below clearly shows this difference. 

    In the figure, some artifacts (unwanted dark and light areas) can be seen on the walls of the entrance of the tunnel. By using advanced settings for the baked lighting, the artifacts can be eliminated. In the finished digital twin, the light baking settings are far from optimal and should be improved, since the lighting is only realistic around the camera. This is shown in Figure \ref{fig:current_lighting} below. The tunnel lights also do not illuminate the walls because of this issue.
    
    \begin{figure}[H]
        \centering
        \includegraphics[width=.6\textwidth]{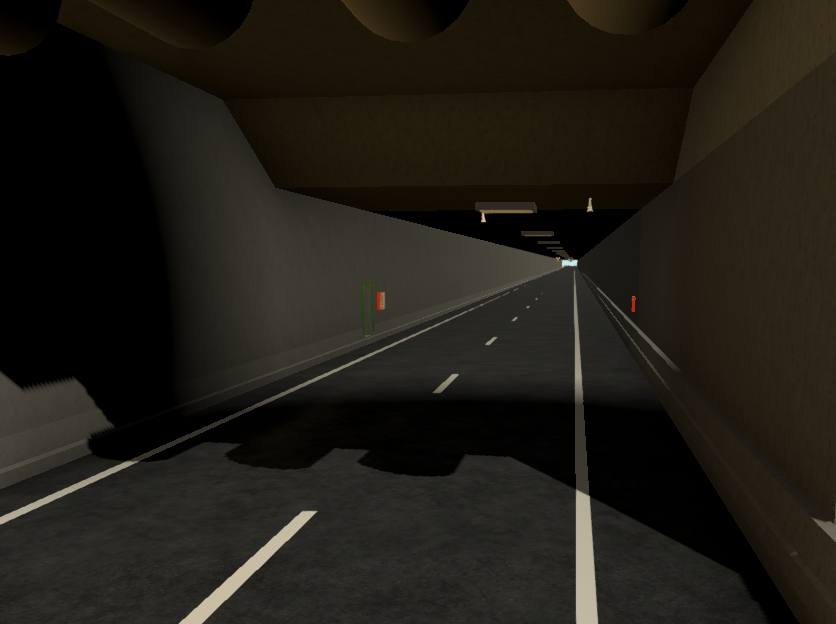}
        \caption{Current lighting situation in the digital twin.}
        \label{fig:current_lighting}
    \end{figure}

\subsection{Smoke}\label{sec:smoke}

\begin{figure}[ht]
    \centering
    \includegraphics[width=.3\textwidth]{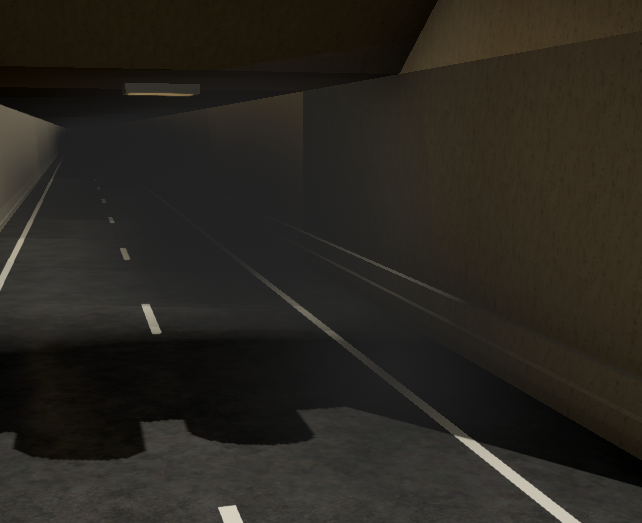}
    \caption{Smoke in a tunnel tube.}
    \label{fig:smoke_in_tube}
\end{figure}

In order for the smoke detection components in the traffic tube to work, smoke objects are added to the digital twin. Smoke is created by a particle system, which is a component in Unity that generates particles, which are small 2D-images floating through space. Two smoke particle systems are present, one for each traffic tube, and they fill the whole traffic tube with smoke when activated. Figure \ref{fig:smoke_in_tube} shows a cloud of smoke at the entrance of a traffic tube. The effect is more apparent when comparing this figure with Figure \ref{fig:current_lighting} above. Because the smoke detection unit can detect smoke in levels ranging from 0 to 8, the smoke `intensity' is varied by varying the alpha component of the smoke color. A lower alpha component makes the smoke more transparent, so changing this value changes the visual density of the smoke.\\

\section{Controlled Entities}\label{sec:controlled_entities_swalmentunnel}
    
    In this section, all controlled entities in the digital twin are described and the way in which the actuator actions and sensing behavior is modeled is explained. For components with a similar behavior, similar modeling methods are used. The components are structured according to the tunnel section where they are situated, so in the same manner as the components were introduced in Chapter \ref{sec:overview_swalmen_tunnel}. The way of handling input and output signals is implemented with IO-scripts, as described in Section \ref{sec:controlled_entities_setup}. Furthermore, the warning box that is used to indicate faulty supervisory controller signals is implemented for all suitable entities. A table with all input and output variable names for all components in the tunnel is given in Appendix \ref{sec:appendixA}. 

\subsection{Traffic Tube Entities}
\vspace{.5em}
\subsubsection{Aid Cabinets}
    
    \begin{figure}[ht]
        \centering
        \includegraphics[width=.3\textwidth]{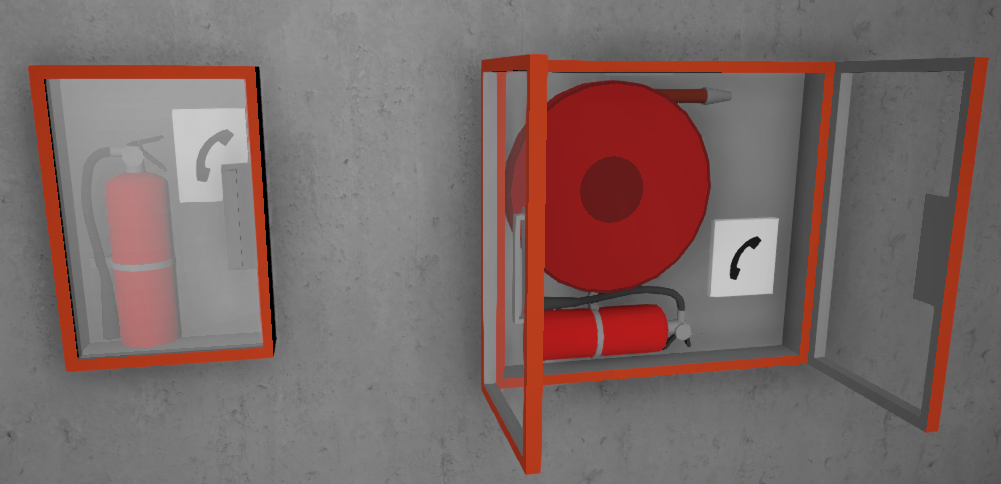}
        \caption{Aid cabinets C (left) and A (right).}
        \label{fig:aid_cabinets}
    \end{figure}
    
    Figure \ref{fig:aid_cabinets} shows the aid cabinets in the digital twin. These entities only have sensors that tell whether the doors of the cabinet are opened and whether the systems housing in the aid cabinets are active. The doors can interactively be opened and closed in the digital twin by clicking on them as explained in Section \ref{sec:interaction_mouse_keyboard}. When the cursor hovers over an interactive part, the mesh of these objects gets a pink outline that indicates that they can be clicked. The components in the aid cabinets, for instance the fire extinguisher, can also be clicked to activate them. This simulates the action of taking the extinguisher out of the cabinet. When the components are activated, a green box is shown around them. Clicking a component while it is active deactivates it again.
    
\subsubsection{Boom Barriers}

    For the boom barrier objects, a detailed description of the setup is already given in Chapter \ref{sec:set_up_digital_twin}. Here, the way in which the motion of the barrier beam is realized is explained. In Unity, a rotational action lets a GameObject rotate around a certain point in space, which is the pivot point of the GameObject. This point is defined for each GameObject in Unity and is the middle of all geometry in a GameObject by default. In non-circular parts, this is often not the point that coincides with the desired rotation axis, meaning that the pivot point should be moved. Prespective includes a tool that can be used to move the pivot point, but it is also possible to insert an empty GameObject as a parent of the rotating part, for which the pivot point is in its center, as it does not have any geometry. This is illustrated in Figure \ref{fig:pivot_GO} below.
    
    \begin{figure}[H]
        \centering
        \includegraphics[width=.8\textwidth]{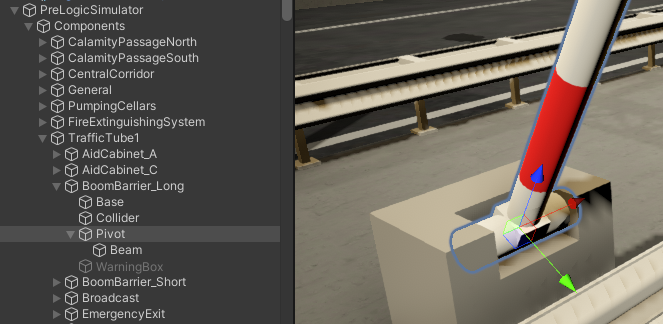}
        \caption{Using an empty GameObject to define the pivot position.}
        \label{fig:pivot_GO}
    \end{figure}
    
    Prespective includes a \textit{Wheel Joint} component and a \textit{Limited Servo Motor} component that can be used to let the barrier open and close realistically. However, when using this, problems were encountered that are related to setting the desired rotation of the beam. The desired rotation that was set and the actual rotation differed by 90 [$\circ$]. Therefore, a new simple motor script is written that linearly interpolates between the current angle and the desired angle, which makes the beam move quite realistically. This script can be found in Appendix \ref{listing:Actuator_RotationMotor}. An additional setting needed for the sensors is an offset for the position measurement. This is added because when the motor interpolates the angle linearly, the 90 [$\circ$] angle is reached asymptotically. With the allowed offset, the angle needed for the sensor \textit{s\_open} to be set to \textit{true} is $(90 - \textrm{offset})$ [$\circ$]. From experiments, it is concluded that an offset value of 1 [$\circ$] feels realistic. \\

    In addition to the sensors and actuators used for opening and closing the barriers, two sensor signals tell whether a vehicle is below the barriers, to prevent them from closing when this is the case. This is modeled using two colliders that function the same as the colliders from the vehicles as described in Section \ref{sec:traffic}. These sensor boxes are shown below in Figure \ref{fig:boom_barriers_with_colliders} as the largest boxes. The smaller colliders are exactly the same size as the beams. These are the colliders that can be sensed by the vehicles.
    
    \begin{figure}[H]
        \centering
        \includegraphics[width=.8\textwidth]{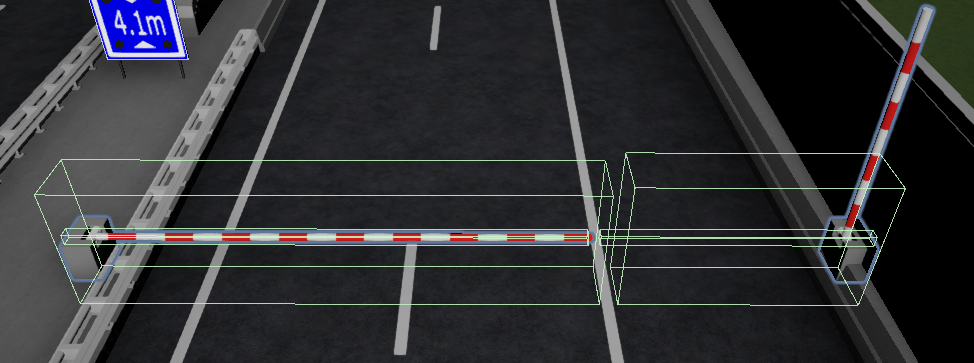}
        \caption{Boom barriers 1 (closed) and 2 (opened) in a traffic tube with all colliders shown.}
        \label{fig:boom_barriers_with_colliders}
    \end{figure}

\subsubsection{CCTV Systems}

    The CCTV system adds an element to the digital twin that can be modeled much more realistically in Unity than it was modeled in CIF. This is because Unity supports cameras that can render to a texture, which can be projected onto a plane. This means that camera GameObjects can be positioned in the scene and the views can be displayed on TV screens in a control room in the digital twin.\\
    
    This implementation is shown in Figure \ref{fig:CCTV_example} below, where a screenshot with four camera views as seen from the control room is shown. On the right in the figure, a (floating) camera is shown that renders the view on the top right screen. The white thin lines coming from the camera object on the right represent the viewport of the CCTV camera.
    
    \begin{figure}[H]
        \centering
        \includegraphics[width=\textwidth]{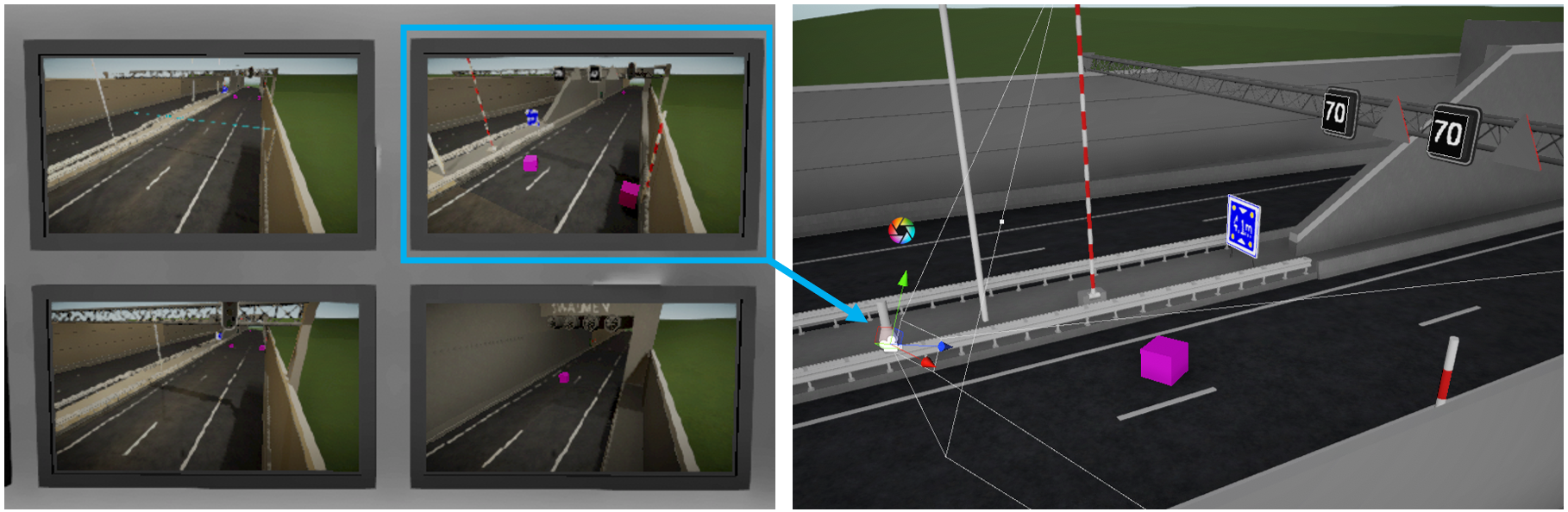}
        \caption{Example of four TV screen views rendered from CCTV cameras in the digital twin (left). The camera shown on the right renders the view on the top right screen.}
        \label{fig:CCTV_example}
    \end{figure}
    
    After adding a few CCTV cameras with screens, the digital twin started running much less smoothly, indicating that the addition of these cameras had a negative impact on CPU usage. Unity has a built-in tool for monitoring CPU usage, which is called the \textit{Profiler}. In Figure \ref{fig:CPU_Performance_CCTV} below, a comparison of CPU usage while rendering six CCTV screens (top) and without rendering any CCTV screens (bottom) is shown. From this, it can be concluded that the CPU impact is significant and can be problematic when rendering many CCTV cameras, which is the case in the Swalmen tunnel.
    
    \begin{figure}[H]
        \centering
        \includegraphics[width=.8\textwidth]{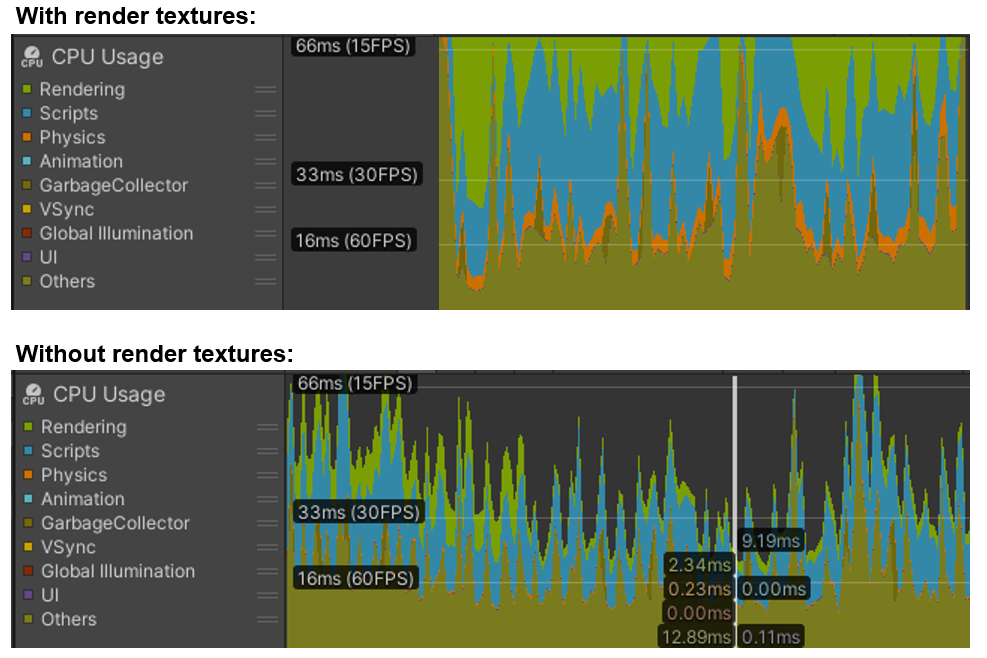}
        \caption{Comparison of the CPU performance while running the digital twin in the Unity Editor with and without rendering CCTV views.}
        \label{fig:CPU_Performance_CCTV}
    \end{figure}
    
    To solve this problem, a script is added that controls the rendering behavior of the camera components. The function shown in Listing \ref{listing:render_one_screen} below runs in \textit{FixedUpdate()}, meaning that the lines are run at a default rate of 50 times per second. Here, the array \textit{\_CCTVCameras} contains all cameras and \textit{NumberOfCameras} is the length of this array. The '\%'-character represents the modulo operator. This script ensures that every 0.02 [s] only one of the cameras renders its view, such that not more than two cameras (including the camera rendering the view of the digital twin itself to the user) are rendered at the same time. This implementation solves the performance problem and allows usage of CCTV screens in the digital twin.
    
    \vspace{3mm}
    \begin{lstlisting}[caption={Function that renders only one CCTV view per frame.}, label={listing:render_one_screen},basicstyle=\tiny,frame=single]
        void FixedUpdate()
    	{
    		_CCTVCameras[counter % NumberOfCameras].Render();
    		counter += 1;
    	}
    \end{lstlisting}
    
    The downside of this solution is that the frame-rate of the CCTV feed goes down when more cameras are added. For instance, with 50 cameras, each feed is refreshed only once per second. For the 9 cameras that are now implemented, which are the cameras for one traffic tube, the lower frame-rate is not a problem. A better solution to the problem would be to render static objects once. Moving objects are then rendered on top of this static image each frame. Using this technique, the tunnel environment only needs to be rendered once for each camera and the cars for instance are drawn over this image, which reduces the rendering costs for the CPU and GPU. This solution is not implemented yet.\\
    
    The IO-script for the CCTV system is written, but since the system still requires some further development, the actuator actions are also not implemented yet. These actions include reactions to operator input regarding CCTV selection that is regulated by the supervisory controller.

\subsubsection{Emergency Exit, Contour Lighting and Sound Beacon}
    
    \begin{figure}[ht]
        \centering
        \includegraphics[width=.3\textwidth]{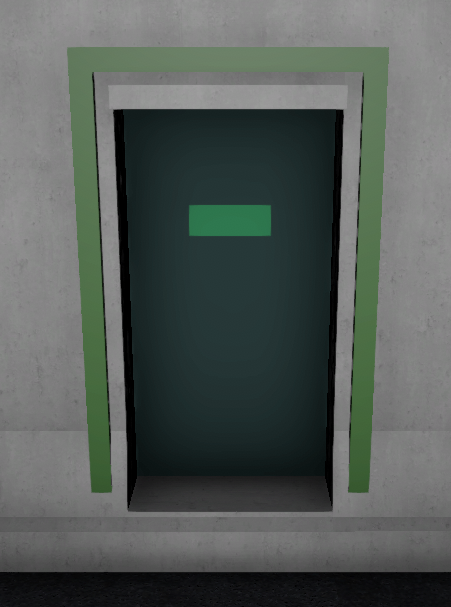}
        \caption{Emergency exit door.}
        \label{fig:emergency_exit_door}
    \end{figure}
     
     An image of the emergency exit door in the digital twin is shown in Figure \ref{fig:emergency_exit_door}. The emergency exit has one sensor that indicates whether the door is opened. The door is interactable in the same way as the aid cabinets, so it can be opened by clicking on it. The contour lighting and sound beacon are activated with actuator Booleans. The green part around the door is a contour lighting strip that is turned off. There is also a contour lighting component that can be turned on, which has a material on it that illuminates the surroundings with a green light. Only one of these two GameObjects, either the contour lighting that is turned on or the contour lighting that is turned off, is shown at the same time in order to get the effect of the light turning on and off. The sound beacon is modeled as an audio source that repeatedly plays an audio clip that says \textit{``Nooduitgang``}, which is Dutch for 'Emergency Exit'. By enabling 3D audio for the audio source, the sound is in stereo and really comes from the physical location of the emergency exit, which means that this system closely matches reality.

\subsubsection{Height Detection}
    
    The height detection system measures whether a vehicle is too high to enter the tunnel and works with light beams above the road in the real tunnel. The Prespective plugin includes a beam sensor component that is suitable for modeling the height detection system. The beam sensor consists of an emitter and a receiver. An instance of the system is shown in Figure \ref{fig:height_detection} below, where the emitter is placed on the right of the blue line and the receiver, which is not visible, is placed on the left. The visual component of the blue line here is deleted when the digital twin starts.
    
    \begin{figure}[H]
        \centering
        \includegraphics[width=\textwidth]{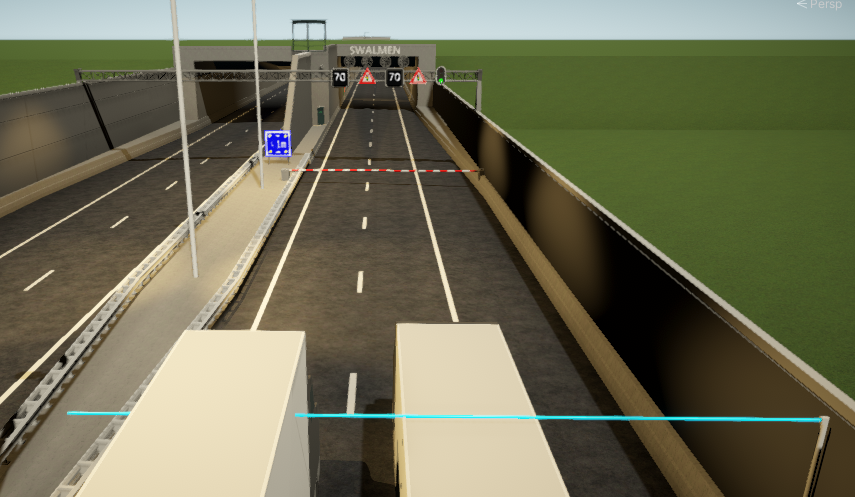}
        \caption{Height detection system with a truck that is too high (left) and a truck that is allowed to enter the tunnel (right).}
        \label{fig:height_detection}
    \end{figure}
    
    The receiver can transmit a high or low signal, based on whether a collider intersects the beam. Based on this signal, the sensor Boolean of the height detection system is determined. The blue cylinder representing the light beam is deleted from the scene when the digital twin is started. The beam is placed at a height of 4.1 [m] from the road surface, since this is the maximum height of a vehicle to be allowed to enter the tunnel. The left truck in the figure intersects the beam and will therefore activate the height detection sensor. The Swalmen tunnel contains a total of 5 height detection systems. One is located at a highway ramp. Because of the simplifications in the environment, this instance of the height detection system never encounters traffic, but since it works identically to the other height detection systems, this still allows for correct validation of the supervisory controller.

\subsubsection{Broadcast with Intercom and HF Installation}

    \begin{figure}[ht]
        \centering
        \includegraphics[width=.2\textwidth]{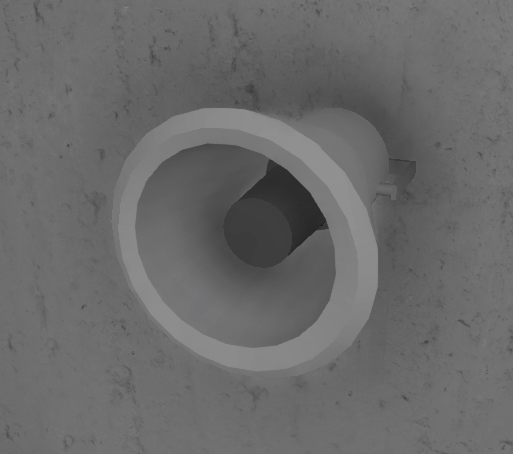}
        \caption{Speaker representing the broadcast sound source.}
        \label{fig:speakser_broadcast}
    \end{figure}

    The broadcast system allows for communication between the tunnel operator and users of the tunnel via audio messages. This is implemented in the digital twin with an audio source as described in the subsection about the sound beacon belonging to the emergency exit. The speaker model in Figure \ref{fig:speakser_broadcast} represents the location of the broadcast system audio source.\\
    
    In Unity, custom audio clips can easily be played on speakers with scripts. This enables implementation of the original pre-recorded messages used in Tunnels maintained by RWS. Now, the implemented spoken messages report what type of messages is being played. The HF system also delivers messages, but does this through the radio system of cars. Since implementing this exact behavior would not aid the user for validation of the supervisory controller, the HF system messages are also generated from the speaker, but the spoken messages specifies by means of some extra words that the HF system sends the message.

\subsubsection{Traffic Lights and Traffic Signs}

    The main components related to traffic signaling are positioned on a metal structure above the highway. Figure \ref{fig:traffic_signage} below shows these components in the digital twin. The matrix board can provide information about speed limits and closed driving lanes. In the supervisory controller that is running on the PLC, the speed limits are not taken into account. The matrix board is therefore either off or displays a red cross, meaning that the driving lane is closed. The red cross is displayed when the J32-sign is turned on. This J32-sign, that can be turned on and off electronically in real life, is modeled as the sign with a gray cover that is shown below the sign in the figure. This cover is placed over the sign when it is turned off. When the J32-sign actuator is then turned on, the cover plate is deactivated.

    \begin{figure}[H]
        \centering
        \includegraphics[width=.5\textwidth]{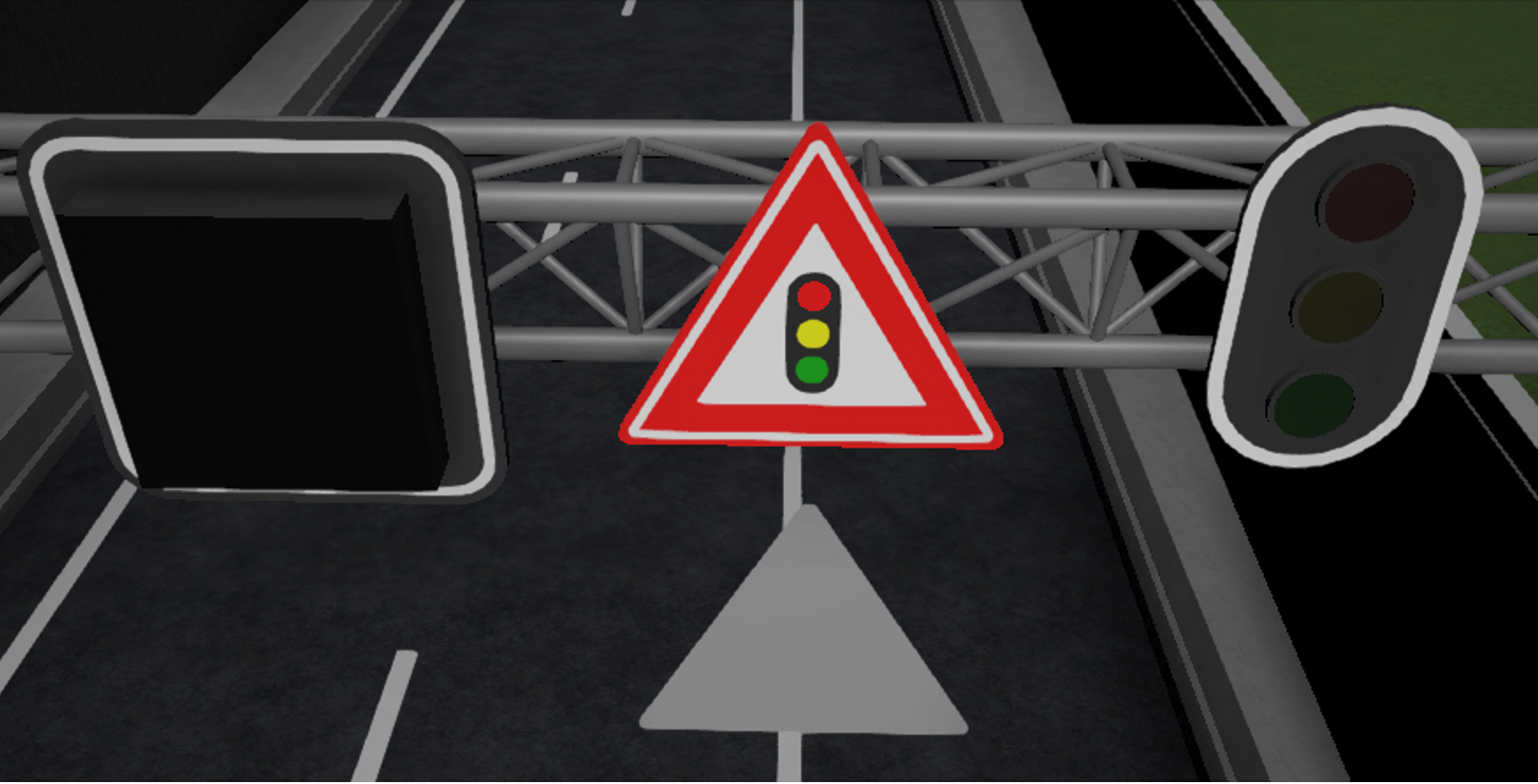}
        \caption{From left to right: a matrix sign, a J32-sign with its cover plate that simulates it turning off and a traffic light.}
        \label{fig:traffic_signage}
    \end{figure}
    
    On the right of the J32-sign, a traffic light is shown. This is a generic traffic light with a red, a yellow and a green light. There are 5 output Booleans that determine the behavior of the traffic lights belonging to a traffic tube. Either the green or red light can be on, the yellow light can be flashing or all lights can be off. In order to model the separate lights, the same technique as used for the contour lighting of the emergency exit is applied.

\subsubsection{Lights and Light Sensors}

    The tunnel lights, of which one unit is shown in Figure \ref{fig:light_unit_tube} below, are fixed on the ceiling and can be in 9 different states, ranging from state 0 to state 8. For each of these states, an output Boolean tells whether the light is in this state. Only one of these Booleans should be \textit{true} at all times. The current light intensity level is then tracked with an integer value, which then is used to determine the light intensity of the light source by multiplying it with an intensity factor $f_i$. This means that the intensity of the lights ranges from 0 to $8f_i$.

    \begin{figure}[H]
        \centering
        \includegraphics[width=.3\textwidth]{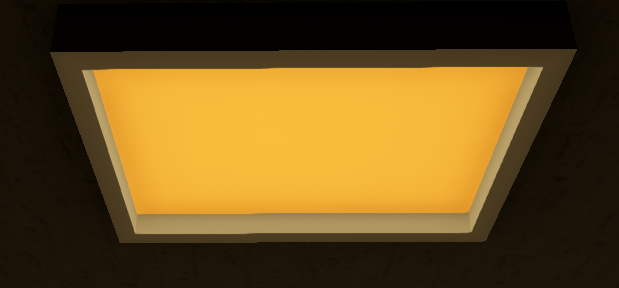}
        \caption{Light unit on the ceiling in a traffic tube.}
        \label{fig:light_unit_tube}
    \end{figure}
    
    As described in Section \ref{sec:simplifications_assumptions}, it suffices to only have one light in each tunnel tube for validation of the supervisory controller. However, this is not very clear, since the whole traffic tube should light up when the lights are turned on. To easily create and control multiple lights, a parent GameObject is set up that handles the actuator signal. This parent initially has only one light unit as a child object. This light unit reads the actuator state from its parent. When the digital twin is started, a script on the parent spawns a row of a specified number of light units in a given direction where all light units are a given distance apart. This allows to easily change the number of light units and the distance between light units.\\
    
    The light sensor in the traffic tube essentially works as the inverse of a light unit, since it measures the light intensity from the directional light source that represents the sun and maps this value to a level ranging from 0 to 8. This is done by linearly interpolating between a minimum and maximum value according to Equation \ref{eq:lightLevelInterpolation}, where \textit{RoundToInt()} is a function that rounds its argument, which is a float variable, to an integer. $I$ denotes light intensity. The input Boolean corresponding to this light level is then set to \textit{true}.
    
    \begin{equation}
        \textrm{Level} = \textrm{RoundToInt}\Bigg(\frac{I_{measured} - I_{min}}{I_{max} - I_{min}} \cdot 8 \Bigg)
        \label{eq:lightLevelInterpolation}
    \end{equation}

\subsubsection{Smoke Detection System}

    The smoke detection system works almost exactly the same as the light sensor. The difference is that the smoke detector measures the smoke intensity by reading the alpha level of the smoke color as explained in Subsection \ref{sec:smoke}. Therefore, the values of $I_{min}$ and $I_{max}$ in the linear interpolation (Equation \ref{eq:lightLevelInterpolation}) are equal to respectively 0 and 1 for the smoke detector.

\subsubsection{SOS System}

    The SOS system has input Booleans that tell whether a speeding vehicle, a wrong way driver or a stationary vehicle is present in the traffic tube. These traffic types can be spawned from the GUI of the digital twin, as explained later in this chapter. Because these vehicles are spawned as a child object of their spawners, the sensor values can be determined by checking whether the number of child object of the respective spawner is greater than 0. The three vehicle types use the same driving behavior scripts as described in Section \ref{sec:traffic}, where the speeding vehicle has a top speed of 160 [km/h] and the stationary vehicle has a speed and acceleration of 0. The wrong way driving vehicle is spawned at the end of a driving lane and drives against the traffic flow. This vehicle has its own destroyer which works the same as the normal vehicle destroyer but uses a different tag.    
    
\subsubsection{Traffic Tube Control}

    The term traffic tube control here concerns a collection of two systems. One of these systems has three input Booleans that communicate whether both traffic lights in the tube are off, flashing or red at the same time. The other system provides sensor information on whether both boom barriers in the traffic tube are opening, opened, stopped, closing or closed at the same time. The information of these systems is used by the supervisory controller as an additional safety factor. The IO-script of this system reads the state of the traffic lights and boom barriers.

\subsubsection{Ventilation}

    The ventilation units in the tunnel are modeled as shown in Figure \ref{fig:ventilation_units} below. The blades of the ventilation unit are actuated with a DC-motor component from the Prespective plugin. The rotational velocity of the blades is set from the IO-script and a negative velocity makes the blades rotate in the other direction. In each traffic tube, four units are located at the entrance and two ventilation units are located in the halfway point of the tunnel.

    \begin{figure}[H]
        \centering
        \includegraphics[width=.3\textwidth]{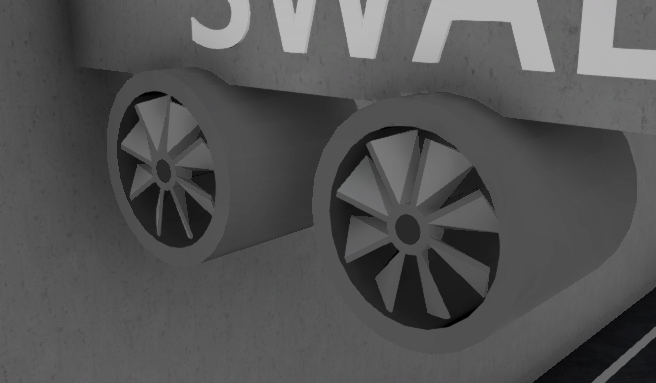}
        \caption{Ventilation units at the entrance of a traffic tube.}
        \label{fig:ventilation_units}
    \end{figure}

\subsection{Central Corridor}
\vspace{.5em}
\subsubsection{Central Channel Head Door}

    \begin{figure}[ht]
        \centering
        \includegraphics[width=.2\textwidth]{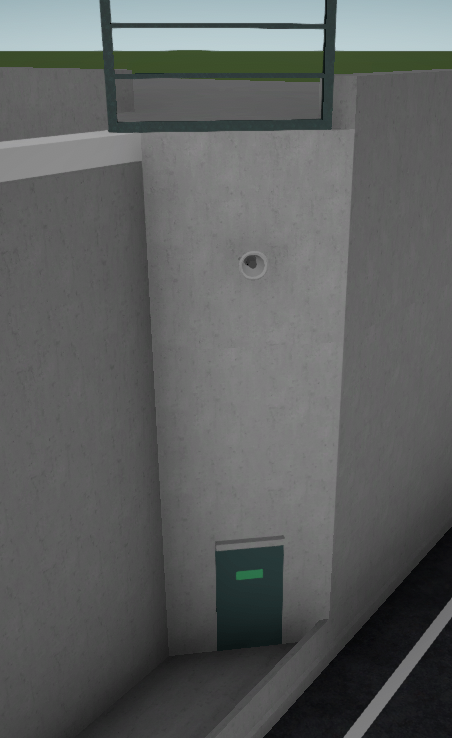}
        \caption{Closed head door leading to the central corridor at the tunnel entrance.}
        \label{fig:head_door_central_corridor}
    \end{figure}

    The central head door uses the same model as the emergency door in the traffic tube, as shown in Figure \ref{fig:head_door_central_corridor}. This door can also be opened by clicking on it and has only two sensors that indicate whether the door is opened or closed.

\subsubsection{Dynamic Escape Route Indication}

    In the central corridor, sets of two dynamic escape route indication signs are placed on the walls. One of these indicates that the optimal escape route is to the right and the other one points to the left, as shown in Figure \ref{fig:dvi_signs} below. Three output Booleans determine whether both signs are off or if the right or left sign should be active. Activating the signs works in the same way as with the contour lighting of the emergency exits, namely by having two signs in the same location of which one is lit and one is off. Based on the values of the output Booleans, the correct objects signs are shown and hidden.

    \begin{figure}[H]
        \centering
        \includegraphics[width=.3\textwidth]{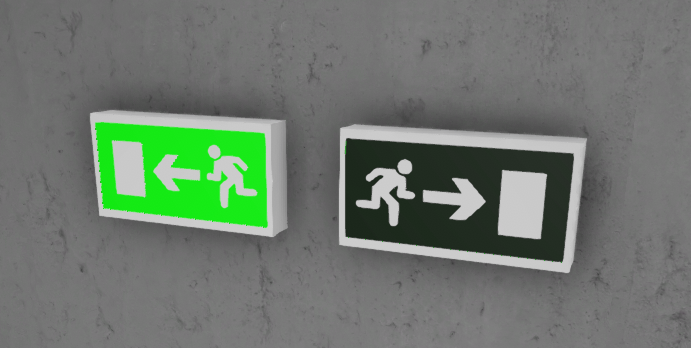}
        \caption{Dynamic escape route indication signs of which the left route indicator is active.}
        \label{fig:dvi_signs}
    \end{figure}

\subsubsection{Overpressure Systems}

\begin{figure}[H]
        \centering
        \includegraphics[width=.67\textwidth]{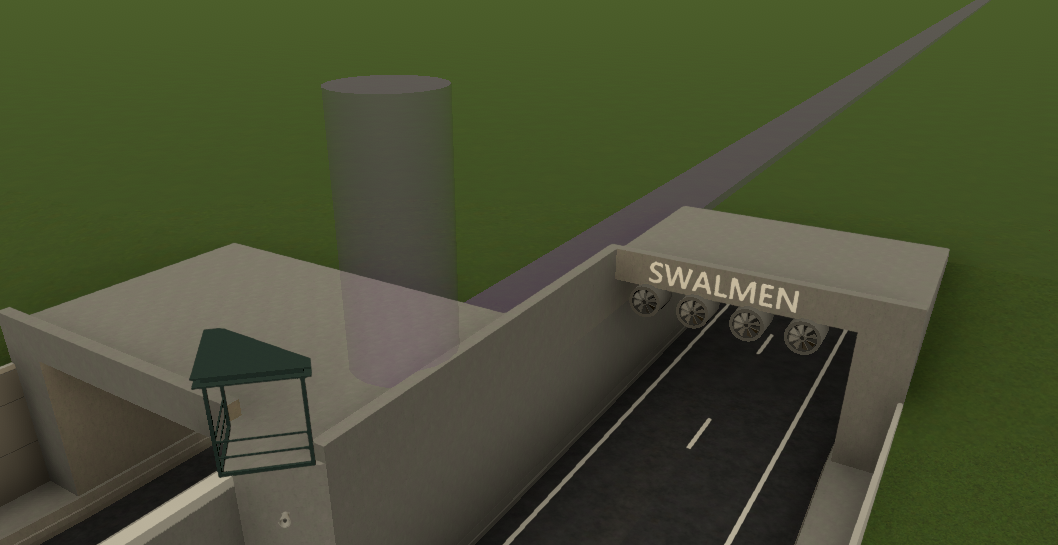}
        \caption{A purple box that represents the overpressure being active and a purple cylinder that represents the overpressure air coming from the entrance shown.}
        \label{fig:overpressure_system_boxes}
    \end{figure}

    Since in real life, it is not possible to see or hear overpressure, this system cannot be represented in a realistic way in the digital twin. Therefore, the overpressure in the central corridor is represented by a large purple box along the length of the central corridor that is visible when the overpressure system is active. The overpressure system has three states, namely left, right and off. Purple cylinders on the right or left entrance show whether the air that is used to maintain the overpressure is pumped in from the left or right side (by which the two entrances or exits of the traffic tubes are meant). An example is shown in Figure \ref{fig:overpressure_system_boxes} below where the overpressure system is active and in the `left' state.

\subsection{Other Systems}
\vspace{.5em}
\subsubsection*{Emergency Passages}
    
    The emergency passages that are placed in the railing between the highway, as shown in Figure \ref{fig:calamity_passage} below, are controlled in the same way as the boom barriers and therefore use the same behavior scripts. The only difference is that the boom barriers have obstacle detection, while the emergency passage does not have this, but the obstacle detection is implemented in a separate script such that the barrier scripts can be reused. Another difference is that the emergency passage is closed by default, while the boom barriers are opened by default. This is however simply changed by altering the orientation in Unity.
    
    \begin{figure}[H]
        \centering
        \includegraphics[width=.67\textwidth]{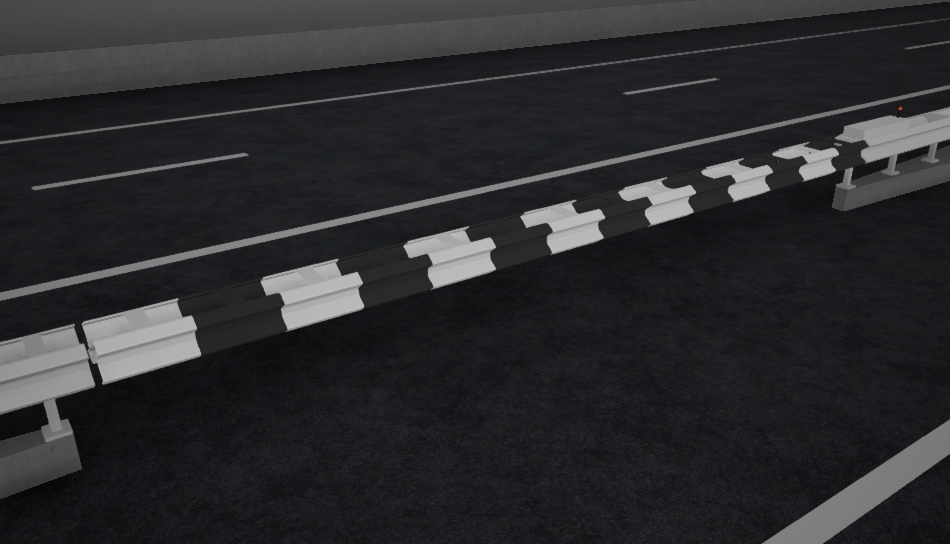}
        \caption{A closed emergency passage barrier.}
        \label{fig:calamity_passage}
    \end{figure}

\subsubsection{Pumping Cellars and Fire Extinguishing System}

    The model of the pumping cellars is shown in Figure \ref{fig:pumping_cellars} below. The water bodies are translucent cubes. The water level is varied by moving the cubes up and down. Output Booleans determine whether the pumps that empty the basins are on or off. When the pumps are turned on, the box is translated down with a predetermined velocity. Input Boolean values are determined by measuring the water levels of the clean and dirty water basins by measuring the $y$-position (the $y$-axis in Unity is the vertical axis) of the water body cubes.

    \begin{figure}[H]
        \centering
        \includegraphics[width=.8\textwidth]{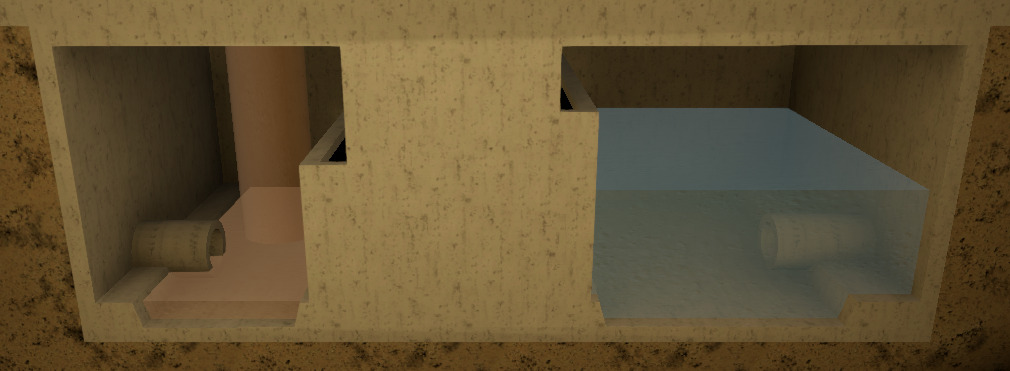}
        \caption{The pumping cellars with the dirty water basin that is being filled (left) and the clean water basin (right).}
        \label{fig:pumping_cellars}
    \end{figure}

    \begin{figure}[ht]
        \centering
        \includegraphics[width=.3\textwidth]{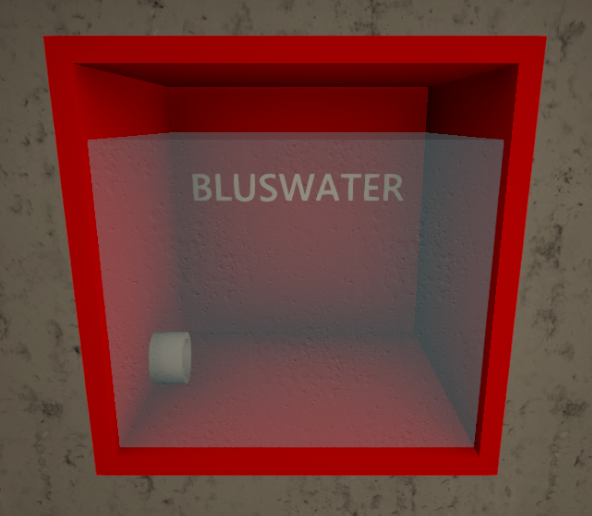}
        \caption{Basin of the fire extinguishing system.}
        \label{fig:fire_extinguishing_system}
    \end{figure}

    The fire extinguishing system also has a basin that stores the water used by the fire hoses from the aid cabinets. Figure \ref{fig:fire_extinguishing_system} shows this basin in the digital twin. The system is very similar to the pumping cellar systems, since sensors measure the water level and there is one actuator, being the pump. Because the number of sensors is different, a separate IO-script is written for each pumping cellar and for the fire extinguishing system, but the code is very similar and can easily be copied and adjusted.

\subsection{Broadcast Synchronization}

As was the case with the CCTV system, the broadcast synchronization system is not fully functional yet in the digital twin. The skeleton of the IO-script is set up and a timer is implemented, but the actuator actions are not implemented yet. A reason for this is that the HF system does not work correctly in the PLC code as is explained in Chapter \ref{sec:validation}.

\section{Prefab Creation}

As stated before, modularity of the digital twin is important regarding evolvability of the system in the design process of both the supervisory controller and the plant. The logic component generation method is implemented for this purpose, and the controlled entities are set up in a modular way as explained in Section \ref{sec:controlled_entities_setup}. Therefore, the controlled entities of the digital twin of the Swalmen tunnel are saved as prefabs, as shown on the left of Figure \ref{fig:prefabs_controlled_entities_with_list} below. These prefabs can be dragged into a scene where they can be positioned and altered. It is important to unpack the prefab before changing it, because the changes would otherwise affect the prefab. The strings representing the PLC variable names on the IO-script can then be set to the correct variable names to let the unpacked prefab communicate with the correct PLC signals. The scene on the right in the figure shows an overview of some prefabs that are dragged into a scene. This, together with the modular way in which the tunnel environment is set up as described in Section \ref{sec:static_environment}, makes the digital twin modular. The non-controllable entities, such as the vehicles and smoke, are also saved as prefabs.

\begin{figure}[H]
    \centering
    \includegraphics[width=\textwidth]{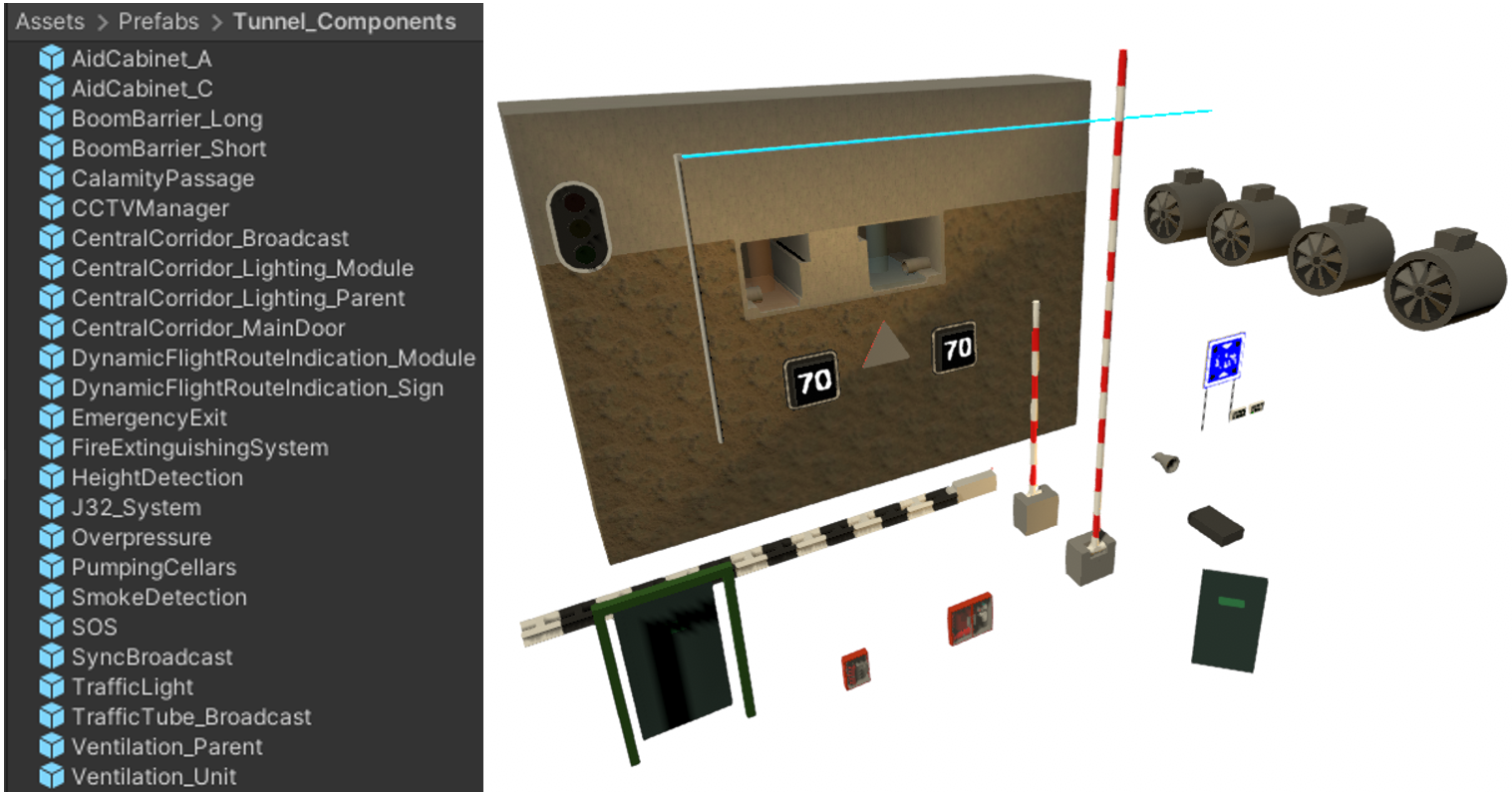}
    \caption{Prefab list of the controlled entities in the digital twin (left) and a collection of these prefabs in a scene (right).}
    \label{fig:prefabs_controlled_entities_with_list}
\end{figure}

\section{Interaction}\label{sec:interaction_dt}

\begin{figure}[H]
    \centering
    \includegraphics[width=\textwidth]{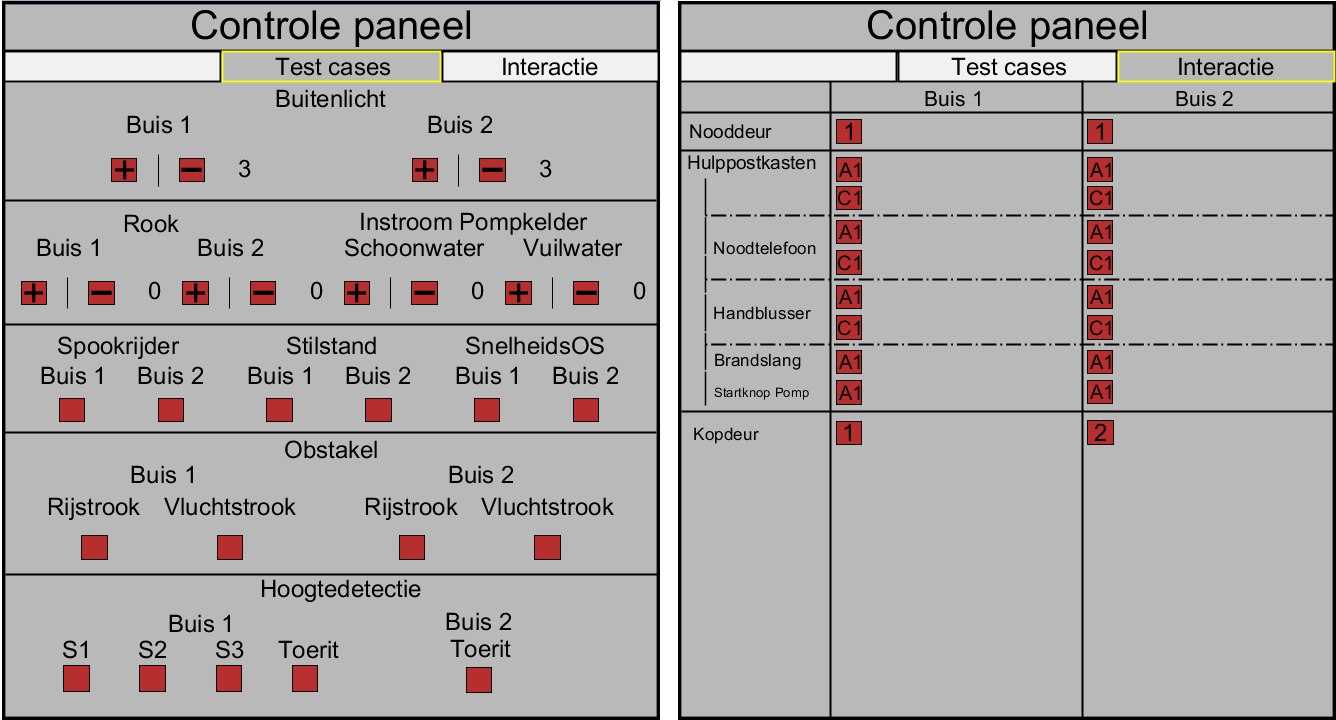}
    \caption{Control panel from the 2D CIF model.}
    \label{fig:control_panel_cif_model}
\end{figure}

For the application of supervisory controller validation, the digital twin should have interactive components. These are used to enable certain test cases, to interact with tunnel entities, and to move through the scene. The full control panel from the hybrid model implemented in the existing CIF model of the Swalmen tunnel is shown below (Figure \ref{fig:control_panel_cif_model}). Here, the part on the left contains buttons that are used to set test scenarios. For example, light levels can be changed, smoke can be generated, pumping cellars can be filled and different vehicle types can be spawned. On the right, buttons are shown that can be used to interact with entities in the tunnel, such as doors and components of the aid cabinets.

\vspace{3mm}

\subsection{Direct Interaction}

First, direct interaction between the user and tunnel entities is discussed. As described earlier in the chapter, the doors and aid cabinet parts can be activated by clicking on them in the digital twin using the raycasting technique described in Section \ref{sec:interaction_mouse_keyboard}. This is implemented as a replacement of all buttons on the right part of the CIF-model control panel shown in Figure \ref{fig:control_panel_cif_model}. The advantage of this direct interaction is that testing scenarios become much more intuitive, since clicking the real components feels much more natural than clicking buttons.

\vspace{3mm}
\begin{figure}[H]
    \centering
    \includegraphics[width=.5\textwidth]{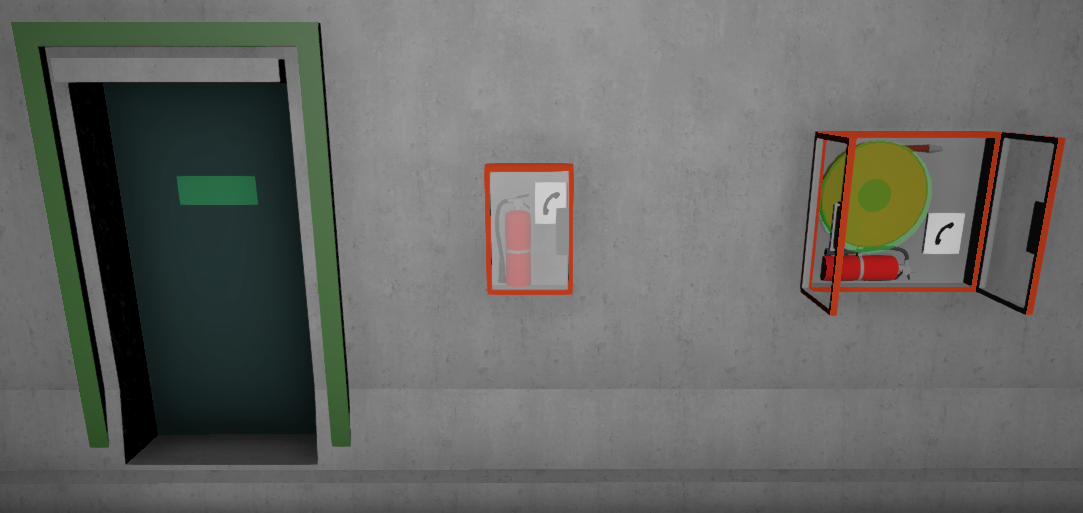}
    \caption{Examples of directly interactive components in the digital twin.}
    \label{fig:control_panel_dt}
\end{figure}

\subsection{Digital Twin GUI}\label{sec:dt_gui}

The left side of the CIF control panel in Figure \ref{fig:control_panel_cif_model} is used for test cases. These buttons are also added to the Digital Twin as a GUI, together with some other features. The GUI of the digital twin is shown in Figure \ref{fig:control_panel_dt} below. The red boxes divide the GUI into five distinct parts that have different functions. The GUI is written in Dutch, because RWS communicates primarily in Dutch. The sections in the GUI are explained below using the numbers from the figure.

\begin{figure}[H]
    \centering
    \includegraphics[width=\textwidth]{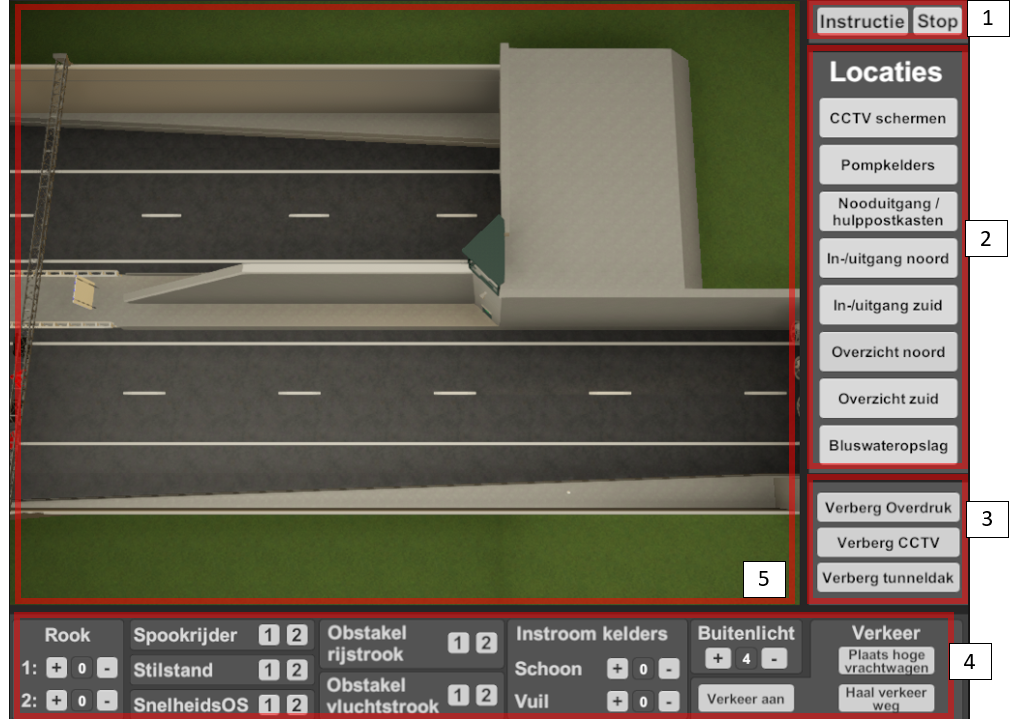}
    \caption{Control panel GUI in the digital twin.}
    \label{fig:control_panel_dt}
\end{figure}

\begin{enumerate}
    \item The right upper part of the GUI contains two buttons. The left button shows a screen with instructions for using the digital twin, such that no separate document is needed for this purpose. The button on the right takes the program to its starting menu, from which the digital twin scene can be reloaded again. This can be useful when the PLC restarts, since the digital twin scene should be reloaded then.
    \item The next section contains several buttons that are used to move the viewport camera to certain locations in the digital twin. This uses the script given as an example in Section \ref{sec:GUI_button_example}. For instance, the button ``\textit{Pompkelders}'' (Dutch for `pumping cellars') moves the camera to the pumping cellars. Additionally, there is another method of moving around in the digital twin, which is described for part 5 of the GUI.
    \item Section 3 contains buttons that can be used to hide some components in the digital twin. For the entities that can be hidden here, the parent GameObjects are deactivated. This means that these entities are not shown, but do also do not function. This is not a problem, for the entities that can be hidden here, since these do not have sensors or interact with other entities in the scene. When the boom barriers are hidden in this way, this could cause problems, so it is preferred to then only deactivate the mesh components in the boom barrier instead of deactivating the whole boom barrier entity. The buttons that hide the overpressure indicator and the tunnel roof are used to get a better view inside the tunnel, because the overpressure system makes the whole tunnel purple and the roof can block a good view of the tunnel as well. The button that hides the CCTV system is implemented to counteract performance issues that are explained in Section \ref{sec:controlled_entities_swalmentunnel}, because the improved CCTV system still causes performance issues sometimes. When the CCTV system is deactivated, meaning that the control room and all cameras are deactivated, all performance issues caused by this system are gone.
    \item The largest button section on the GUI is essentially the same as the left control panel shown in Figure \ref{fig:control_panel_cif_model}. Here, all test scenarios can be enabled. An addition compared to the hybrid model is that traffic can be started and stopped. This enables or disables the vehicle spawners described in Section \ref{sec:traffic}. Furthermore, a truck that is too large to enter the tunnel can be instantiated in this menu. Finally, all traffic can be deleted by pressing ``\textit{Haal verkeer weg}'' (Dutch for `delete traffic'), which also deletes all stationary vehicles that are obstacles for test scenarios. In this part of the GUI, some feedback is also given. An example is the number in between the smoke level buttons that provides information about the current smoke level in a traffic tube. Also, the button that says ``\textit{Verkeer aan}'' (traffic on) changes to ``\textit{Verkeer uit}'' (traffic off) when the traffic stream is started. This GUI section is not fully complete yet, as the high trucks can now only be spawned in one position in the whole tunnel. An improvement would be to enable the user to spawn a high truck in a desired place to enable more testing flexibility. This could for instance be implemented by enabling the user to choose a position on the road by clicking in the scene when the button is pressed.
    \item The last part of the viewport of the digital twin is actually not a part of the GUI, but can be used for an alternative method of moving through the scene. When the user presses the F-key on the keyboard, the cursor disappears and the user can look around by moving the mouse. It is then also possible to move around in the scene with a basic button layout as used in many first person games. The movement scheme is shown in Table \ref{tab:movement_keys} below.
    
\end{enumerate}    
    \begin{table}[H]
        \centering
        \begin{tabular}{|l|l|}
            \hline
            \textbf{Key}        & \textbf{Action}                       \\\hline \hline
            ``W''               & move forward                          \\\hline
            ``S''               & move backward                         \\\hline
            ``A''               & move left                             \\\hline
            ``D''               & move right                            \\\hline
            space               & move up                               \\\hline
            left ctrl           & move down                             \\\hline
            left shift          & hold to move more quickly             \\\hline
            ``F''               & enable or disable free movement mode  \\\hline
        \end{tabular}
        \caption{Free movement controls in the digital twin. All movement directions are relative to the camera orientation.}
        \label{tab:movement_keys}
    \end{table}
    
\section{Conclusion}

In this chapter, all important aspects of the design process of the digital twin of the Swalmen tunnel have been described using the techniques for digital twin development from Chapter \ref{sec:set_up_digital_twin}. This includes modeling the environment, traffic streams, controllable components and adding interactive elements in the form of a GUI. Some areas that require more development are discussed, such as the CCTV system and the broadcast synchronization system. Furthermore, the interaction GUI can still be expanded, such that all important locations can be reached with the GUI buttons and to enable more test scenarios. This chapter, together with Chapter \ref{sec:PLC_control} concludes the development process of the digital twin, such that the validation process can start.
    
    \chapter{Digital Twin Validation}\label{sec:validation}

In this chapter, the validation process of the digital twin is described. In the development of the digital twin, small scale tests are carried out for separate controllable entities and separate sensor and actuator signals. The small scale tests also include the use of a test controller for validating the interface between the PLC and the digital twin. The test setup used for validating the full digital twin is described next and the test results are presented.

\section{Small Scale Tests}

The first tests are carried out for each implemented sensor and actuator signal. The tests are easily executed using the BoolToggle scripts on the logic components. When the PLC is not connected, the output BoolToggle Booleans can be toggled by clicking the box at the setting \textit{``Boolean''} on the inspector view of the BoolToggle script shown in Figure \ref{fig:logicComponent_output_a_open}. For the input logic components, the box displays a check mark when the Boolean is set to \textit{true}. This way, it can be checked inside Unity whether the input and output signals from an IO-script function correctly. These tests are done for each new IO-script that is written.\\

\begin{lstlisting}[caption={\textit{TestController\_BoomBarrier.cif}.}, label={listing:TestController_BoomBarrier},basicstyle=\scriptsize,frame=single]
automaton BoomBarrier:
  cont t der 1;
  uncontrollable u_closed , u_opened ;
  location closing :
    initial ;
    edge u_closed when t >= 10 do t := 0.0 goto opening ;
  location opening :
    edge u_opened when t >= 10 do t := 0.0 goto closing ;
  end
automaton HW_Boombarrier :
  // === ACTUATORS ===
  disc bool a_open , a_close ;
  // === SENSORS ===
  input bool s_opened , s_closed ;
  location:
    initial;
    // === ACTUATORS ===
    edge when a_open != BoomBarrier opening do a_open := BoomBarrier.opening ;
    edge when a_close != BoomBarrier.closing do a_close := BoomBarrier.closing ;
    // === SENSORS ===
    edge BoomBarrier.u_closed when s_closed ;
    edge BoomBarrier.u_opened when s_opened ;
end
\end{lstlisting}

For testing the functionality of the logic component GameObjects in combination with the logic component generation script, a test controller is written that uses a simplified IO-script for a boom barrier. The CIF specification of this test controller is given in Listing \ref{listing:TestController_BoomBarrier} above. Lines 1-12 model a controller and lines 14-32 contain the hardware mapping automaton. There are two sensors that tell whether the barrier is fully opened or fully closed. Two actuators tell whether the barrier should be opening or closing. A timer is implemented such that the barrier is opened when it is fully closed and 10 seconds have elapsed. The timer is then reset and when the barrier is fully opened and 10 more seconds have elapsed, the door starts closing again.\\

PLC code of this CIF specification was used and loaded into TwinCAT. The whole process explained in Chapter \ref{sec:PLC_control} was then followed to verify the logic component GameObject generation script and the interface between Unity and TwinCAT. Furthermore, two sensors and actuators of the barrier are tested. According to the results, the interface between the digital twin in Unity and TwinCAT works fully.\\

\section{Full System Control Validation}

After verifying that the actuator and sensor signals for each IO-script and the control signal interface between Unity (Prespective) and TwinCAT function correctly, the full system is tested. This section first discusses the experimental setup for this testing process.

\subsection{Experimental Setup}

Figure \ref{fig:set_up_tests} below shows the control stack with the elements of the test setup. The digital twins functions as a model of the plant and the supervisory controller runs in TwinCAT, which functions as a virtual PLC. As mentioned before, it is important that the interface between the digital twin and TwinCAT is almost identical to the interface between the digital twin and a hardware PLC. For the operator interface, a test GUI is made. The reason why the test GUI is made and the development of this test GUI is described in the next subsection.

\vspace{-3mm}
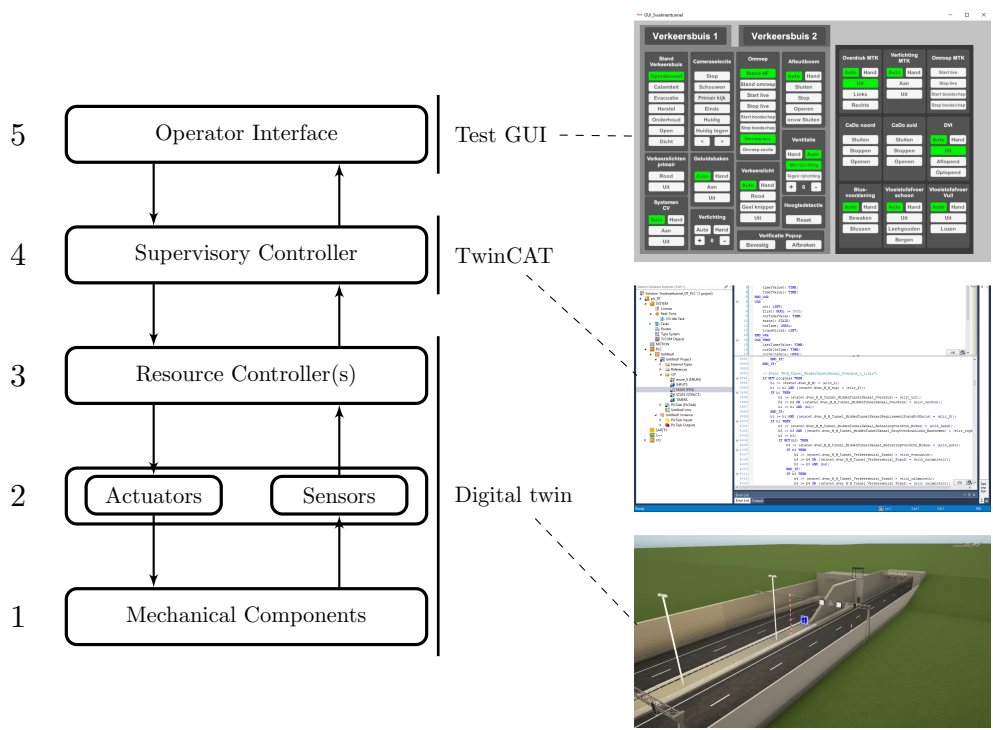
\begin{figure}[H]
    \centering
    \begin{tikzpicture}[node distance=1cm, auto, scale=1.1, transform shape]  
    		\tikzset{
    	    mynode/.style={rectangle,rounded corners,draw=black, top color=white, bottom color=white!50,very thick, inner sep=0.75em, minimum size=2.5em, minimum width=16em, text 		centered},
    	    mynode2/.style={rectangle,rounded corners,draw=black, top color=white, bottom color=white!50,very thick, inner sep=0.5em, minimum size=1em, minimum width=6em, text centered},
    	    mynode3/.style={rectangle,rounded corners,draw=none, top color=white, bottom color=white!50,very thick, inner sep=0.5em, minimum size=1em, minimum width=1em, text centered},
    	    myarrow/.style={->, >=latex', thick},
    	    mylabel/.style={text width=7em, text centered}
    		}  
    		
    		\definecolor{airforceblue}{rgb}{0.36, 0.54, 0.66}
    		\definecolor{navyblue}{rgb}{0.0, 0.0, 0.5}
    		\definecolor{royalblue}{rgb}{0.25, 0.41, 0.88}
    
      			\node[mynode,scale=0.7] (q1) {Operator Interface}; 
    		\node[mynode3, left=0.25cm  of q1, scale=1]   (n5) {5};  
    		\node[mynode3, right=0.25cm of q1, scale=1]   (n5) {};

      			\node[mynode,  below=0.75cm of q1, scale=0.7] (q2) {Supervisory Controller};   
    		
    		\node[mynode3, left=0.25cm  of q2, scale=1]   (n4) {4};

      			\node[mynode, below=0.75cm of q2, scale=0.7] (q3) {Resource Controller(s)};
      			\node[mynode, below=0.75cm of q3, scale=0.7] (q6) {};
      			\node[mynode2, below=0.85cm of q3, xshift=-1.11cm, scale=0.7] (q4) {Actuators};  
    			\node[mynode2, below=0.85cm of q3, xshift= 1.11cm, scale=0.7] (q5) {Sensors};  
      			\node[mynode, below=0.75cm of q6, scale=0.7] (q7) {Mechanical Components};
      		
      		\node[mynode3, left=0.25cm  of q3, scale=1]   (n3) {3}; 
      		\node[mynode3, left=0.25cm  of q6, scale=1]   (n2) {2};  
    		\node[mynode3, left=0.25cm  of q7, scale=1]   (n1) {1};

    			\node[mynode,draw=none, above right= -2cm and 1.5cm  of q7] (q3fig) {\includegraphics[width=.25\columnwidth]{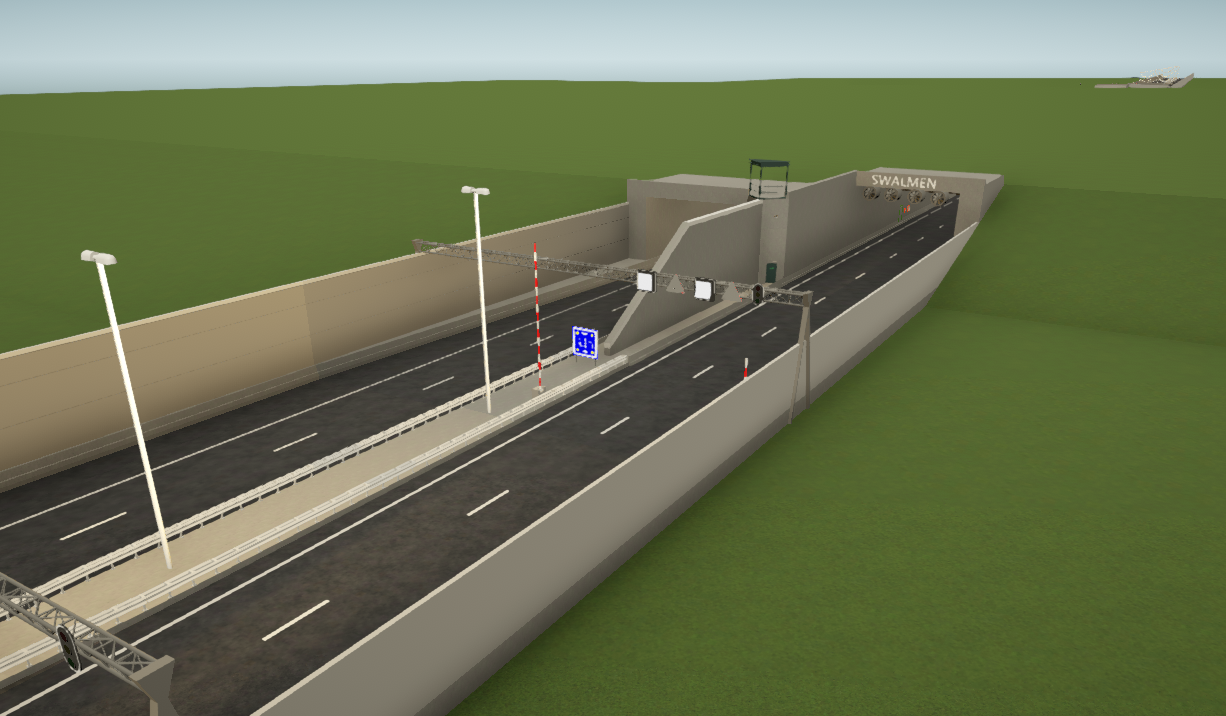}};
    		
    			\node[mynode,draw=none, right= 1.5cm  of q3] (q2fig) {\includegraphics[width=.25\columnwidth]{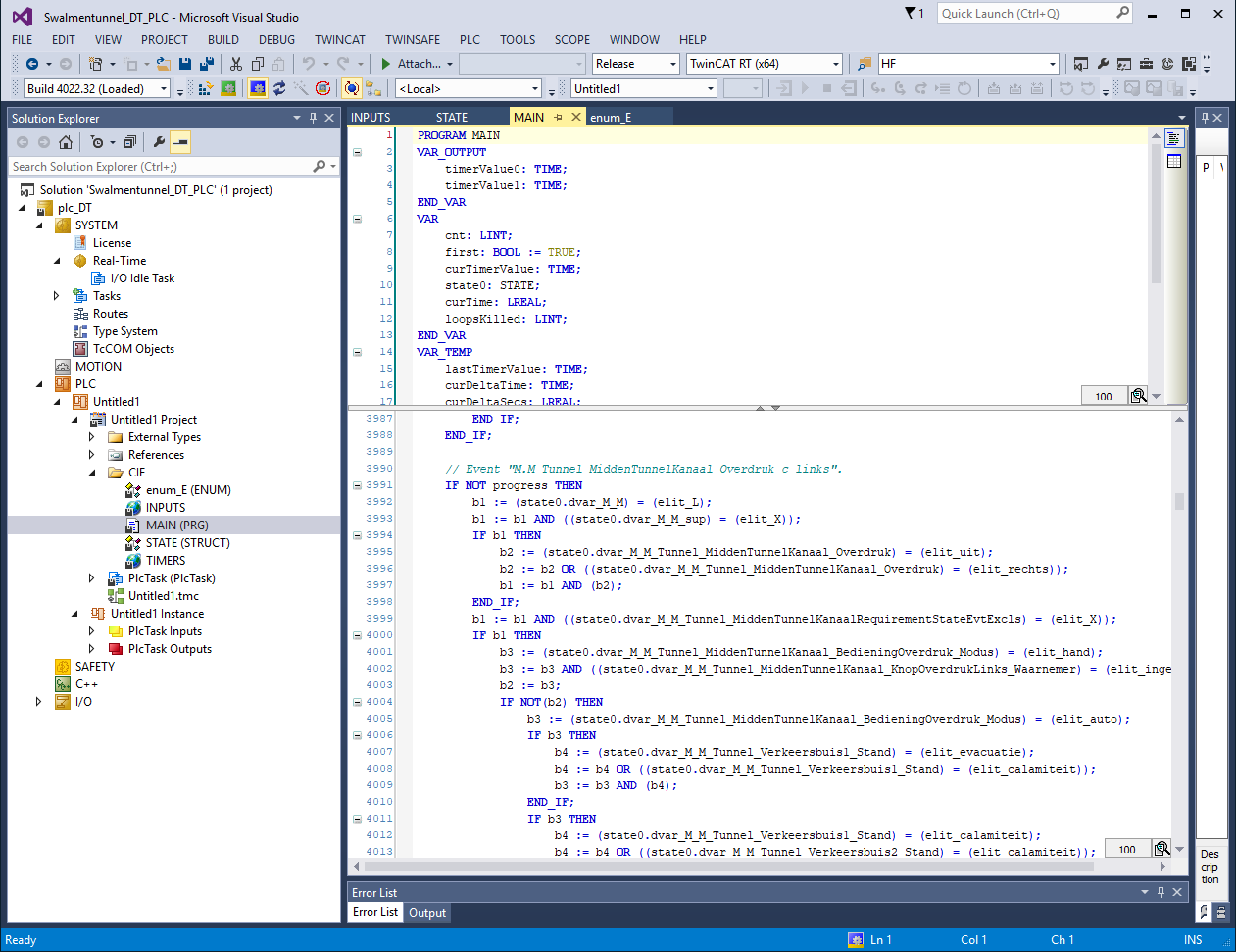}};
    		
    			\node[mynode,draw=none, below right= -2.15cm and 1.5cm  of q1] (q1fig) {\includegraphics[width=.25\columnwidth]{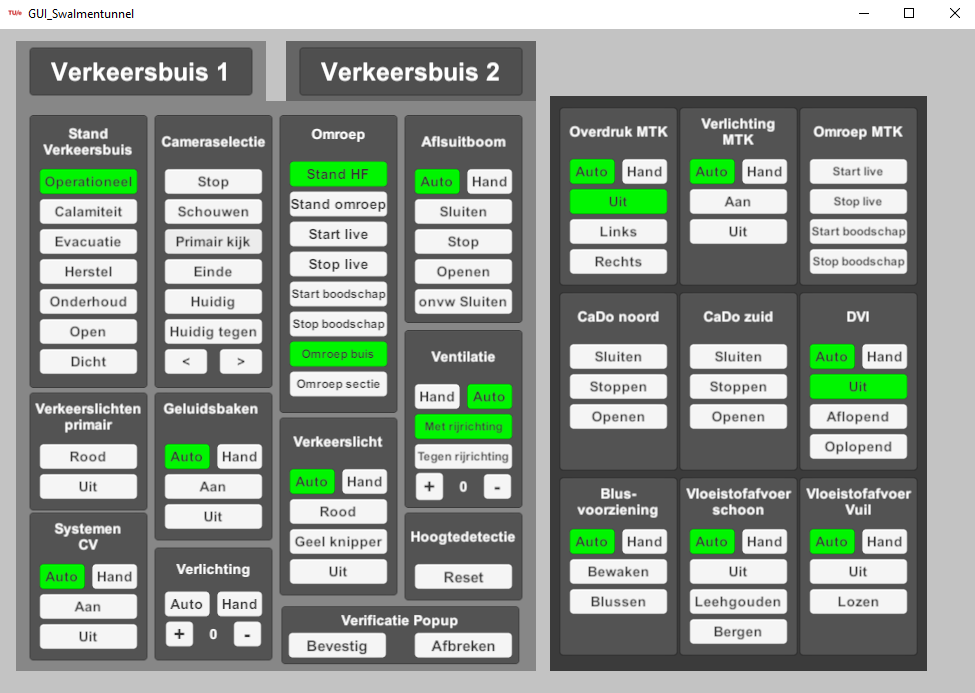}};
    			
    		\draw[myarrow] (q4.north |- q1.south) -| (q4.north |- q2.north);
    		\draw[myarrow] (q4.north |- q2.south) -| (q4.north |- q3.north);
    		\draw[myarrow] (q4.north |- q3.south) -| (q4.north);
    		\draw[myarrow] (q4.south) -| (q4.south |- q7.north);
    		
    		\draw[myarrow] (q5.north |- q2.north) -| (q5.north |- q1.south);
    		\draw[myarrow] (q5.north |- q3.north) -| (q5.north |- q2.south);
    		\draw[myarrow] (q5.north) -| (q5.north |- q3.south);
    		\draw[myarrow] (q5.south |- q7.north) -| (q5.south);	
    		
    		\node [fit=(q3) (q7)] (fit) {}; 
      		
      			\node [mynode3, right=0.05cm  of fit, scale=.7]   (plant) {Digital twin};    
      			\draw [decorate, line width=1pt] (fit.south east) -- (fit.north east);
      		
      		\node [fit=(q2)] (fit2) {};
      		
      			\node [mynode3, right=0.05cm  of fit2, scale=.7]   (controller) {TwinCAT};      
      			\draw [decorate, line width=1pt] (fit2.south east) -- (fit2.north east);
      			
      		\node [fit=(q1)] (fit3) {};
      			
      			\node [mynode3, right=0.05cm  of fit3, scale=.7]   (Test GUI) {Test GUI};    
      			\draw [decorate, line width=1pt] (fit3.south east) -- (fit3.north east);
      			
      			\draw[dashed] (Test GUI) -- ($(q1fig.west)+(1,0)$);	
      			\draw[dashed] (controller) -- ($(q2fig.west)+(1,0)$);	
      			\draw[dashed] (plant) -- ($(q3fig.west)+(1,0)$);	
      		
    	\end{tikzpicture}
    \caption{Schematic overview of the test setup for the digital twin.}
    \label{fig:set_up_tests}
\end{figure}

\newpage
\subsection{Test Operator Interface GUI}

For testing the digital twin, ideally an operator interface that conforms to the LTS is used. This operator interface GUI is not to be confused with the digital twin GUI as described in Section \ref{sec:dt_gui}. In \cite{Wang_2021}, an operator interface for the Eerste Heinenoordtunnel is made in Ignition SCADA that resembles the LTS operator interface that would be used in the real Eerste Heinenoordtunnel. However, for the Swalmen tunnel, such an operator interface is not available. The Ignition SCADA operator interface for the Eerste Heinenoordtunnel can be altered to acquire an operator interface for the Swalmen tunnel, but this is yet to be developed and implemented.\\

In order to test the digital twin, a test GUI is developed as shown in Figure \ref{fig:test_gui_dt} below. This GUI is a model of the operator interface of the Swalmen tunnel. It is developed in Unity using the same techniques for the input and output signals as described in Chapter \ref{sec:PLC_control} using a GUI that is made similarly to the GUI in the digital twin of the Swalmen tunnel (Section \ref{sec:dt_gui}). All buttons present on the LTS operator interface are present in the supervisor PLC code as input signals. The test GUI program sends these input signals when a button is pressed. Additionally, feedback is added to the test GUI in the form of output signals from the PLC code, for which an additional hardware mapping CIF specification section is written. The output signals enable feedback through colored buttons in the test GUI as shown in the figure.\\

\begin{figure}[H]
    \centering
    \includegraphics[width=.85\textwidth]{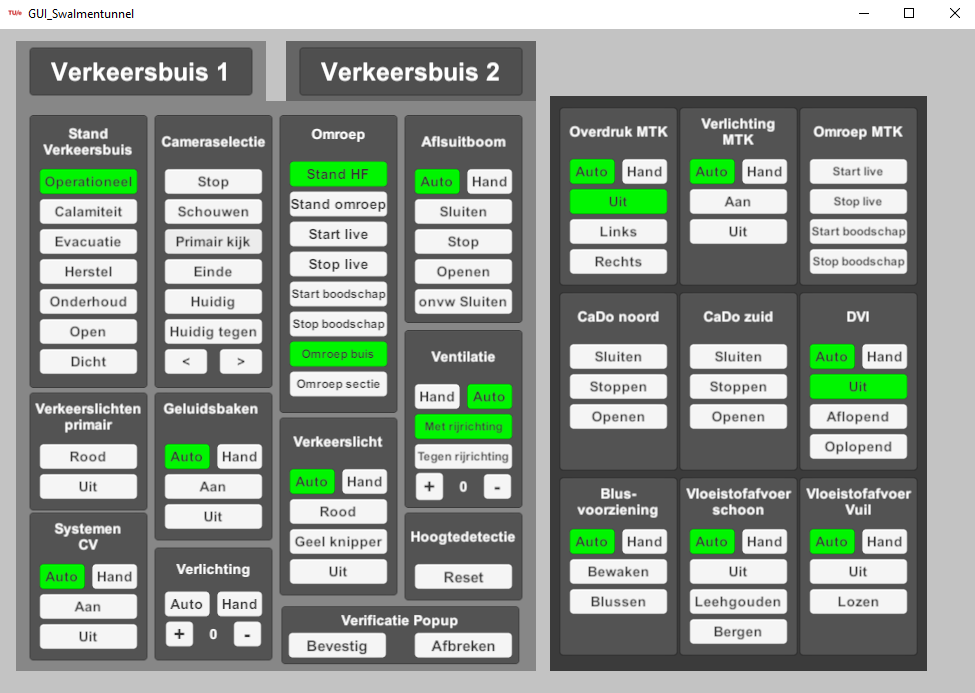}
    \caption{Test GUI digital twin.}
    \label{fig:test_gui_dt}
\end{figure}

Both the test GUI and the digital twin of the Swalmen tunnel communicate simultaneously with the PLC code of the supervisory controller in TwinCAT. Since the test GUI is entirely separated from the digital twin of the Swalmen tunnel and the buttons in the test GUI are modeled to work exactly like the buttons on the `real' operator interface, the test GUI can be used to carry out the validation tests of the digital twin. For validation of the supervisory controller however, some aspects are missing when using the test GUI. These mainly concern the operator interface windows that are also controlled by the PLC code. Examples are that some buttons are only shown in particular situations in separate pop-ups, which is not implemented in the test GUI.\\ \newpage

Below in Listing \ref{eq:lightLevelInterpolation}, the IO-script used for button input is given. This IO-script is used for all normal buttons in the test GUI. The function \textit{ButtonPressed()} is called when a button is clicked. This starts the function in lines 18-24 that sets the input signal to \textit{true} for 0.05 [s] and then sets it to \textit{false} again. This models the function of the LTS operator interface buttons that also provide short pulses to the PLC for button press signals. 

\vspace{3mm}    
\begin{lstlisting}[caption={IO-script for the buttons in the test GUI.}, label={listing:IO_buttons_test_gui},basicstyle=\scriptsize,frame=single]
public class DTGUI_IO_Button : MonoBehaviour
{
    [SerializeField, Tooltip("Name of the button logic component belonging to the button")]
    private string buttonString;

    private BoolToggle ButtonBoolToggle;

    private void Start()
    {
        ButtonBoolToggle = DTGUI_Static_Functions.FindBoolToggle(buttonString);
    }
    
    public void ButtonPressed() 
    {
        StartCoroutine(ButtonPressedCoRoutine());
    }
    
    IEnumerator ButtonPressedCoRoutine()
    {
        ButtonBoolToggle.Boolean = true;
        yield return new WaitForSeconds(0.05f);
        ButtonBoolToggle.Boolean = false;
    }
}
\end{lstlisting}

\subsection{PLC Code}

The PLC code used for validating the digital twin comes from the supervisory controller that is synthesized with the CIF3 Toolbox in an internship at RWS, as described in Section \ref{sec:prev_work}. This supervisory controller is already validated using model simulation, as explained in Subsection \ref{sec:synthesis_based_controller_design} with a hybrid model in the CIF3 Toolbox, but not in a hardware-in-the-loop test. The hardware mapping for the supervisor is written and the PLC code is generated as explained in Section \ref{sec:PLC_code_generation} and loaded into TwinCAT as described in Section \ref{sec:plc_cif_twincat}. Because the supervisory controller has not been validated using hardware-in-the-loop tests, errors can seem to occur in the digital twin that are actually caused by mistakes in the PLC code. This needs to be taken into account when analyzing errors occurring in the validation tests of the digital twin.

\subsection{Test Results}

For the first full scale tests, the digital twin and test GUI are started and the PLC is activated. The first problem is found in the PLC code that had become stuck in a loop, meaning that no control is possible in the digital twin. Apparently, the requirements used for synthesizing the supervisory controller section for the high frequency broadcast system in the supplied CIF model were incomplete, allowing for an event in the PLC code to always be possible. The parts of the PLC code controlling the HF-system were deleted, which solves the problem. This does not have any consequences for the validation process of the digital twin, since the non-HF broadcasting system is still tested. This error in the PLC code was not found because of the digital twin, because just running the PLC yielded the error. This means that the error would also have been revealed in a hardware-in-the-loop test with the hybrid CIF model. \\ \newpage

With the working PLC code, further validation tests were done by carrying out several test scenarios and observing the behavior of the digital twin by pressing buttons on the test GUI. This observed behavior is then compared to the behavior in the hybrid CIF model and the requirements of the systems in the Swalmen tunnel. Since not all systems have fully been implemented yet, some button signals do not result in any actions in the digital twin. Because the missing systems do not have behavior that is entirely different from other implemented systems, the testing method is still valid. \\

While running the testing scenarios and testing the systems, some minor modeling mistakes for sensor values and actuator behavior were revealed and fixed. However, no fundamental problems were found and all testing scenarios yielded the desired behavior of all systems. A problem was found in the test GUI, where the PLC code would sometimes get stuck in a loop when two buttons on the operator interface are pressed in very quick succession ($\Delta t << 1$ [s]). This however is inherent to the setup of the test GUI, for which the goal is not to validate it and the problem does not occur in normal use. Therefore, this problem was not fixed. A possible solution is to lower the time duration for which a button input signal is set to \textit{true} in Listing \ref{listing:IO_buttons_test_gui}. \\

Apart from all controlled entities, the function of all interactive components in the digital twin itself were tested thoroughly and worked as intended after some small alterations. Hyperlinks to videos of several test scenarios and tests with interactive components can be found in Appendix \ref{sec:test_results}. These include test scenarios with interaction between boom barriers and traffic, a test with the height detection system when a truck nears the tunnel that is too large and a test with commands for the pumping cellars. Furthermore, the CCTV-feeds are illustrated and a test with direct interaction in the digital twin by clicking objects is done.

\chapter{Conclusion and Recommendations}\label{sec:evaluation_conclusion}

Having developed and tested the digital twin of the Swalmen tunnel, some conclusions can be drawn from the development process and the applicability of digital tunnel twins. First, the three research objectives from Chapter \ref{sec:introduction} are discussed and conclusions are drawn based on findings from previous chapters. This includes the applicability of digital tunnel twins and the place of a digital twin in the supervisory controller development process from Chapter \ref{sec:controller_design}. Afterwards, the methods used and the digital twin developed in this report are evaluated and recommendations are given.


\section{Conclusions}

The first research objective concerns the development of a digital twin of the Swalmen tunnel for validation of a supervisory controller:

\begin{itemize}
    \item[] \textit{1. Develop a digital twin of the Swalmen tunnel that can be used to test and validate a supervisory controller.}
\end{itemize}

From the test results of Chapter \ref{sec:validation}, it can be concluded that the digital twin of the Swalmen tunnel is suitable for supervisory controller validation. As mentioned, not all systems were implemented to their full extent in the digital twin. However, the methods used for modeling other systems in the digital twin that are successfully validated are similar. In the recommendations, further development and validation guidelines are given. The fact that the I/O interface between the virtual PLC in TwinCAT and the digital twin in Unity is almost identical to the I/O interface between the digital twin and a hardware PLC proves that the digital twin can be used for hardware-in-the-loop simulations to validate a supervisory controller. This brings forward the second research objective:

\begin{itemize}
    \item[] \textit{2. Investigate applications of digital tunnel twins.}
\end{itemize}

First, the application of supervisory controller validation is discussed, since this is the application for which the digital twin of the Swalmen tunnel is developed primarily. Figure \ref{fig:controller_development_process_with_dt} below shows the schematic overview of the synthesis-based supervisory controller process that is discussed in Section \ref{sec:synthesis_based_controller_design}. The difference here is that the digital twin is added as a separate plant model between the hybrid plant and the realized plant. The role of the digital twin shown in the figure is not necessarily the optimal role, because the whole development is not yet executed with a digital twin model, but this is assumed to be a good place from experience and knowledge gained from testing the digital twin from the Swalmen tunnel. The first controller validation step in Figure \ref{fig:controller_development_process_with_dt} is still model simulation, where the synthesized controller directly controls the hybrid model of the plant. This step is useful for quickly identifying fundamental mistakes in the formalized requirements for the controller or in the model of the plant used in the synthesizing steps. The second validation method is still hardware-in-the-loop simulation, but now the digital twin is used instead of the hybrid model. The reason is that having an additional step where the hardware-in-the-loop simulation is carried out with the hybrid plant would not have additional value. The reason is that potential problems that are found with the PLC implementation are also found when running the digital twin in a hardware-in-the-loop-simulation. Developing a digital twin costs some additional effort, but the advantages of the use of a digital twin described here outweigh this additional work. Furthermore, the digital twin of the Swalmen tunnel has been designed in a relatively short time by one person and the modular nature of the digital twin further decreases the amount of work needed.

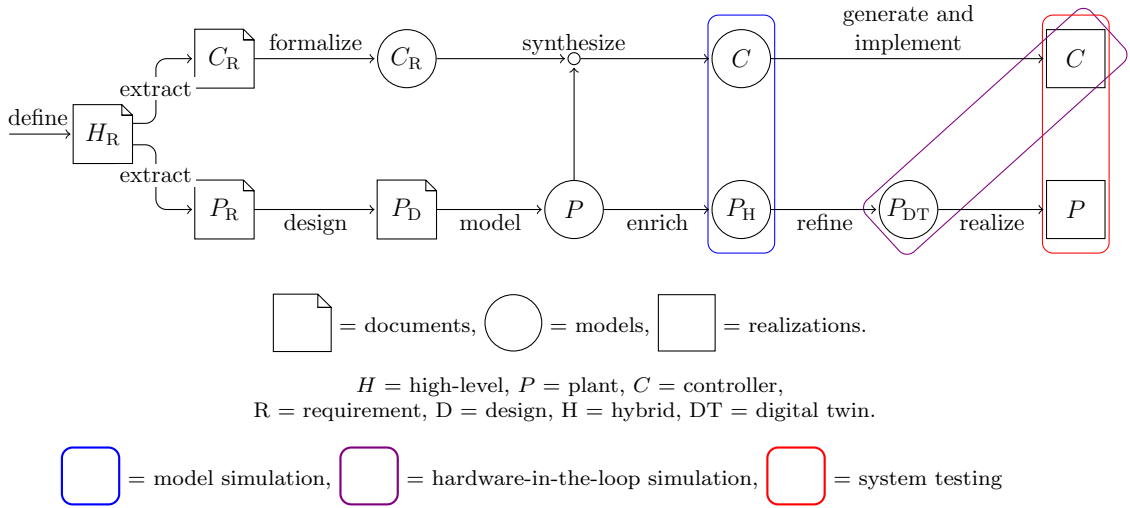
\begin{figure}[H]
    \centering
    \vspace{-2mm}
    \begin{tikzpicture}[fmbe,scale=.8,transform shape, node distance=2.75cm ]
		\node[document]                                    	(D)       	{$H_\mathrm{R}$};
		\node[document,above right=1.25cm and 2cm of D]		(R1)      	{$C_\mathrm{R}$};
		\node[model,right=3cm of R1]                     	(D1)      	{$C_\mathrm{R}$};
		\node[right of=D1,circle,draw,inner sep=2pt] 	    (t)       	{};
		\node[model,right of=t]                           	(M1)     	{$C$};
		\node[realization,right=5.5cm of M1]                (Z1)      	{$C$};		
		\node[document,below right=1.25cm and 2cm of D]     (R2)      	{$P_\mathrm{R}$};
		\node[document,right=3cm of R2]                  	(D2)      	{$P_\mathrm{D}$};
		\node[model,right of=D2]                       		(M3)      	{$P$};
		\node[model,right of=M3]                            (M35)       {$P_\mathrm{H}$};
		\node[model,right of=M35]                           (M4)      	{$P_\mathrm{DT}$};
		\node[realization,right of=M4]                 	   	(Z2)      	{$P$};
		
		\definecolor{evalcolor}{named}{blue} \path (M35) -- node[eval,minimum width=2.85em,minimum height=10.25em,sloped,rotate=-90] (MZ) {} (M1);
		
		
		\definecolor{evalcolor}{named}{violet} \path (M4) -- node[eval,minimum width=3.0em,minimum height=13em, xshift=.4em, sloped,rotate=-90] (MZ) {} (Z1);
		
		\definecolor{evalcolor}{named}{red} \node[eval,minimum width=2.85em,minimum height=10.25em] (Z) at (D -| Z1)  {};	
		
		\draw[act] (D.20)  -- +(0:1em) |- node[actlb',pos=.25] {extract}
		(R1);
		\draw[act] (D.-20) -- +(0:1em) |- node[actlb',pos=.25] {extract} (R2);
		\path[act]
		([xshift=-4em]D.center) 
		      edge node[pos=.45]       {define}                     (D)
		(R1)  edge node                {formalize}                  (D1)
		(R2)  edge node[below]                {design}                     (D2)
		(D1)  edge                                                  (t)
		(M3)  edge                                                  (t)
		(t)   edge                                                  (M1)
		(D1)  --   node                {synthesize}                 (M1)
		(M1) edge node[align=center]   {generate and \\implement}   (Z1)
		(D2)  edge node[below]                {model}                      (M3)
		(M3)  edge node[pos=.5][below]        {enrich}                     (M35)
		(M35)  edge node[pos=.5][below]       {refine}                     (M4)
		(M4)  edge node[below]                {realize}                    (Z2)
		;
		
		\pgfusepath{use as bounding box}
		
		\end{tikzpicture}
		
    	\scriptsize{
    		\begin{center}
    			\tikz[fmbe,baseline=(x.base)]{\node[document] (x) {$\phantom{X}$};} = documents,
    			\tikz[fmbe,baseline=(x.base)]{\node[model](x){$\phantom{Y}$};} = models,
    			\tikz[fmbe,baseline=(x.base)]{\node[realization](x){$\phantom{Z}$};} = realizations.
    			
    			\vspace{1em}
    			$H$ = high-level, $P$ = plant, $C$ = controller, \\ $\mathrm{R}$ = requirement, $\mathrm{D}$ = design, $\mathrm{H}$ = hybrid, $\mathrm{DT}$ = digital twin.		
    			\vspace{1em}
    			
    			\resizebox{.8\columnwidth}{!}{
    			\tikz[fmbe,baseline=(x.base)]{\node[realization, blue, rounded corners, thick] (x) {$\phantom{X}$};} = model simulation,
    			\tikz[fmbe,baseline=(x.base)]{\node[realization, violet, rounded corners, thick] (x) {$\phantom{X}$};} = hardware-in-the-loop simulation,
                \tikz[fmbe,baseline=(x.base)]{\node[realization, red, rounded corners, thick] (x) {$\phantom{X}$};} = system testing\hspace{10mm}
    			}
    			
    		\end{center}
    	}
    	\vspace{-5mm}
    \caption{Supervisory controller development with a digital twin.}
    \label{fig:controller_development_process_with_dt}
\end{figure}

As described in Section \ref{sec:digital_tunnel_twins_COB}, the additional value of a digital tunnel twin in the design of a supervisory controller for a tunnel lies in early, high-fidelity simulation for validation of the system in several levels of detail. This is essential for tunnels, since this reduces the time duration of tunnel closure needed for testing purposes. Additionally, the controller validation testing process is more accessible due to the 3D nature of the digital twin and its resemblance to the real system. Also, test scenarios can be more elaborate than in the hybrid model, since traffic flows can be modeled and the additional spatial dimension (2D to 3D) enables more realistic testing. An example is the height detection system. The improved testing possibilities are also realized by the interactive nature of the digital twin program, of which the realization is described in Section \ref{sec:interaction_dt}. The accessible nature of the digital twin enables experienced tunnel operators to also be involved in the controller validation process, which is beneficial, because operators are the users of the final operator interface that interacts with the controller and they can provide useful feedback. Other differences that make the digital twin useful in hardware-in-the-loop simulation is that systems such as the broadcast system and the CCTV system are modeled much more realistically, using sound and real-time footage in the tunnel. This even enables determining the optimal orientation and location of the CCTV cameras in the tunnel environment.\\

Another interesting application mentioned in Section \ref{sec:digital_tunnel_twins_COB} is operator training. This application already arises from the involvement of operators in the digital twin controller validation process, because of the accessibility of the digital twin. With some small additions and changes, a digital twin that is used for supervisory controller validation can also be used for operator training. What makes this possible is that a 3D model of the plant with a functioning controller and operator interface is already present. The inclusion of CCTV feeds, which are already implemented partially in the digital twin of the Swalmen tunnel, is an important aspect in operator training, since operators usually only use the camera views when operating a tunnel. Figure \ref{fig:comparison_dt_real_tunnel} below shows a comparison between the entrance of the real Swalmen tunnel and the same location in the digital twin. Here, it is evident that the digital twin resembles the real tunnel, but not on a detailed level. For instance, the entrance is not exactly the same size and some railings and signs are not modeled in the digital twin. Furthermore, the location of the boom barriers and the matrix signs is simplified. For high-fidelity operator training, some of these simplifications should be omitted to create a 3D model of the tunnel that resembles the physical tunnel in more detail.

    \begin{figure}[H]
        \centering
        \includegraphics[width=.9\textwidth]{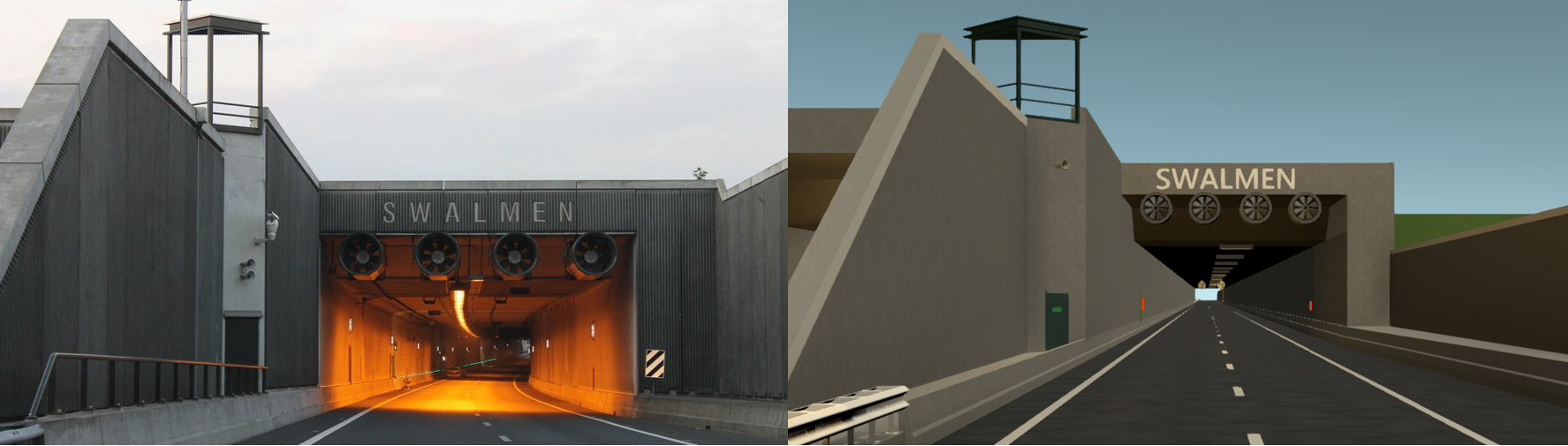}
        \caption{Comparison between the real Swalmen tunnel (left) and the digital twin of the Swalmen tunnel (right).}
        \label{fig:comparison_dt_real_tunnel}
    \end{figure}

The third and final research objective is to provide a manual for the development of digital twins:

\begin{itemize}
    \item[] \textit{3. Describe the method of developing digital twins of systems similar to the Swalmen tunnel.}
\end{itemize}

Three chapters of the report are devoted to the development of digital tunnel twins. These are all aimed on development of digital twins that serve as a tool for supervisory control validation, as stated in research objective 1. The first of these chapters is Chapter \ref{sec:set_up_digital_twin}, where the general methods used for developing a digital twin that can be connected to a PLC with Boolean signals are described. Chapter \ref{sec:PLC_control} includes a detailed description of the methods used to connect the digital twin with a (virtual) PLC and the methods used to generate PLC code from a supervisory controller that is synthesized in the CIF Toolbox. Finally, Chapter \ref{sec:digital_twin_swalmentunnel} describes how the methods described in the previous chapters are applied to model the different aspects of the Swalmen tunnel. Together, these three chapters form a manual that can be used to develop digital twins for infrastructural systems in Unity using the Prespective plugin. Apart from describing methods, several methods have been developed that can support the development process of a digital twin. Examples are the automatic generation of logic component GameObjects, as described in Section \ref{sec:logic_component_generation}, and the development of the digital twin of the Swalmen tunnel itself, from which modular parts can be used to develop other digital tunnel twins.

\section{Recommendations}

Concerning the development of the digital twin of the Swalmen tunnel, some entities in the tunnel are not fully implemented yet, as mentioned in Chapter \ref{sec:digital_twin_swalmentunnel}. This includes the broadcast synchronization system and the CCTV system. The CCTV system should also be improved in terms of performance. In order to formally confirm that the digital twin can be used to validate the supervisory controller, the digital twin should be finished and tested fully. This includes fixing the problem in the generated PLC-code concerning the HF-system, such that all tunnel systems can be tested in the digital twin. Additionally, an RWS operator interface that conforms to the LTS should be implemented in the test setup, such that a valid hardware-in-the-loop test can be carried out. In theory, this should not yield results different from the tests with the soft PLC, but in the controller design process, a hardware PLC is used. Implementing the RWS operator interface also enables the user to test the GUI control elements of the PLC-code. It also removes the problem of PLC getting stuck in a loop when buttons on the test GUI are pressed in quick succession. Furthermore, as explained in Section \ref{fig:current_lighting}, the lighting is not correct yet. This should also be fixed, since this allows for better testing of the light systems in the traffic tubes and the central corridor. Correct lighting also improves the visibility of entities inside the tunnel.\\

The first section of Chapter \ref{sec:digital_twin_swalmentunnel} mentions some simplifications that were implemented to simplify the development process of the digital twin. Here, it was assumed that these simplifications and assumptions would not have a negative effect on validation of the controller. While this is the case for the test scenarios that were also carried out in the hybrid model, some more advanced testing scenarios can not be fully tested, meaning that the controller cannot be validated for these cases. This includes more intricate vehicle behavior for instance, including lane switching and collisions. Another example is the simplification for which some entities are not in the correct locations. This is for example the case for the traffic lights, matrix signs, boom barriers and the height detection systems. For the latter, this is the case because the high way entrance before the Northern entrance of the tunnel is not modeled, but a height detection system is located here in reality. For more complex situations, for instance interaction between vehicles around height detection systems and boom barriers and the formation of traffic jams over a larger distance in front of the tunnel, these simplifications might still allow for unwanted behavior of the supervisory controller that is not detected. It is therefore important to evaluate in further research whether the assumptions and simplifications are sufficient for full validation of the supervisory controller. In this consideration, it should also be taken into account that the tunnel operator also plays a role when these complex situations emerge. \\

\vspace{-2mm}

Regarding control validation in the digital twin, improvements can be made in order to give more feedback. Now, the warning boxes are only activated when multiple actuator signals are \textit{true} when this should not be possible and additional cases can be added. Furthermore, warning messages could be added that tell the user for which component a warning box is activated. This prevents warning boxes from being ignored when they are not visible on the screen.\\

\vspace{-2mm}

The tools used for developing the digital twin as mentioned in \ref{sec:tools_DT} have all been useful in developing the digital twin. Unity already contains a lot of features from the realms of game development, such as cameras, lighting, visuals and physics that are used for developing the digital twin. The scripts that are used in Unity are also very useful, since C\# and JavaScript can be used, which are intuitive coding languages for people that have some experience with programming. Another advantage is that the Unity community is very large, so a solution to almost all Unity-related questions can be found on the internet. The Prespective plugin is very useful for the implementation of PLC input and output signals in Unity, but the range of reference material available is not very wide. Prespective also contains some modeling components to Unity, such as motors and indicator lights. Some of these components do not work intuitively and contain some bugs, so in the development process of the digital twin, these have not been used extensively. Most of the modeling, for instance the barrier movement, is implemented with self-written scripts.\\

\vspace{-2mm}

Concerning the whole synthesis-based controller design process, from synthesizing the controller in the CIF Toolbox to developing and running the digital twin, some steps can be improved. Firstly, the hardware mapping process can be improved by setting up the CIF model in a more structured way, as described in the last paragraph of Section \ref{sec:hardware_mapping}. This eliminates the need to write the hardware mapping manually afterwards as a separate element, because the process is tied in the setup of the whole model. A second point lies in the language difference between the CIF model and the digital twin. This sometimes leads to confusion and can be ambiguous, because most terms used by RWS do not translate into English one-to-one. It would be preferred to also write the CIF model that is used for synthesis of the supervisory controller in English, such that the signal names can be linked directly to the structure of the CIF model. A final remark regards the way in which the signals for entities such as the ventilation units are realized. These rotate at a velocity that links to a level from 0 up to 8. These levels are now all represented by Boolean signals, such that it is always the case that exactly one Boolean is \textit{true} and eight are \textit{false}. It would be more efficient to use an integer-type signal to communicate this level variable. This is already implemented in \cite{mats_digital_twin_brug} and could also be used for the light units in the traffic tubes and the central corridor.\\

Because the digital twin is designed for the application of supervisory control validation, many physical sensors and actuators are modeled at a global level. For other applications, such as training mechanics and validating the plant at a more detailed level, some entities in the tunnel can be modeled with more fidelity. An example is the movement of a barrier beam. This is now done by linearly interpolating the angle, but the real control software of the barrier that lets the beam rotation follow a set point could also be implemented. This could be useful for studying the behavior of the resource controllers. For the sake of operator training, an additional more realistic feature is to improve the broadcast messages. The prerecorded messages should then be replaced with authentic prerecorded messages that are also used in tunnels maintained by RWS. A possible addition is then to implement the option of recording messages with a microphone connected to the computer that runs the digital twin. This way, live broadcasting can also be enabled.\\

As mentioned in the conclusions, the model of the tunnel in the digital twin needs to be more detailed for some operator training applications. A solution for new bridges is to implement 3D models of the tunnel structure that are modeled for construction purposes, for example CAD or BIM models. This ensures that the digital twin exactly resembles the real tunnel in terms of its dimensions. A disadvantage of this method can be that the construction models are too detailed, such that the performance of the digital twin becomes poor. To solve this issue, the models can be altered to get rid of an excessive level of detail and video game techniques can be used to make rendering the mesh of the models more efficient. \\

\addcontentsline{toc}{chapter}{References}

\bibliography{Bibliography}

\appendix 
\chapter{Overview of all Inputs and Outputs}\label{sec:appendixA}

In this appendix, a list of all inputs and outputs of the PLC code that is used to control the digital twin of the Swalmen tunnel is given. For components of which multiple entities are present, the inputs and outputs are given only once. These input and output names are chosen to represent the action or state linked to the variable. The names of the variables in the PLC link to the variable names in the lists below by the structure of the tunnel, for example, the variable ``$dvar\_M\_M\_HW\_TrafficTube\_1\_BoomBarrier\_2\_a\_open$" controls the open actuator Boolean of the second boom barrier in traffic tube 1.

\begin{longtable}[]{|p{0.25\textwidth}||p{.375\textwidth}|p{.375\textwidth}|}

        \hline
        \textbf{Component}          & \textbf{Input Variables}  & \textbf{Output Variables} \\ 
        \hline \hline
        \multirow{10}{*}{\textbf{Aid cabinet A}}
                                    & s\_cabinet\_opened        &                           \\ 
                                    & s\_cabinet\_closed        &                           \\ 
                                    & s\_emergencyPhone\_on     &                           \\ 
                                    & s\_emergencyPhone\_off    &                           \\ 
                                    & s\_handExtinguisher\_on   &                           \\ 
                                    & s\_handExtinguisher\_off  &                           \\ 
                                    & s\_fireHose\_on           &                           \\ 
                                    & s\_fireHose\_off          &                           \\ 
                                    & s\_startPump\_on          &                           \\ 
                                    & s\_startPump\_off         &                           \\ 
        \hline \hline
        \multirow{6}{*}{\textbf{Aid cabinet C}}
                                    & s\_cabinet\_opened        &                           \\ 
                                    & s\_cabinet\_closed        &                           \\ 
                                    & s\_emergencyPhone\_on     &                           \\ 
                                    & s\_emergencyPhone\_off    &                           \\ 
                                    & s\_handExtinguisher\_on   &                           \\ 
                                    & s\_handExtinguisher\_off  &                           \\ 
        \hline \hline
        \multirow{7}{*}{\textbf{Boom barrier}}       
                                    & s\_opened                 & a\_noChoice               \\ 
                                    & s\_opening                & a\_open                   \\ 
                                    & s\_stopped                & a\_stop                   \\ 
                                    & s\_closing                & a\_close                  \\ 
                                    & s\_closed                 &                           \\ 
                                    & s\_obst\_on               &                           \\ 
                                    & s\_obst\_off              &                           \\ 
        \hline \hline
        \multirow{6}{*}{\textbf{Boom barrier control}}
                                    & s\_bothOpened             &                           \\ 
                                    & s\_bothOpening            &                           \\ 
                                    & s\_bothStopped            &                           \\ 
                                    & s\_bothClosing            &                           \\ 
                                    & s\_bothClosed             &                           \\ 
                                    & s\_individual             &                           \\
        \hline \newpage \hline
        \multirow{8}{*}{\textbf{Broadcast}}          
                                    & s\_broadcast\_recordingFinished  & a\_HF\_radio       \\
                                    & s\_HF\_messageFinished    & a\_HF\_message            \\
                                    &                           & a\_HF\_mute               \\
                                    &                           & a\_broadcast\_off         \\
                                    &                           & a\_broadcast\_live        \\
                                    &                           & a\_broadcast\_message     \\
                                    &                           & a\_broadcastSection\_tube    \\
                                    &                           & a\_broadcastSection\_section \\
        \hline \hline
        \multirow{5}{*}{\textbf{CCTV system}}        
                                    & s\_autoSwitch\_on         & a\_surveying              \\
                                    & s\_autoSwitch\_off        & a\_surveyingCurrent       \\
                                    &                           & a\_direction\_with        \\
                                    &                           & a\_direction\_against     \\
                                    &                           & a\_registration\_on       \\
        \hline \hline
        \multirow{3}{*}{\textbf{Emergency exit}}     
                                    & s\_opened                 & a\_mute\_on               \\
                                    & s\_closed                 & a\_soundBeacon\_on        \\
                                    &                           & a\_contourLighting\_on    \\
        \hline \hline
        \multirow{3}{*}{\textbf{Height detection}}   
                                    & s\_on                     & a\_lights\_on             \\
                                    & s\_off                    & a\_timerStarted           \\
                                    & s\_timerFinished          & a\_contourLighting\_on    \\
        \hline \hline
        \multirow{9}{*}{\textbf{Lighting}}           
                                    &                           & a\_s0                     \\
                                    &                           & a\_s1                     \\
                                    &                           & a\_s2                     \\
                                    &                           & a\_s3                     \\
                                    &                           & a\_s4                     \\
                                    &                           & a\_s5                     \\
                                    &                           & a\_s6                     \\
                                    &                           & a\_s7                     \\
                                    &                           & a\_s9                     \\
        \hline \hline
        \multirow{9}{*}{\textbf{Light sensor}}       
                                    & s\_0                      &                           \\
                                    & s\_1                      &                           \\
                                    & s\_2                      &                           \\
                                    & s\_3                      &                           \\
                                    & s\_4                      &                           \\
                                    & s\_5                      &                           \\
                                    & s\_6                      &                           \\
                                    & s\_7                      &                           \\
                                    & s\_8                      &                           \\
        \hline \newpage \hline
        \multirow{9}{*}{\textbf{Smoke detection}}    
                                    & s\_0                      &                           \\
                                    & s\_1                      &                           \\
                                    & s\_2                      &                           \\
                                    & s\_3                      &                           \\
                                    & s\_4                      &                           \\
                                    & s\_5                      &                           \\
                                    & s\_6                      &                           \\
                                    & s\_7                      &                           \\
                                    & s\_8                      &                           \\
        \hline \hline
        \multirow{6}{*}{\textbf{SOS system}}         
                                    & s\_wrongwayDriver\_on     &                           \\
                                    & s\_wrongwayDriver\_off    &                           \\
                                    & s\_stationaryVehicle\_on  &                           \\
                                    & s\_stationaryVehicle\_off &                           \\
                                    & s\_speedingDriver\_on     &                           \\
                                    & s\_speedingDriver\_off    &                           \\
        \hline \hline
        \multirow{5}{*}{\textbf{Traffic light}}      
                                    & s\_flashing               & a\_flashing               \\
                                    & s\_red                    & a\_red                    \\
                                    & s\_green                  & a\_green                  \\
                                    & s\_off                    & a\_off                    \\
                                    &                           & a\_noChoice               \\
        \hline \hline
        \multirow{4}{*}{\textbf{Traffic light control}} 
                                    & s\_trafficlights\_off     &                           \\
                                    & s\_trafficlights\_flashing &                          \\
                                    & s\_trafficlights\_yellow  &                           \\
                                    & s\_ttrafficlights\_red    &                           \\
        \hline \hline
        \multirow{1}{*}{\textbf{Traffic signs (J32)}} 
                                    &                           & a\_on                     \\
        \hline \hline
        \multirow{11}{*}{\textbf{Ventilation}}        
                                    &                           & a\_s0                     \\
                                    &                           & a\_s1                     \\
                                    &                           & a\_s2                     \\
                                    &                           & a\_s3                     \\
                                    &                           & a\_s4                     \\
                                    &                           & a\_s5                     \\
                                    &                           & a\_s6                     \\
                                    &                           & a\_s7                     \\
                                    &                           & a\_s8                     \\
                                    &                           & a\_drivingDir             \\
                                    &                           & a\_againstDrivingDir      \\  
        \hline
        
    \caption{List of input and output variables used in the digital twin for the traffic tube.}
    \label{tab:IO_list_traffic_tube}
\end{longtable}      

\vspace{6mm}

\begin{longtable}{|p{0.33\textwidth}||p{.33\textwidth}|p{.33\textwidth}|}

        \hline
        \textbf{Component}          
                                    & \textbf{Input Variables}  & \textbf{Output Variables} \\
        \hline \hline
        \multirow{3}{*}{\textbf{Broadcast}}          
                                    & s\_recordingStopped       & a\_broadcast\_off         \\
                                    &                           & a\_broadcast\_live        \\
                                    &                           & a\_broadcast\_message     \\
        \hline \hline
        \multirow{1}{*}{\textbf{Dynamic escape}}     
                                    &                           & a\_off                    \\
        \multirow{2}{*}{\textbf{route indication system}} 
                                    &                           & a\_ascending              \\
                                    &                           & a\_descending             \\
        \hline \hline
        \multirow{1}{*}{\textbf{Lighting}}           
                                    &                           & a\_on                     \\
        \hline \hline
        \multirow{2}{*}{\textbf{Main door}}          
                                    & s\_open                   &                           \\
                                    & s\_closed                 &                           \\
        \hline \hline
        \multirow{3}{*}{\textbf{Overpressure}}      
                                    &                           &  a\_off                   \\
                                    &                           &  a\_left                  \\
                                    &                           &  a\_right                 \\
        \hline
                                    
    \caption{List of input and output variables used in the digital twin for the central corridor.}
    \label{tab:IO_list_central_corridor}
\end{longtable}  

\begin{longtable}{|p{0.33\textwidth}||p{.33\textwidth}|p{.33\textwidth}|}

        \hline
        \textbf{Component}          & \textbf{Input Variables}  & \textbf{Output Variables} \\
        \hline \hline
        \multirow{3}{*}{\textbf{Broadcast synchronization}} 
                                    & s\_timerGB\_timeout       & a\_off                    \\
                                    &                           & a\_reset                  \\ 
        \hline \hline
        \multirow{5}{*}{\textbf{Emergency passage}}   
                                    & s\_opened                 & a\_noChoice               \\
                                    & s\_opening                & a\_open                   \\
                                    & s\_stopped                & a\_stop                   \\
                                    & s\_closing                & a\_close                  \\
                                    & s\_closed                 &                           \\ 
        \hline \hline
        \multirow{4}{*}{\textbf{Fire extinguishing system}} 
                                    & s\_low\_on                & a\_pump\_on               \\
                                    & s\_low\_off               &                           \\
                                    & s\_high\_on               &                           \\
                                    & s\_high\_off              &                           \\ 
        \hline \hline
        \multirow{10}{*}{\textbf{Pumping cellar clean}} 
                                    & s\_low\_on                & a\_pump\_on               \\
                                    & s\_low\_off               &                           \\
                                    & s\_start\_on              &                           \\
                                    & s\_start\_off             &                           \\
                                    & s\_maxStart\_on           &                           \\
                                    & s\_maxStart\_off          &                           \\
                                    & s\_lowHigh\_on            &                           \\
                                    & s\_lowHigh\_off           &                           \\
                                    & s\_highHigh\_on           &                           \\
                                    & s\_highHigh\_off          &                           \\ 
        \hline \hline
        \multirow{2}{*}{\textbf{Pumping cellar dirty}}
                                    & s\_low\_on              & a\_pump\_on               \\
                                    & s\_low\_off               &                           \\
        \hline
                                    
    \caption{List of input and output variables used in the digital twin for the other systems.}
    \label{tab:IO_list_central_corridor}
\end{longtable}        

\chapter{Scripts}\label{sec:appendix_scripts}

Here, the most important scripts (C\# and Tooldef) that are used in the digital twin of the Swalmen tunnel or for generating the PLC code are shown. Not all C\# scripts are given, since the IO scripts for the controlled entities in the digital twin are similar in structure.

\section{IO-scripts}
    
\begin{lstlisting}[caption={\textit{IO\_Barrier.cs}.}, label={listing:IO_Barrier},basicstyle=\tiny,frame=single]
using System.Collections.Generic;
using UnityEngine;

public class IO_Barrier : MonoBehaviour
{
    [Header("Output logic component names")]

        [SerializeField, Tooltip("No instruction is given")]
        private string a_noChoice;

        [SerializeField, Tooltip("Open the barrier")]
        private string a_open;

        [SerializeField, Tooltip("Close the barrier")]
        private string a_close;

        [SerializeField, Tooltip("Stop moving the barrier")]
        private string a_stop;

    [Header("Input logic component names")]

        [SerializeField, Tooltip("The barrier is fully opened")]
        private string s_opened;

        [SerializeField, Tooltip("The barrier is opening")]
        private string s_opening;

        [SerializeField, Tooltip("The barrier has stopped moving")]
        private string s_stopped;

        [SerializeField, Tooltip("The barrier is closing")]
        private string s_closing;

        [SerializeField, Tooltip("The barrier is fully closed")]
        private string s_closed;

    [Header("Settings")]

        [SerializeField, Tooltip("Warning box that shows when multiple actuators are turned on")]
        private GameObject WarningBox;

        [SerializeField, Tooltip("Collider that prevents vehicles from driving through the barrier")]
        private GameObject BarrierCollider;

        [SerializeField, Tooltip("Accepted offset of the motor position sensing [deg]")]
        private float SensorOffsetPos;

    private Actuator_RotationalMotor motor;
    private Vector3 colliderStartPosition;

    [HideInInspector]
    public Dictionary<string, BoolToggle> IO_dict = new Dictionary<string, BoolToggle>();

    private void Start()
    {
        colliderStartPosition = BarrierCollider.transform.position;

        // Components to communicate
        motor = GetComponentInChildren<Actuator_RotationalMotor>();

        // Filling the I/O dictionary

            // Actuators
            IO_dict.Add("a_noChoice", Static_Functions.FindBoolToggle(a_noChoice));
            IO_dict.Add("a_open", Static_Functions.FindBoolToggle(a_open));
            IO_dict.Add("a_close", Static_Functions.FindBoolToggle(a_close));
            IO_dict.Add("a_stop", Static_Functions.FindBoolToggle(a_stop));

            // Sensors
            IO_dict.Add("s_opened", Static_Functions.FindBoolToggle(s_opened));
            IO_dict.Add("s_opening", Static_Functions.FindBoolToggle(s_opening));
            IO_dict.Add("s_stopped", Static_Functions.FindBoolToggle(s_stopped));
            IO_dict.Add("s_closing", Static_Functions.FindBoolToggle(s_closing));
            IO_dict.Add("s_closed", Static_Functions.FindBoolToggle(s_closed));
    }

    private void Update()
    {
        // Check whether multiple actuators are on and activate a red box signal
        if (Static_Functions.CountTrueActuators(IO_dict) > 1)
        {
            WarningBox.SetActive(true);
        }

        // --- Enable events based on the actuator states ---
        if (IO_dict["a_noChoice"].Boolean)
        {
            // Do nothing
        }
        else if (IO_dict["a_open"].Boolean)
        {
            motor.RotationDirection = 1;
        }
        else if (IO_dict["a_close"].Boolean)
        {
            motor.RotationDirection = -1;
        }
        else if (IO_dict["a_stop"].Boolean)
        {
            motor.RotationDirection = 0;
        }
        else
        {
            // Default state
            motor.RotationDirection = 0;
        }

        // --- Set the sensor values based on states of the components ---
        IO_dict["s_opened"].Boolean = 
            Mathf.Abs(motor.OpenRotation - motor.currentRotation) < SensorOffsetPos;
        IO_dict["s_opening"].Boolean =
            Mathf.Abs(motor.ClosedRotation - motor.currentRotation) < SensorOffsetPos;
        IO_dict["s_stopped"].Boolean =
            motor.RotationDirection == 0;
        IO_dict["s_closing"].Boolean =
            motor.RotationDirection == 1;
        IO_dict["s_closed"].Boolean =
            motor.RotationDirection == -1;

        // Move the collider out of the way when the barrier is opened
        if (IO_dict["s_opened"].Boolean)
        {
            BarrierCollider.transform.position = Vector3.Lerp(transform.position, colliderStartPosition 
            	+ Vector3.up * 20, Time.deltaTime * 100);
        }
        else
        {
            BarrierCollider.transform.position = Vector3.Lerp(transform.position, colliderStartPosition, 
            	Time.deltaTime * 100);
        }
    }
}
\end{lstlisting}

\section{CIF3 Toolbox}
    
\begin{lstlisting}[caption={PLC code generation script in CIF.}, label={listing:ST_DT_PLC.cif},basicstyle=\tiny,frame=single]
from "lib:cif3" import *;

// Create directory for generated files.
mkdir("generated_files", force=true);

// Eliminate the event condition requirements.
cif3cif("ST_DT_HW.cif",
    "-o", "generated_files/ST_DT_controller.cif",
    "-t elim-state-evt-excl-invs");
outln("Removed event exclusion invariants.");

// Remove algebraic variables.
cif3cif("generated_files/ST_DT_controller.cif",
    "-o", "generated_files/ST_DT_controller.cif",
    "-t elim-alg-vars,simplify-values");
outln("Removed algebraic variables.");

// Linearize the model.
cif3cif("generated_files/ST_DT_controller.cif",
    "-o", "generated_files/ST_DT_controller.cif",
    "-t linearize-merge,simplify-values");
outln("Linearized specification.");

// Generate PLC code.
outln("PLC code: generating...");
cif3plc("generated_files/ST_DT_controller.cif",
        "-o generated_files/ST_DT_PLC.xml",
        "-t plc-open-xml");
outln("PLC code: generated.");
\end{lstlisting}

\section{Logic Component Scripts}
    
\begin{lstlisting}[caption={\textit{Input\_Bool\_Logic.cs}.}, label={listing:Input_Bool_Logic},basicstyle=\tiny,frame=single]
using System;
using System.Collections.Generic;
using u040.prespective.prelogic;
using u040.prespective.prelogic.component;
using u040.prespective.prelogic.signal;
using UnityEngine;

public class Input_Bool_Logic : PreLogicComponent
{
    #if UNITY_EDITOR || UNITY_EDITOR_BETA
        [HideInInspector] public int toolbarTab;
    #endif

    [Header("Logic and I/O")]

        [Tooltip("BooleanToggle component for the in or output")]
        public BoolToggle BooleanToggle;
        [Tooltip("Input Boolean (used for input to the PLC)")]
        public bool InputBool;

    public override List<SignalDefinition> SignalDefinitions
    {
        get
        {
            return new List<SignalDefinition>
            {
                new SignalDefinition(gameObject.name, PLCSignalDirection.INPUT, SupportedSignalType.BOOL, "", 
                gameObject.name, null, null, false),
            };
        }
    }
    protected override void onSimulatorUpdated(int _simFrame, float _deltaTime, float _totalSimRunTime, 
    	DateTime _simStart)
    {
        readComponent();
    }
    void readComponent()
    {
        if (BooleanToggle.Boolean != InputBool)
        {
            InputBool = BooleanToggle.Boolean;
            WriteValue(gameObject.name, InputBool);
        }
    }
}
\end{lstlisting}    
    
\begin{lstlisting}[caption={\textit{Output\_Bool\_Logic.cs}.}, label={listing:Output_Bool_Logic},basicstyle=\tiny,frame=single]
using System;
using System.Collections.Generic;
using u040.prespective.prelogic;
using u040.prespective.prelogic.component;
using u040.prespective.prelogic.signal;
using UnityEngine;

public class Output_Bool_Logic: PreLogicComponent
{
#if UNITY_EDITOR || UNITY_EDITOR_BETA
    [HideInInspector] public int toolbarTab;
#endif

    [Header("Logic and I/O")]

        [Tooltip("BooleanToggle component for the in or output")]
        public BoolToggle BooleanToggle;
        [Tooltip("Input Boolean (used for input to the PLC)")]
        public bool OutputBool;

    public override List<SignalDefinition> SignalDefinitions
    {
        get
        {
            return new List<SignalDefinition>
            {
                new SignalDefinition(gameObject.name, PLCSignalDirection.OUTPUT, SupportedSignalType.BOOL, "",
                "Value", onSignalChanged, null, false),
            };
        }
    }

    void onSignalChanged(SignalInstance _signal, object _newValue, DateTime _newValueReceived, 
    	object _oldValue, DateTime _oldValueReceived)
    {
        if (_signal.definition.defaultSignalName == gameObject.name) 
        {
            OutputBool = (bool)_newValue;
            if (BooleanToggle.Boolean != OutputBool)
            {
                BooleanToggle.Boolean = OutputBool;
            }
        }
        else 
        {
            Debug.LogWarning("Unknown Signal received:" + _signal.definition.defaultSignalName);
        }
    }
}
\end{lstlisting}

\begin{lstlisting}[caption={\textit{LogicComponentGeneration.cs}.}, label={listing:LogicComponentGeneration},basicstyle=\tiny,frame=single]
using UnityEngine;

public class LogicComponentGeneration : MonoBehaviour
{
    [Header("References")]

        [SerializeField, Tooltip("Prefab object for the actuator logic components")]
        private GameObject actuatorPrefab;
        [SerializeField, Tooltip("Prefab object for the sensor logic components")]
        private GameObject sensorPrefab;
        [SerializeField, Tooltip("Prefab object for the button logic components")]
        private GameObject buttonPrefab;

        [SerializeField, Tooltip("Parent object of the actuator logic components")]
        private Transform actuatorParent;
        [SerializeField, Tooltip("Parent object of the sensor logic components")]
        private Transform sensorParent;
        [SerializeField, Tooltip("Parent object of the button logic components")]
        private Transform buttonParent;

    [Header("Settings")]

        [SerializeField, Tooltip("Option bool true if GUI logic objects also need to be generated")]
        private bool GenerateGUILogicComponents = false;
    
    // Text file paths
    private string outputNames;
    private string inputNames;

    // Arrays of input and output file lines
    private string[] outputList;
    private string[] inputList;

    public void GenerateNewLogicComponents()
    {
        Destroy_Sensors_Actuators_Buttons();
        Generate_Sensors_Actuators_Buttons();
    }

    private void Destroy_Sensors_Actuators_Buttons()
    {
        // Loop through the children of the parents and destroy the child objects
        foreach (Transform child in actuatorParent)
        {
            DestroyImmediate(child.gameObject);
        }

        foreach (Transform child in sensorParent)
        {
            DestroyImmediate(child.gameObject);
        }

        foreach (Transform child in buttonParent)
        {
            DestroyImmediate(child.gameObject);
        }

        // Check if all child objects are destroyed and run the function again if this is not the case
        if (!(actuatorParent.childCount == 0 && sensorParent.childCount == 0 && buttonParent.childCount == 0))
        {
            Destroy_Sensors_Actuators_Buttons();
        }
    }

    private void Generate_Sensors_Actuators_Buttons()
        {

        // Find the file paths and read in the name lists
        outputNames = System.IO.Path.Combine(Application.dataPath, "_outputNames.txt");
        inputNames = System.IO.Path.Combine(Application.dataPath, "_inputNames.txt");
        outputList = System.IO.File.ReadAllLines(outputNames);
        inputList = System.IO.File.ReadAllLines(inputNames);

        // Create the actuators, sensors and buttons

        foreach (string outputLine in outputList)
        {
            // varName is the first 'word' on the the line
            string varName = outputLine.Trim().Split((" ").ToCharArray()[0])[0];

            // Instantiate an actuator if the line contains a correct variable name of the correct type
            if (varName.StartsWith("dvar") && !outputLine.Contains("enum_E") && outputLine.Contains("BOOL"))
            {
                if (varName.Contains("GUI"))
                {
                    if (GenerateGUILogicComponents)
                    {
                        InstantiateLogicComponent(varName, buttonPrefab, buttonParent);
                    }
                }
                else
                {
                    InstantiateLogicComponent(varName, actuatorPrefab, actuatorParent);
                }
            }
        }

        Debug.Log("Outputs Generated");

        foreach (string inputLine in inputList)
        {
            // varName is the first 'word' on the the line
            string varName = inputLine.Trim().Split((" ").ToCharArray()[0])[0];

            // Instantiate an actuator if the line contains a correct variable name
            if (varName.StartsWith("ivar"))
            {
                // Decide it the variable name represents a button or sensor input
                if (varName.Contains("button"))
                {
                    if (GenerateGUILogicComponents)
                    {
                        InstantiateLogicComponent(varName, buttonPrefab, buttonParent);
                    }
                }
                else
                {
                    InstantiateLogicComponent(varName, sensorPrefab, sensorParent);
                }
            }
        }

        Debug.Log("Inputs Generated");
    }

    void InstantiateLogicComponent(string ComponentName, GameObject PrefabToInstantiate, 
    	Transform ComponentParent)
    {
        // Instantiate the logic object, set the desired name and assign it to the desired parent
        GameObject newLogicComponent = Instantiate(PrefabToInstantiate, Vector3.zero, Quaternion.identity);
        newLogicComponent.name = ComponentName;
        newLogicComponent.transform.parent = ComponentParent;
    }
}

\end{lstlisting}

\begin{lstlisting}[caption={\textit{LogicComponentGeneration\_Editor.cs}.}, label={listing:LogicComponentGeneration_Editor},basicstyle=\tiny,frame=single]
using UnityEngine;
using UnityEditor;

[CustomEditor(typeof(LogicComponentGeneration))]
public class LogicComponentGeneration_Editor : Editor
{
    public override void OnInspectorGUI()
    {
        DrawDefaultInspector();

        LogicComponentGeneration myScript = (LogicComponentGeneration)target;
        if (GUILayout.Button("Generate new logic components"))
        {
            myScript.GenerateNewLogicComponents();
        }
    }
}
\end{lstlisting}
    
\begin{lstlisting}[caption={\textit{Static\_Functions.cs}.}, label={listing:Static_Functions},basicstyle=\tiny,frame=single]
using System.Collections.Generic;
using UnityEngine;

public class Static_Functions : MonoBehaviour
{
    public static int CountTrueActuators(Dictionary<string, BoolToggle> IO_dict)
    {
        int value = 0;

        foreach (var dict_entry in IO_dict)
        {
            if (dict_entry.Key.StartsWith("a") && dict_entry.Value.Boolean)
            {
                value++;
            }
        }

        return value;
    }

    public static BoolToggle FindBoolToggle(string IOName)
    {
        if (IOName.StartsWith("dvar"))
        {
            // The variable is an output
            Transform actuatorParent = GameObject.Find("ActuatorLogicComponents").transform;
            return actuatorParent.Find(IOName).GetComponent<BoolToggle>();
        }
        else if (IOName.StartsWith("ivar"))
        {
            // The variable is an input
            if (IOName.Contains("button"))
            {
                // The variable is a button input
                Transform buttonParent = GameObject.Find("ButtonLogicComponents").transform;
                return buttonParent.Find(IOName).GetComponent<BoolToggle>();
            } 
            else 
            {
                // The variable is a sensor input
                Transform sensorParent = GameObject.Find("SensorLogicComponents").transform;
                return sensorParent.Find(IOName).GetComponent<BoolToggle>();
            }
        }
        else
        {
            Debug.LogWarning(IOName + " is not an input or output variable");
            return null;
        }
    }
}

\end{lstlisting}
    
    \section{Behavior Scripts}
\begin{lstlisting}[caption={\textit{vehicle\_behavior.cs}.}, label={listing:vehicle_behavior},basicstyle=\tiny,frame=single]
using UnityEngine;
using System.Collections;

public class vehicle_behavior : MonoBehaviour
{

    [Header("Driving Properties")]

        [SerializeField, Tooltip("Maximum speed [km/h]")]
        private float max_speed = 1f;

        [SerializeField, Tooltip("Acceleration of the vehicle [km/h/s]")]
        private float acc = 1f;

        [SerializeField, Tooltip("Braking decelleration of the vehicle [km/h/s]")]
        private float dec = 1f;

        [SerializeField, Tooltip("Tag of the collider that destroys this vehicle")]
        private string DestroyTag = "vehicle_destroy";

    private float speed = 0;
    private bool driving = true;

    private void Start() 
    {
        // Start with half the maximum speed
        speed = max_speed/2;
    }

    private void FixedUpdate() 
    {
        // Determine whether to accelerate or decellerate
        if(driving){
            if(speed < max_speed){
                speed += acc/3.6f*Time.fixedDeltaTime;
            }
        } else {
            if(speed > 0){
                speed -= dec/3.6f*Time.fixedDeltaTime;
            }
        }

        // Prevent the vehicle from going backwards
        if(speed < 0){
            speed = 0;
        }

        // Move forward with the current speed
        transform.Translate(speed / 3.6f * Vector3.forward * Time.fixedDeltaTime, Space.Self);
    }

    private void OnTriggerEnter(Collider other)
    {
        // Check if there is an obstacle in front of the vehicle
        if (other.gameObject.CompareTag("vehicle_stop"))
        {
            driving = false;
        }

        // Check if the vehicle destroyer is in front of the vehicle
        if (other.gameObject.CompareTag(DestroyTag))
        {
            // Warp the vehicle such that it exits all triggers before being destroyed
            StartCoroutine(WarpAndDestroy());
        }
    }

    private void OnTriggerExit(Collider other)
    {
        // Check if there is an obstacle in front of the vehicle
        if (other.gameObject.CompareTag("vehicle_stop"))
        {
            driving = true;
        }
    }

    public void WarpAndDestroyVehicle()
    {
        // Warp the vehicle such that it exits all triggers before being destroyed
        StartCoroutine(WarpAndDestroy());
    }

    IEnumerator WarpAndDestroy()
    {
        max_speed = 10000;
        speed = 10000;
        yield return new WaitForSeconds(0.5f);
        Destroy(this.gameObject);
    }
}

\end{lstlisting}

\begin{lstlisting}[caption={\textit{vehicle\_spawner.cs}.}, label={listing:vehicle_spawning_script},basicstyle=\tiny,frame=single]
using UnityEngine;

public class vehicle_spawner : MonoBehaviour
{
    [Header("Vehicle Prefab Objects")]

        [SerializeField, Tooltip("Array of vehicles that this spawner generates")]
        private GameObject[] spawn_vehicles;

    [Header("Spawner Settings")]

        [SerializeField, Tooltip("Minimum time between spawining vehicles [s]")]
        private float t_inter_min = 2;

        [SerializeField, Tooltip("Maximum time between spawining vehicles [s]")]
        private float t_inter_max = 6;

        [SerializeField, Tooltip("Unoccupied distance in front of the spawner needed for spawning 
        	a new vehicle [m]")]
        private float min_spawn_dist;

    private float timer;
    private GameObject new_vehicle;
    private float last_dist = 50;
    [HideInInspector]
    public bool SpawnerOn = false;

    private void Start()
    {
        // Set the timer to a random value in the bounds
        timer = Random.Range(t_inter_min, t_inter_max);
    }

    private void Update()
    {
        timer -= Time.deltaTime;

        // Determine whether to span a new gameobject
        if (timer < 0 && last_dist > min_spawn_dist && SpawnerOn)
        {
            timer = Random.Range(t_inter_min, t_inter_max);
            Spawn_vehicle();
        }

        // Obtain the distance from the spawner to the vehicle that was spawned last
        // If the spawner is off, set the last distance to enable spawning when it starts again
        if (SpawnerOn)
        {
            last_dist = Vector3.Magnitude(transform.position - new_vehicle.transform.position);
        }
        else
        {
            last_dist = min_spawn_dist + 1;
        }
    }

    private void Spawn_vehicle()
    {
        // Spawn a random vehicle from the vehicles array
        new_vehicle = Instantiate(spawn_vehicles[Random.Range(0, spawn_vehicles.Length)], 
        	transform.position, transform.rotation);
        new_vehicle.transform.parent = transform.parent;
    }
}

\end{lstlisting}
    
\begin{lstlisting}[caption={\textit{Actuator\_RotationMotor.cs}.}, label={listing:Actuator_RotationMotor},basicstyle=\tiny,frame=single]
using UnityEngine;

public class Actuator_RotationalMotor : MonoBehaviour
{
    [Header("Settings")]

        [SerializeField, Tooltip("Rotational velocity [deg/s] around local x-axis")]
        private float rotVel;

        [Tooltip("Rotation in the opened state [deg]")]
        public float OpenRotation;

        [Tooltip("Rotation in the closed state [deg]")]
        public float ClosedRotation;

        [Tooltip("True: initial state is opened")]
        public bool InitialOpen;

        [Tooltip("Horizontal direction vector")]
        public Vector3 horizontalDir;

    // Rotation direction, 1: forward, 0: stopped, -1:backward
    [HideInInspector]
    public int RotationDirection = 0;
    //[HideInInspector]
    public float currentRotation;
    private Vector3 SignedAngleAxis;

    private void Start()
    {
        SignedAngleAxis = Vector3.Cross(horizontalDir, transform.forward);
    }

    private void FixedUpdate()
    {
        currentRotation = Vector3.SignedAngle(horizontalDir, transform.forward, SignedAngleAxis);

        if (RotationDirection == 1 && currentRotation < OpenRotation)
        {
            transform.Rotate(Vector3.right * rotVel * RotationDirection * Time.fixedDeltaTime, Space.Self);
        }
        else if(RotationDirection == -1 && currentRotation > ClosedRotation)
        {
            transform.Rotate(Vector3.right * rotVel * RotationDirection * Time.fixedDeltaTime, Space.Self);
        }

    }
}
\end{lstlisting}

\chapter{Test Results and Demo Videos}\label{sec:test_results}

The links below lead to demo videos in which parts of the validation process of the digital twin are shown. This includes supervisory controller tests with boom barriers, height detection system and pumping cellars and demonstration of the CCTV feeds and the direct interaction options with doors and aid closets. In the controller tests, the functioning of the test GUI is also demonstrated.

    \vspace{3mm}
    \begin{enumerate}
        \item \textbf{Boom barriers (\color{blue}\underline{\href{https://youtu.be/M3vwjcTznH4}{video link}}\color{black}):} a demonstration of the boom barriers in a traffic tube that interact with traffic. A scenario is simulated where a car is stationary below one of the boom barriers at an entrance of the tunnel.
        \item \textbf{Height detection system (\color{blue}\underline{\href{https://youtu.be/ncSWChP2crc}{video link}}\color{black}):} a demonstration of the height detection system. A scenario is shown where a truck drives towards the tunnel while this truck is too high to be allowed to enter. This triggers the height detection system that needs to be reset manually after the high truck is removed.
        \item \textbf{CCTV feeds (\color{blue}\underline{\href{https://youtu.be/BCNfd0UhHiY}{video link}}\color{black}):} a demonstration of the CCTV feeds. Here, the effect of the lower frame rate as described in Chapter 6 is shown. A camera object in the scene is also highlighted, which can easily be moved around in the scene to show other camera angles.
        \item \textbf{Direct interaction (doors and aid closets) (\color{blue}\underline{\href{https://youtu.be/YJOJv9LK_QA}{video link}}\color{black}):} a demonstration where doors and aid cabinets are opened by clicking them. The activation of elements inside the aid cabinets is also demonstrated, after which it can be noticed that the lights in the tunnel get a bit brighter.
        \item \textbf{Pumping cellars (\color{blue}\underline{\href{https://youtu.be/KMbX_VepdRQ}{video link}}\color{black}):} a demonstration of a pumping cellar scenario. Here, the effect of heavy rain in the tunnel is simulated by filling the pumping cellar. Two modes are demonstrated; one where the water level is kept low and one where the cellar is filled to its full capacity.
    \end{enumerate}

\end{document}